\documentclass[draftclsnofoot, onecolumn]{IEEEtran}

\usepackage{graphicx}
\usepackage{amsmath}
\usepackage{amssymb}
\usepackage{graphics}
\usepackage{latexsym}
\usepackage[dvips]{epsfig}
\usepackage[nospace]{cite}
\usepackage{subfigure}
\usepackage{setspace}
\usepackage{color}
\usepackage{tabularx}

\newtheorem{Definition}{Definition}
\newtheorem{Lemma}{Lemma}
\newtheorem{Corollary}[Lemma]{Corollary}
\newtheorem{Proposition}[Lemma]{Proposition}
\newtheorem{Theorem}{Theorem}

\newtheorem{Remark}{Remark}

\allowdisplaybreaks[1]

\def\Pr{{\mathrm{Pr}}}
\def\E{{\mathbb E}}
\def\Var{{\mathrm {Var}}}

\DeclareMathOperator*{\argmin}{arg\,min}

\begin{document}
%
\title{\huge On Achievable Rates of AWGN Energy-Harvesting Channels with Block Energy Arrival and Non-Vanishing Error Probabilities}
%
%
%


\author{Silas~L.~Fong, Vincent~Y.~F.~Tan, and Ayfer \"{O}zg\"{u}r
\thanks{S.~L.~Fong is with the Department of Electrical and Computer Engineering, National University of Singapore, Singapore 117583 (e-mail: \texttt{silas\_fong@nus.edu.sg}).}%
\thanks{ V.~Y.~F.~Tan is with the Department of Electrical and Computer Engineering, National University of Singapore, Singapore 117583, and also with the Department of Mathematics, National University of Singapore, Singapore 119076 (e-mail: \texttt{vtan@nus.edu.sg}).}%
\thanks{A.~\"{O}zg\"{u}r is with the Department of Electrical Engineering, Stanford University, CA 94305, USA (email: \texttt{aozgur@stanford.edu}).}%
}

\maketitle

\begin{abstract}
This paper investigates the achievable rates of an additive white Gaussian noise (AWGN) energy-harvesting (EH) channel with an infinite battery. The EH process is characterized by a sequence of blocks of harvested energy, which is known causally at the source. The harvested energy remains constant within a block while the harvested energy across different blocks is characterized by a sequence of independent and identically distributed (i.i.d.) random variables. The blocks have length~$L$, which can be interpreted as the coherence time of the energy arrival process.
If $L$ is a constant or grows sublinearly in the blocklength~$n$, we fully characterize the first-order term in the asymptotic expansion of the maximum transmission rate subject to a fixed tolerable error probability $\varepsilon$. The first-order term is known as the $\varepsilon$-capacity. In addition, we obtain lower and upper bounds on the second-order term in the asymptotic expansion, which reveal that the second order term scales as $\sqrt{\frac{L}{n}}$ for any $\varepsilon$ less than $1/2$.
The lower bound is obtained through analyzing the save-and-transmit strategy. If $L$ grows linearly in~$n$, we obtain lower and upper bounds on the $\varepsilon$-capacity, which coincide whenever the cumulative distribution function (cdf) of the EH random variable is continuous and strictly increasing. In order to achieve the lower bound, we have proposed a novel adaptive save-and-transmit strategy, which chooses different save-and-transmit codes  across different blocks according to the energy variation across the blocks.
\end{abstract}


\IEEEpeerreviewmaketitle

\flushbottom

\section{Introduction} \label{Introduction}
The energy-harvesting (EH) channel consists of one source equipped with an energy buffer (also called battery), and one destination.  For simplicity, we assume in this paper that the buffer has infinite capacity. At each discrete time $k \in \{1,2,\ldots\}$, a random amount of energy $E_k\in [0, \infty)$ arrives at the buffer and the source transmits a symbol $X_k\in (-\infty, \infty)$ such that
 \begin{equation}
 \sum_{i=1}^k X_i^2 \le \sum_{i=1}^k E_i  \qquad\mbox{almost surely}. \label{EHResultIntro}
 \end{equation}
 This implies that the total harvested energy $\sum_{i=1}^k E_i$ must be no smaller than the energy of the codeword $\sum_{i=1}^k X_i^2$  at every discrete time $k$ for transmission to take place successfully. The knowledge of~$E_k$ is available at the source at time~$k$ before encoding $X_k$, and the destination has no access to the energy-arrival process.
\\ \indent We assume that $\{E_i\}_{i=1}^\infty$ arrive at the buffer in a block-by-block manner as follows: For each $\ell\in\mathbb{N}$, let
 \begin{equation}
 b_\ell \triangleq (\ell-1)L \label{blockIndex}
 \end{equation}
such that $b_\ell+1$ is the index of the first channel use within the $\ell^{\text{th}}$ block of energy arrival, where~$L$ denotes the length of each block. In other words, the $\ell^{\text{th}}$ block of energy arrival starts at the $(b_\ell+1)^{\text{th}}$ channel use. The EH random variables that mark the starting points of the blocks (i.e., $\{E_{b_\ell+1}\}_{\ell=1}^\infty$) are assumed to be independent and identically distributed (i.i.d.) random variables where\footnote{If the constraint $\E[E_1^3]<+\infty$ is replaced with the less stringent one $\E[E_1^2]<+\infty$, all the achievability results in this paper continue to hold. In fact, the only place that requires $\E[E_1^3]<+\infty$ is the use of the Berry-Ess\'een theorem in Section~\ref{subsecLemmaConverse} in the course of proving the converse of Theorem~\ref{thmMainResult}.} $\E[E_1^3]<+\infty$ and $\E[E_1]=P$ for some $P>0$. A large class of distributions of practical interests satisfy the third-moment assumption including those with well-defined moment generating function. In addition, we assume
 \begin{equation}
E_{b_\ell+1}=E_{b_\ell+2}=\ldots = E_{b_{\ell}+L} \label{assumptionEH}
 \end{equation}
 for all $\ell\in\mathbb{N}$. In other words, the harvested energy in each channel use within a block remains constant while the harvested energy across different blocks is characterized by a sequence of i.i.d.\ random variables with mean equal to~$P$. This block-by-block energy-arrival assumption is useful for modeling practical scenarios when the energy-arrival process evolves at a slower timescale compared to the transmission process \cite[Sec.~II-C]{ZhangLau14}. This is often the case for most natural energy processes, such as solar energy or wind energy. The block i.i.d.\ EH process can model, for example, a solar panel which harvests energy from the sun, and the appearance
of clouds can change randomly and block certain amounts
of sunshine for a certain period of time. Similarly, this is a
good model for a device which harvests RF energy from other
transmitting devices in its environment. Such transmitting
devices typically transmit continuously for certain periods of
time and are silent for the remaining periods (as in TDMA,
for example), which warrants a block i.i.d.\ model. Most importantly, the block i.i.d.\
model provides a simple way to study the impact of correlations in the EH process on the system capacity. Such block i.i.d. models are  popularly used in wireless communication as a means to
capture correlations in the channel fading process by a simple model.
In that context, the block length $L$~is called the coherence time of the channel,
which corresponds to the time duration over which the channel
remains approximately constant~\cite[Sec.~2.3]{davidTseBook}. Analogously, we refer to~$L$ as the coherence time of the energy arrival process.

 The channel noise of the EH channel is modeled as an \emph{additive white Gaussian noise (AWGN)}, which is described as follows. In each time slot $k\in\mathbb{N}$, after the source has transmitted $X_k$, the destination receives
\begin{equation}
 Y_k = X_k + Z_k
 \end{equation}
where $\{Z_k\}_{k=1}^{\infty}$ are i.i.d.\ standard normal random variables. The above EH channel is referred to as the \textit{AWGN EH channel}. It was shown by Ozel and Ulukus~\cite{ozel12} that the capacity of the AWGN EH channel for the case $L=1$ is
 \begin{equation}
\mathrm{C}(P)\triangleq\frac{1}{2}\log (1+P) \qquad\mbox{bits per channel use}, \label{eqn:capacity_awgn}
 \end{equation}
 where $P=\E[E_1]$ is the expectation of the harvested energy for each energy arrival.
 In this paper, we assume that $L$ can grow with~$n$ and would like to investigate how the growth rate of~$L$ affects the first- and second-order terms of the asymptotic expansion of the maximum transmission rate. The first-order term is also known as the $\varepsilon$-capacity \cite[Sec.~3.4]{Han10}, and the second-order term divided by an appropriate scaling (which is $\frac{1}{\sqrt{n}}$ in many cases including the AWGN channel) is known as the second-order coding rate \cite{Hayashi09}. The following two cases regarding the growth rate of~$L$ will be investigated in this paper:
  \begin{enumerate}
  \item[(i)] $L$ is a constant or $L$ grows sublinearly in~$n$. The latter statement means
   \begin{equation}
\omega(1)= L = o(n). \label{LsublinearN}
 \end{equation}
 \item[(ii)] $L$ grows linearly in~$n$.
  \end{enumerate}
\indent Note that in practice $L$ and $n$ are two independent parameters. The first one is dictated by the nature of the EH process and the second one is a design parameter typically dictated by the delay and reliability requirements of the application and the complexity constraints at the transmitter and the receiver. Depending on how fast the EH process changes over time, $L$ can be significantly smaller than $n$ or comparable to $n$. In order to reveal the impact of the interplay between these two parameters on the second-order term, we couple these parameters in different ways, say $L=n^\gamma$, and study the limiting case when both $L$ and~$n$ approach $\infty$ for different couplings, i.e., different values of $\gamma$ in $[0,1]$ where $\gamma$ captures how large $L$ is with respect to $n$. This allows us to identify how $L$ and $n$ together impact the second-order term. In particular, we conclude that it is the ratio of the two that determines the second-order term. The impact of correlation in the EH process can be interpreted as effectively decreasing the blocklength by a factor of $L$. Note that keeping $L$ constant while taking $n$ to infinity, i.e., considering only the special case $\gamma=0$,  would lead to a degenerate regime where the correlation in the EH process does not play a significant role. The approach we take here has been extensively used in the wireless information theory literature to obtain asymptotic results in multi-parameter problems where the problem involves multiple independent parameters that can be large or small with respect to each other. (See for example the notion of generalized degrees of freedom in \cite{DavidIC} and follow-up work, or Section 3.1 of \cite{Ayfermono} for a detailed discussion of a similar formulation in the context of scaling laws for wireless networks.)

 \subsection{Main Contribution} \label{sectionContribution}
We use $O(\cdot)$, $o(\cdot)$, $\omega(\cdot)$ and $\Theta(\cdot)$ denote standard asymptotic Bachmann-Landau notations except our convention that they must be non-negative. The contributions of this paper are summarized in the following:
\\
\begin{figure}
\centering
\includegraphics[width=3.7 in, height=1.6 in, bb = 37 126 837 466, angle=0, clip=true]{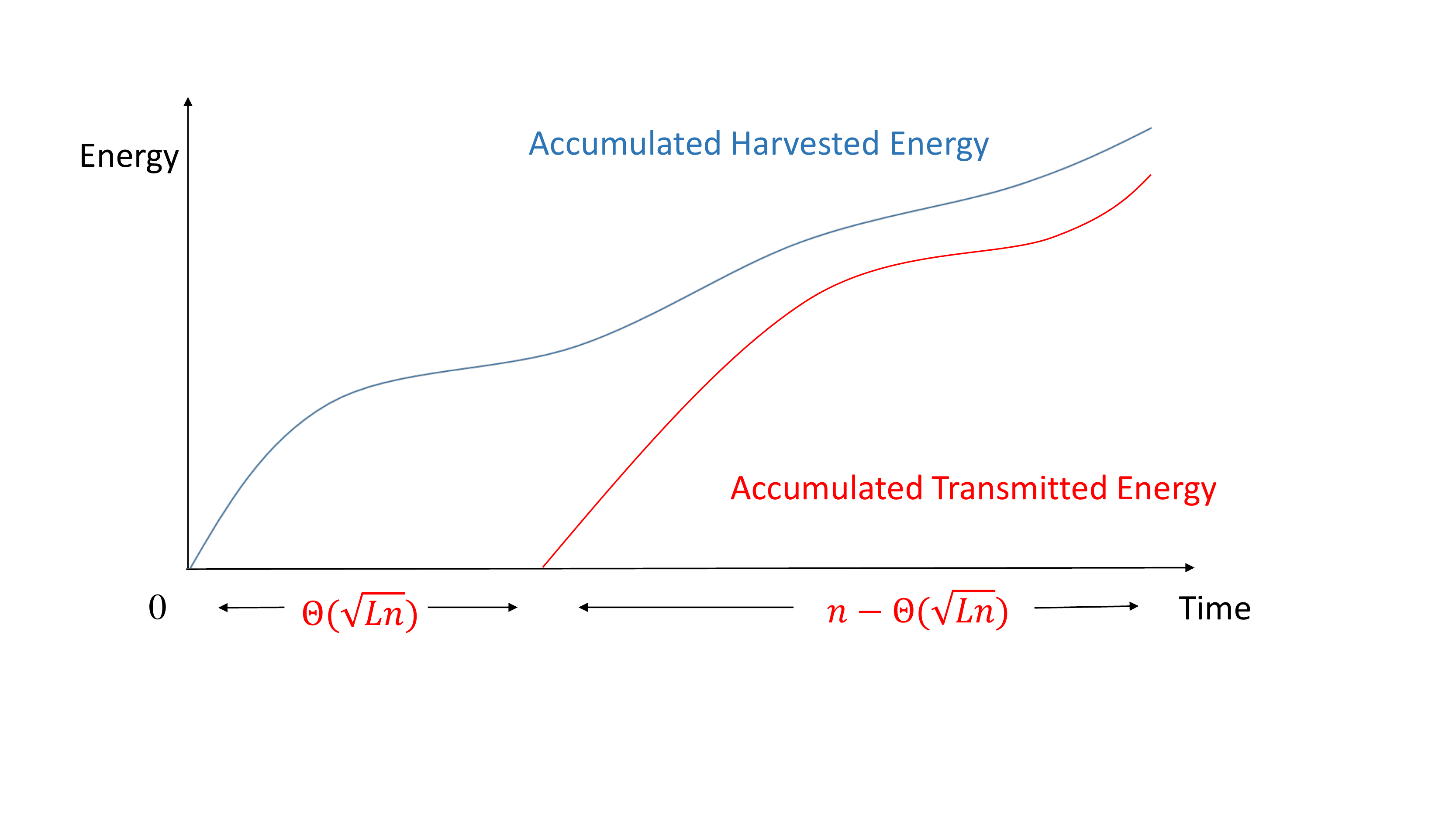}
\caption{Save-and-transmit strategy for $L=o(n)$.}
\label{figSaveAndTransmit}
\end{figure}
\textbf{Case (i): When $L$ is a constant or grows sublinearly in~$n$}
\begin{enumerate}
\item[1.] First, we prove an achievable finite blocklength bound based on the save-and-transmit strategy of~\cite{ozel12}. During the saving phase of the save-and-transmit strategy, energy is saved for a certain number of time slots. During this period, no information is transmitted. Subsequently, during the transmission phase, the source uses a Gaussian codebook to send information. In order to analyze the performance of this save-and-transmit strategy, we construct a single sequence of random variables that characterizes the probability of the available energy being insufficient to support the Gaussian codeword (i.e., $\sum_{i=1}^k E_i < \sum_{i=1}^k X_i^2$ for all $k$) and derive a concentration bound related to the random sequence. Our analysis reveals that the backoff from capacity~$\mathrm{C}(P)$ for the optimal length-$n$ code with error probability less than~$\varepsilon$ is no larger than $O\Big(\sqrt{\frac{L}{n}}\Big)$. More specifically, the maximum alphabet size of the message we can transmit over $n$ channel uses with average probability of error no larger than $\varepsilon\in(0,1/2)$, denoted by $M_{n,\varepsilon}^*$, satisfies
\begin{equation}
\frac{1}{n}\log M_{n,\varepsilon}^* \ge \mathrm{C}(P) + V_\varepsilon^-\sqrt{\frac{L}{n}} - o\bigg(\sqrt{\frac{L}{n}}\bigg) \label{introResult1}
\end{equation}
where $V_\varepsilon^- < 0$ is some constant that does not depend on~$n$.
 We also identify the implied constant $V_\varepsilon^-$ in Theorem~\ref{thmMainResult} in Section~\ref{sectionMainResult}. The qualitative interpretation of $V_\varepsilon^-$ will be given in Remark~\ref{remark1} in Section~\ref{subsecRemark}. The lower bound~\eqref{introResult1} is obtained by choosing the lengths of the saving phase and transmission phase to be $\Theta(\sqrt{Ln})$ and $n-\Theta(\sqrt{Ln})$ respectively, which is illustrated in Figure~\ref{figSaveAndTransmit} where the accumulated harvested energy is always above the accumulated transmitted energy due to the EH constraints~\eqref{EHResultIntro}.

\item[2.] Second, we prove a non-asymptotic upper bound on achievable rates by simplifying the type-II error of a carefully chosen binary hypothesis test. The first-order term of the upper bound is~$\mathrm{C}(P)$ and the second-order term is proportional to $-\sqrt{\frac{L}{n}}$ for all $\varepsilon \in (0,1/2)$. More specifically, for all $\varepsilon\in(0,1/2)$, we have
\begin{equation}
\frac{1}{n}\log M_{n,\varepsilon}^* \le \mathrm{C}(P) + V_\varepsilon^+\sqrt{\frac{L}{n}} + o\bigg(\sqrt{\frac{L}{n}}\bigg) \label{introResult2}
\end{equation}
where $V_\varepsilon^+ <0$ is some constant that does not depend on~$n$.  We also identify the implied constant $V_\varepsilon^+$ in Theorem~\ref{thmMainResult} in Section~\ref{sectionMainResult}. The qualitative interpretation of $V_\varepsilon^+$ will be given in Remark~\ref{remark2} in Section~\ref{subsecRemark}. Note that \eqref{introResult1} and \eqref{introResult2} together reveal that the back-off from $\mathrm{C}(P)$ for the optimal length-$n$ code with error probability less than~$\varepsilon$ is of the order $\sqrt{\frac{L}{n}}$. Therefore, the impact of correlation in the EH process can be interpreted as effectively decreasing the blocklength by a factor of $L$. In other words, to achieve the same reliability, one needs to increase the blocklength by a factor equal to the coherence time of the EH process.
\end{enumerate}
It is readily seen from~\eqref{introResult1} and~\eqref{introResult2} that for any fixed $\varepsilon\in(0,1/2)$, the $\varepsilon$-capacity is~$\mathrm{C}(P)$ and the second-order term in the asymptotic expansion for the maximum achievable rate is proportional to $- \sqrt{\frac{L}{n}}$. In addition, define
\begin{equation}
V_\varepsilon \triangleq \sup\left\{S\in\mathbb{R}\left|\,
\liminf_{n\rightarrow \infty}\,\frac{\log M_{n,\varepsilon}^* - n \mathrm{C}(P)}{\sqrt{Ln}} \ge S
\right.\right\}
\end{equation}
to be the second-order coding rate \cite{Hayashi09}. We can see from~\eqref{introResult1} and~\eqref{introResult2} that for any fixed $\varepsilon\in(0,1/2)$, the second-order coding rate is sandwiched between $V_\varepsilon^-$ and $V_\varepsilon^+$.\\
\begin{figure}
\centering
\includegraphics[width=3.7 in, height=1.6 in, bb = 37 116 837 466, angle=0, clip=true]{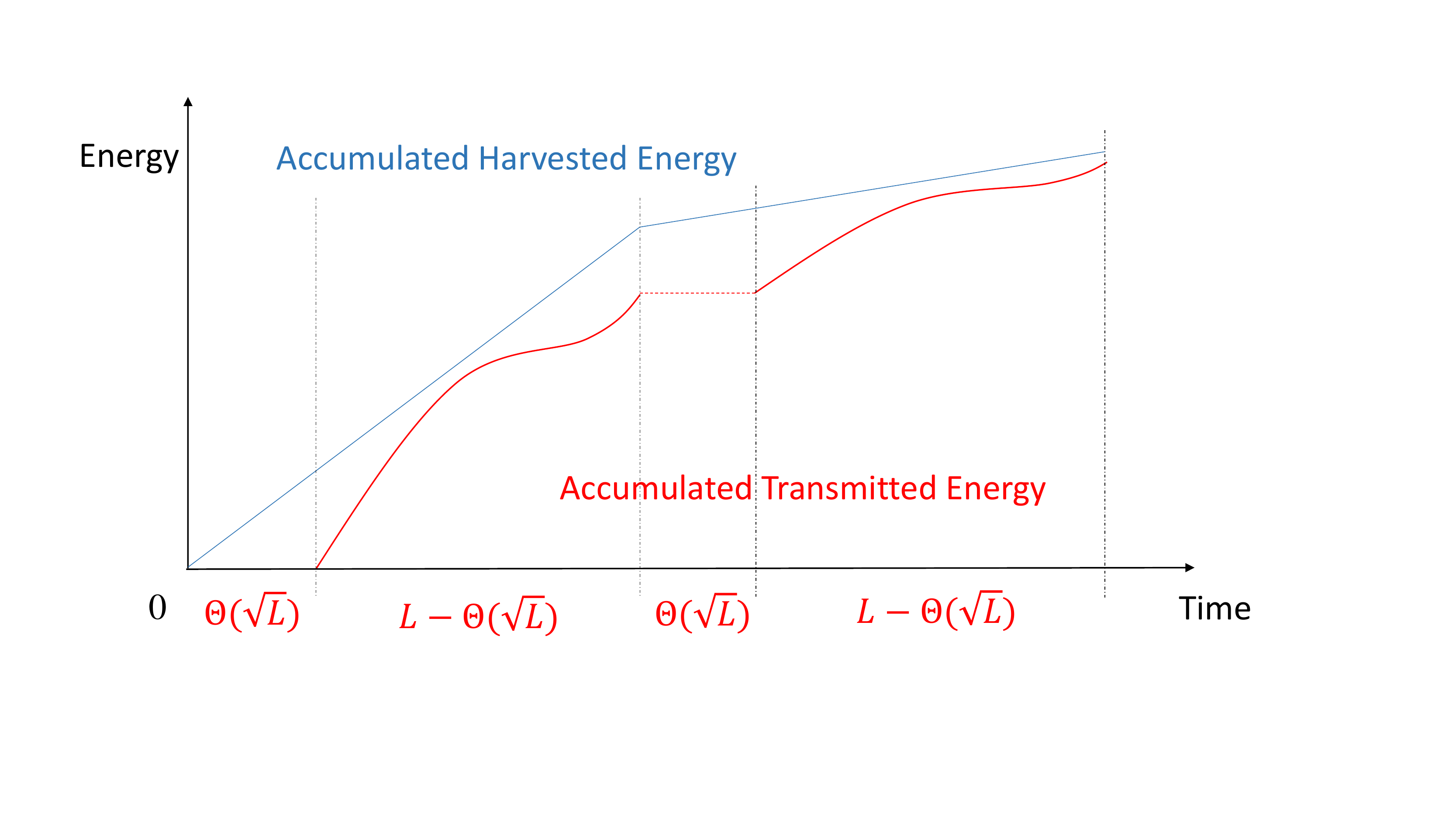}
\caption{Adaptive save-and-transmit strategy for $L=n/2$.}
\label{figAdaptiveSaveAndTransmit}
\end{figure}
\textbf{Case (ii): When $L$ grows linearly in~$n$}
\begin{enumerate}
\item[1.] First, we prove a lower bound on the $\varepsilon$-capacity, as shown in Theorem~\ref{thmArrivalBlockLinearInN} in Section~\ref{sectionMainResult}, based on a modified version of the save-and-transmit strategy called the \textit{adaptive save-and-transmit strategy}. Under the adaptive save-and-transmit strategy which is described in Section~\ref{subsecAchProofLinearIn}, different save-and-transmit codes are used across different blocks. In each block~$\ell$, the coding rate is adapted to the corresponding EH random variable $E_{b_\ell+1}$ so that it is close to $\mathrm{C}(E_{b_\ell+1})$. In addition, the lengths of the saving phase and transmission phase for block~$\ell$ are chosen to be $\Theta(\sqrt{L})$ and $L-\Theta(\sqrt{L})$ respectively as illustrated in Figure~\ref{figAdaptiveSaveAndTransmit}.
    \item[2.] 
        Second, we prove an upper bound on the $\varepsilon$-capacity (Theorem~\ref{thmArrivalBlockLinearInN}). We do so by considering a typical set of sequences of EH random variables followed by simplifying the type-II error of a binary hypothesis test conditioned on the aforementioned typical set.
\end{enumerate}
For any EH process whose EH random variable has a continuous and strictly increasing cumulative density function (cdf), the upper and lower bounds in Theorem~\ref{thmArrivalBlockLinearInN} coincide and hence the $\varepsilon$-capacity is fully characterized. See Remark~\ref{remark3} in Section~\ref{subsecRemark} for a detailed discussion. Case~(ii) is useful for modeling the scenario where the energy-harvesting rate changes slowly such that the number of energy-arrival blocks stays constant as~$n$ increases. Since the number of energy-arrival blocks stays constant and the length of each energy-arrival block grows with~$n$, it is first-order optimal to choose an appropriate save-and-transmit scheme that achieves the maximum coding rate for each block according to the energy level in that block. Therefore, we need an adaptive save-and-transmit scheme rather than the conventional non-adaptive one to achieve the overall maximum coding rate.
\subsection{Related Work}
The channel capacity was characterized for the AWGN channel with an i.i.d.\ EH process in~\cite{ozel12} and with a stationary ergodic EH process in~\cite{RSV14}. The aforementioned studies showed that with an unlimited battery, the capacity of the AWGN channel with stochastic energy constraints is equal to the capacity of the same channel under an average power constraint as long as the average power equals the average recharge rate of the battery. In this paper, we focus on the AWGN channel with a block EH process, where the energy arrivals remain constant for a block of duration $L$ and are independent across blocks drawn from an arbitrary distribution. A similar block i.i.d.\ EH model has been recently considered in \cite{shavivOzgur16-1, ShavivOzgur16-2} concurrently with the current paper. However, these papers focus on the power control problem for EH communications with finite battery at the transmitter. In this paper, we rather consider the information-theoretic capacity of the channel and with infinite battery at the transmitter. Characterizing the information theoretic capacity of the channel with a finite battery is known to be a difficult problem even for an i.i.d.\ model for the energy arrivals and in general remains an open problem. It has been studied in several recent works~\cite{MaoHassibi2013,Jog:ISIT:2014,SNOzgur2016, SOzgur2017} and the most recent ones~\cite{SNOzgur2016, SOzgur2017} characterize the capacity within a constant gap. Due to the lack of a complete characterization of the capacity under a finite battery assumption, in this paper we focus on the AWGN EH channel with infinite battery and develop bounds on the first- and second-order terms in the asymptotic expansion of the maximum transmission rate.

For a fixed tolerable error probability~$\varepsilon$, Fong et al.~\cite{FTY15} recently performed a finite blocklength analysis of save-and-transmit schemes proposed in~\cite{ozel12} and obtained a non-asymptotic achievable rate for the AWGN channel with an i.i.d.\ EH process. The first-, second- and third-order terms of the non-asymptotic achievable rate presented in~\cite[Th.~1]{FTY15} are equal to the capacity, $-c_1\sqrt{\frac{\log n}{n}}$ and $-c_2\sqrt{\frac{2+\varepsilon}{n \varepsilon}}$ respectively where $c_1$ and $c_2$ are some positive constants that do not depend on~$n$ and~$\varepsilon$. Subsequently, Shenoy and Sharma~\cite{ShenoySharma16} refined the analysis in~\cite{FTY15} and improved the second-order term to $-\frac{c_3}{\sqrt{n \varepsilon}}$ where $c_3$ is some positive constant that does not depend on~$n$ and~$\varepsilon$. 
This paper further improves the second-order term to $-c_4\sqrt{\frac{\log(1/\varepsilon)}{n}}$ for any $\varepsilon\in(0,1/2)$ where $c_4$ is some positive constant that does not depend on~$n$ and $\varepsilon$ (see Remark~\ref{remark1*}). The aforementioned improvements are due to better analyses of the ``energy outage" probability for the same save-and-transmit strategy, where the ``energy outage" occurs when the source cannot output the desired codeword due to energy shortage.

\subsection{Paper Outline}
This paper is organized as follows. The notation used in this paper is described in the next subsection. Section~\ref{sectionDefinition} states the formulation of the AWGN EH channel with block energy arrival. Section~\ref{sectionMainResult} presents our two main results --- the first result fully characterizes the $\varepsilon$-capacity and provides lower and upper bounds on the second-order coding rate when $L$ is a constant or grows sublinearly in~$n$; the second result presents lower and upper bounds on the $\varepsilon$-capacity when $L$ grows linearly in~$n$, where the two bounds coincide for random variables with continuous and strictly increasing cdf. In Section~\ref{sectionAchConvThm}, we present the proof of the first main result, which relies on a save-and-transmit achievability lemma and a converse lemma. The proofs of the achievability and converse lemmas are provided respectively in Sections~\ref{sectionSaveAndTransmit} and~\ref{sectionConverse}, which are briefly described as follows. Section~\ref{sectionSaveAndTransmit} describes the save-and-transmit strategy which is the key to the achievability part of the first result. More specifically, we use Shannon's achievability bound~\cite{sha57} to prove a non-asymptotic achievable rate for the save-and-transmit strategy. Section~\ref{sectionConverse} proves the converse part of the first result, and the proof technique involves simplifying a non-asymptotic bound derived from the type-II error of a binary hypothesis test.
In Section~\ref{sectionArrivalBlockLinearInN}, we provide the proof of the second result when~$L$ grows linearly in~$n$.
 Concluding remarks are provided in Section~\ref{sec:conclusion}.

 \subsection{Notation} \label{sectionNotation}
The sets of natural, real and non-negative real numbers are denoted by $\mathbb{N}$, $\mathbb{R}$ and $\mathbb{R}_+$ respectively. We let $\boldsymbol{1}\{\mathcal{E}\}$ be the indicator function of the set  $\mathcal{E}$.
An arbitrary (discrete or continuous) random variable is denoted by an upper case letter (e.g., $X$), and the realization and alphabet of the random variable are denoted by the corresponding lower case letter (e.g., $x$) and calligraphic letter (e.g., $\mathcal{X}$) respectively.
We use $X^n$ to denote the random tuple $(X_1, X_2, \ldots, X_n)$. 

The following notations are used for any arbitrary random variables~$X$ and~$Y$ and any real-valued function $g$ with domain $\mathcal{X}$. We let $p_{X,Y}$ and $p_{Y|X}$ denote the probability distribution of $(X,Y)$
 and the conditional probability distribution of $Y$ given $X$ respectively. More specifically, $p_{X,Y}$ is the Radon-Nikodym derivative of a measure with respect to the Lebesgue measure in an appropriate Euclidean space.
We let $p_{X,Y}(x,y)$ and $p_{Y|X}(y|x)$ be the evaluations of $p_{X,Y}$ and $p_{Y|X}$ respectively at $(X,Y)=(x,y)$. 
To make the dependence on the distribution explicit, we let $\Pr_{p_X}\{ g(X)\in\mathcal{A}\}$ denote $\int_{\mathcal{X}} p_X(x)\mathbf{1}\{g(x)\in\mathcal{A}\}\, \mathrm{d}x$ for any set $\mathcal{A}\subseteq \mathbb{R}$.
The expectation and the variance of~$g(X)$ are denoted as
$
\E_{p_X}[g(X)]$ and
$\Var_{p_X}[g(X)]$ respectively.
 We let $\mathcal{N}(\,\cdot\, ;\mu,\sigma^2): \mathbb{R}\rightarrow [0,\infty)$ denote the probability density function of a Gaussian random variable whose mean and variance are $\mu$ and $\sigma^2$ respectively, i.e.,
 \begin{equation}
\mathcal{N}(z;\mu,\sigma^2)\triangleq\frac{1}{\sqrt{2\pi \sigma^2}}\mathrm{e}^{-\frac{(z-\mu)^2}{2\sigma^2} }. \label{eqnNormalDist}
\end{equation}
The cdf of the standard normal distribution is denoted by $\Phi$, i.e.,
 \begin{equation}
\Phi(a) \triangleq \int_{-\infty}^a \mathcal{N}(z;0,1)\mathrm{d}z.
\end{equation}
We will take all logarithms to base~$2$ throughout this paper unless specified otherwise. The logarithm function to base~$2$ is denoted by $\log$, and the natural logirhtm function is denoted by $\ln$. 

\section{Additive White Gaussian Noise Energy-Harvesting Channel with Block Energy Arrival}
\label{sectionDefinition}
\subsection{Problem formulation}
The AWGN EH channel consists of one source and one destination, denoted by $\mathrm{s}$ and $\mathrm{d}$ respectively. Node~$\mathrm{s}$ transmits information to node~$\mathrm{d}$ in $n$ time slots as follows. Node~$\mathrm{s}$ chooses message
$
W
$
 and sends $W$ to node~$\mathrm{d}$, where $W$ is uniformly distributed over $\{1, 2, \ldots, M\}$ for some $M$ that denotes the message size. Then for each $k\in \{1, 2, \ldots, n\}$, node~$\mathrm{s}$ transmits $X_{k}\in \mathbb{R}$ and node~$\mathrm{d}$ receives $Y_k\in \mathbb{R}$ in time slot~$k$. Let $\{E_{b_\ell+1}\}_{\ell=1}^{\infty}$ be i.i.d.\ random variables that satisfy
$
\Pr\{E_1< 0\}=0$ ($b_\ell$ was defined in~\eqref{blockIndex}),
\begin{equation}
\E[E_1]=P \label{expectationE1}
\end{equation}
and
$
\E[E_1^3]<\infty$ (hence $\E[E_1^2]<\infty$) for some $P>0$. Each other $E_k$ is equal to the nearest preceding $E_{b_\ell+1}$ according to~\eqref{assumptionEH}.
In other words, for each $k\in\{1, 2, \ldots, n\}$ and all $e^k\in\mathbb{R}_+^k$,
\begin{equation}
p_{E_k|E^{k-1}}(e_k| e^{k-1})=
\begin{cases}
p_{E_1}(e_k) & \text{if $k=b_\ell+1$ for some $\ell\in\mathbb{N}$,}\\
\mathbf{1}\{e_k=e_{k-1}\} & \text{otherwise.}
\end{cases} \label{defDistEk}
\end{equation}
The knowledge of~$E_k$ is available at the source at time~$k$ before encoding~$X_k$, and the destination has no access to the energy-arrival process.
The length of each energy-arrival block~$L$ is assumed to remain constant, grow sublinearly in~$n$, or grow linearly in~$n$.
We assume the following for each $k\in\{1, 2, \ldots, n\}$:
\begin{enumerate}
\item[(I)] $E_k$ and $(W, X^{k-1}, Y^{k-1})$ are independent when conditioned on $E^{k-1}$, i.e.,
\begin{align}
p_{W, E^{k}, X^{k-1}, Y^{k-1}} = p_{E_k|E^{k-1}}p_{W, E^{k-1}, X^{k-1}, Y^{k-1}}. \label{assumption(i)}
\end{align}
    \item[(II)] Every codeword $X^n$ transmitted by~$\mathrm{s}$ must satisfy the harvested energy constraint
\begin{equation}
\Pr\left\{\left.\sum_{i=1}^k X_i^2 \le \sum_{i=1}^k E_i\right|E^n=e^n, W=w\right\}=1 \label{eqn:eh}
\end{equation}
for each $e^n\in\mathbb{R}_+^n$ and each $w\in\mathcal{W}$.
  \end{enumerate}
 Assumption~(I) is a mathematical statement of the following fact due to the block i.i.d.\ EH process: If $E_k$ is the first energy-arrival random variable in a block, then it is independent of any random variables that are generated before time~$k$. Otherwise, $E_k$ equals $E_{k-1}$. In both cases, $E_k$ and $(W, X^{k-1}, Y^{k-1})$ are independent when conditioned on $E^{k-1}$.

   After~$n$ time slots, node~$\mathrm{d}$ declares~$\hat W$ to be the transmitted~$W$ based on $Y^n$. The standard definitions are formally stated in the following subsection. 
\subsection{Standard definitions}
\begin{Definition} \label{defCode}
An {\em $(n, M)$-code} consists of the following:
\begin{enumerate}
\item A message set
$
\mathcal{W}\triangleq \{1, 2, \ldots, M\}
$
 at node~$\mathrm{s}$. Message $W$ is uniform on $\mathcal{W}$.

\item A sequence of  encoding functions
$
f_k : \mathcal{W}\times \mathbb{R}_+^{k}\rightarrow \mathbb{R}
$
 for each $k\in\{1, 2, \ldots, n\}$, where $f_k$ is the encoding function for node~$\mathrm{s}$ at time slot $k$ for encoding $X_k$ such that
$
X_k=f_k (W, E^{k})
$
and \eqref{eqn:eh} holds.
\item A decoding function
$
\varphi :
\mathbb{R}^{n} \rightarrow \mathcal{W}
$
for decoding $W$ at node~$\mathrm{d}$ where the message estimate $\hat W$ is produced by setting
$
 \hat W \triangleq \varphi(Y^{n})$.
\end{enumerate}
\end{Definition}
\begin{Definition}\label{defAWGNchannel}
The {\em AWGN EH channel} is characterized by $q_{Y|X}\triangleq \mathcal{N}(y-x; 0,1)$. The distribution induced by any $(n, M)$-code used for the AWGN EH channel follows the channel law below: For each $k\in\{1, 2, \ldots, n\}$,
\begin{align}
p_{W, E^k, X^k, Y^k}
 = p_{W, E^k, X^k, Y^{k-1}}p_{Y_k|X_k} \label{memorylessStatement*}
\end{align}
where
\begin{equation}
p_{Y_k|X_k}(y_k|x_k) = q_{Y|X}(y_k|x_k) = \mathcal{N}(y_k-x_k; 0,1) \label{defChannelInDefinition*}
\end{equation}
for all $x_k$ and $y_k$.
Since $p_{Y_k|X_k}$ does not depend on~$k$ by \eqref{defChannelInDefinition*}, the channel is stationary.
\end{Definition}

 For any $(n, M)$-code defined on the AWGN EH channel, let $p_{W,E^n, X^n, Y^n, \hat W}$ be the joint distribution induced by the code. We can use Definition~\ref{defCode}, \eqref{assumption(i)} and \eqref{memorylessStatement*} to factorize $p_{W,E^n, X^n, Y^n, \hat W}$ as follows:
\begin{align}
 p_{W,E^n, X^n, Y^n, \hat W}
&=
p_W p_{E^n} \left(\prod_{k=1}^n p_{X_k|W, E^k} p_{Y_k|X_k}\right)p_{\hat W |Y^n}. \label{memorylessStatement}
 \end{align}
\begin{Definition} \label{defErrorProbability}
For an $(n, M)$-code defined on the AWGN EH channel, we can calculate, according to \eqref{memorylessStatement}, the \textit{average probability of decoding error} defined as $\Pr\big\{\hat W \ne W\big\}$.
We call an $(n, M)$-code with average probability of decoding error no larger than $\varepsilon$ an {\em $(n, M, \varepsilon)$-code}.
\end{Definition}
\begin{Definition} \label{defAchievableRate}
Let $\varepsilon\in (0,1)$ be a real number. A rate $R$ is \textit{$\varepsilon$-achievable} for the AWGN EH channel if there exists a sequence of $(n, M, \varepsilon)$-codes\footnote{Although $M$ always depends on~$n$, it is not explicitly indicated to simplify notation.} such that
\begin{equation}
\liminf_{n\rightarrow \infty}\frac{1}{n}\log M \ge R.
\end{equation}
\end{Definition}

\begin{Definition}\label{defCapacity}
Let $\varepsilon\in (0,1)$ be a real number. The {\em $\varepsilon$-capacity} of the AWGN EH channel, denoted by $C_\varepsilon$, is defined to be
$
C_\varepsilon \triangleq \sup\{R\,| R\text{ is $\varepsilon$-achievable}\}$.
\end{Definition}

\section{Main Results}\label{sectionMainResult}
Section~\ref{subsubsecMainResult1} contains the first main result in this paper, which concerns the $\varepsilon$-capacity and the second-order coding rate when $L$ is a constant or grows sublinearly in~$n$. Section~\ref{subsubsecMainResult2} contains the second main result in this paper, which concerns the $\varepsilon$-capacity when $L$ grows linearly in~$n$.
\subsection{When $L$ is a constant or grows sublinearly in~$n$}\label{subsubsecMainResult1}
In this section, we assume that $L$ is a constant or $\omega(1)=L=o(n)$ so that $\lim\limits_{n\rightarrow\infty}\frac{L}{n}=0$. Our goal in this section is to formalize the results in~\eqref{introResult1} and~\eqref{introResult2}. Before presenting the first main result, we define the second-order achievable rate as follows.
\begin{Definition} \label{defAchDispersion}
Let $\varepsilon\in (0,1)$. A real number $S$ is said to be a \emph{second-order $\varepsilon$-achievable rate} if there exists a sequence of $(n, M, \varepsilon)$-codes such that\footnote{Although $L$ can depend on~$n$, it is not explicitly indicated to simplify notation.}
\begin{equation}
\liminf_{n\rightarrow \infty}\,\frac{\log M - nC_\varepsilon}{\sqrt{Ln}} \ge S. \label{stDefAchDispersion}
\end{equation}
\end{Definition}
 The justification of the choice of $\sqrt{Ln}$ in~\eqref{stDefAchDispersion} will be explained after the following definition concerning the second-order coding rate is presented.
\smallskip
\begin{Definition} \label{defDispersion}
Let $\varepsilon\in (0,1)$. The \emph{$\varepsilon$-second-order coding rate} is defined as
\begin{equation}
V_\varepsilon \triangleq \sup\left\{S\in\mathbb{R}\left|S \text{ is a second-order $\varepsilon$-achievable rate}\right.\right\}.
\end{equation}
\end{Definition}

The choice of $\sqrt{Ln}$ in~\eqref{stDefAchDispersion} can be justified as follows by inspecting~\eqref{st2ThmMainResult} in the main theorem. More specifically, if we replace $\sqrt{Ln}$ in~\eqref{stDefAchDispersion} with any $f(n)>0$ such that $\lim\limits_{n\rightarrow\infty}\frac{\sqrt{Ln}}{f(n)} \in \{0, \infty\}$ and define $V_\varepsilon^*$ as in Definition~\ref{defDispersion}, it will then follow from~\eqref{st2ThmMainResult} that
 \begin{align}
 V_\varepsilon^*=
  \begin{cases}
0  & \text{if $\lim\limits_{n\rightarrow\infty}\frac{\sqrt{Ln}}{f(n)}=0$,}
  \\ \text{$+\infty$ or $-\infty$}& \text{if $\lim\limits_{n\rightarrow\infty}\frac{\sqrt{Ln}}{f(n)}=\infty$.}
  \end{cases}
 \end{align}
 Our choice of $\sqrt{Ln}$ in~\eqref{stDefAchDispersion} is analogous to the choice of $n^{\beta}$ in~\cite[Sec.~II-D]{TomamichelTan14} which studies the $\varepsilon$-second-order coding rate of channels with states.
 \medskip

 We are ready to present the first main result in this paper.
\begin{Theorem}\label{thmMainResult}
Suppose $L$ is a constant or $\omega(1)=L=o(n)$. Fix any $\varepsilon\in(0,1)$. Recalling the definition of~$\mathrm{C}(\cdot)$ in~\eqref{eqn:capacity_awgn}, we have
\begin{equation}
C_\varepsilon =  \mathrm{C}(P). \label{st0ThmMainResult}
\end{equation}
In addition, define
\begin{equation}
\varrho\triangleq 2\left(\frac{\E[E_1^2]}{P^2} +1\right), \label{defVarrho}
\end{equation}
\begin{align}
V_\varepsilon^- \triangleq
\begin{cases}
 - \mathrm{C}(P)\sqrt{\varrho\log \frac{1}{\varepsilon}} & \text{if $\omega(1)=L=o(n)$,} \\ \sup\limits_{\substack{(\varepsilon_1, \varepsilon_2)\in (0,1)^2: \\\varepsilon_1+ \varepsilon_2=\varepsilon }}\left\{- \mathrm{C}(P)\sqrt{\varrho\log \frac{1}{\varepsilon_1}} + \sqrt{\frac{P(\log \mathrm{e})^2}{L(1+P)}}\Phi^{-1}(\varepsilon_2)\right\} & \text{if $L$ is a constant,} \end{cases}
 \label{defV-}
\end{align}
and
\begin{align}
V_\varepsilon^+\triangleq \frac{\log \mathrm{e}}{2(1+P)}\sqrt{2P^2+\E[E_1^2]}\,\Phi^{-1}(\varepsilon). \label{defV+}
\end{align}
 Then, the $\varepsilon$-second-order coding rate $V_\varepsilon$ satisfies
\begin{align}
V_\varepsilon^-\le V_\varepsilon \le V_\varepsilon^+. \label{st1ThmMainResult}
\end{align}
In other words, if we let
\begin{align}
M_{n,\varepsilon}^* \triangleq \sup\{M\in\mathbb{N}\left| \:\text{There exists an $(n, M, \varepsilon)$-code}\right.\} \label{defMn*}
\end{align}
be the maximum alphabet size of the message we can transmit using an $(n, M, \varepsilon)$-code, then
\begin{align}
 \mathrm{C}(P) + V_\varepsilon^-\sqrt{\frac{L}{n}} - o\bigg(\sqrt{\frac{L}{n}}\bigg) \le \frac{1}{n}\log M_{n,\varepsilon}^*  \le  \mathrm{C}(P) + V_\varepsilon^+\sqrt{\frac{L}{n}} + o\bigg(\sqrt{\frac{L}{n}}\bigg). \label{st2ThmMainResult}
\end{align}
\end{Theorem}

Theorem~\ref{thmMainResult} presents a complicated lower bound on~$V_\varepsilon$ as stated in~\eqref{defV-}. The following corollary presents a simpler lower bound, which implies that $V_\varepsilon$ scales as $-O\Big(\sqrt{\log\frac{1}{\varepsilon}}\Big)$. Since the proof of Corollary~\ref{corollaryMainResult} is straightforward, it is relegated to Appendix~\ref{appendixA---}.
\begin{Corollary} \label{corollaryMainResult}
Fix an $\varepsilon\in(0,1/2)$. Following the definitions in Theorem~\ref{thmMainResult}, if we define
\begin{align}
V_\varepsilon^{--} \triangleq \begin{cases}
 - \mathrm{C}(P)\sqrt{\varrho\log \frac{1}{\varepsilon}} & \text{if $\omega(1)=L=o(n)$,} \\ -\left( \mathrm{C}(P)\sqrt{2\varrho} + \sqrt{\frac{4P\log \mathrm{e}}{1+P}}\right) \sqrt{\log\frac{1}{\varepsilon}} & \text{if $L$ is a constant,} \end{cases}
 \label{defV--}
\end{align}
then
\begin{align}
V_\varepsilon^{--} \le V_\varepsilon^{-} \le V_\varepsilon \le V_\varepsilon^+. \label{st1CorollaryMainResult}
\end{align}
\end{Corollary}

The following corollary presents an explicit bound on $V_\varepsilon^+ - V_\varepsilon^-$, whose proof relies on Corollary~\ref{corollaryMainResult} and is relegated to Appendix~\ref{appendixA--}.
\begin{Corollary} \label{corollaryMainResult*}
Fix an $\varepsilon\in (0, \Phi(-1))$ (note that $\Phi(-1)\approx 0.1586$). Following the definitions in Theorem~\ref{thmMainResult}, we have
\begin{align}
V_\varepsilon^{+}-V_\varepsilon^- \le \begin{cases}
  \left(\mathrm{C}(P)\sqrt{\varrho} \, - \sqrt{\frac{(2P^2+\E[E_1^2])\log \mathrm{e}}{2(1+P)^2}}\right)\sqrt{\log \frac{1}{\varepsilon}} & \text{if $\omega(1)=L=o(n)$,} \\ \left( \mathrm{C}(P)\sqrt{2\varrho} + \sqrt{\frac{4P\log \mathrm{e}}{1+P}} - \sqrt{\frac{(2P^2+\E[E_1^2])\log \mathrm{e}}{2(1+P)^2}}\right) \sqrt{\log\frac{1}{\varepsilon}} & \text{if $L$ is a constant.} \end{cases} \label{st1CorollaryMainResult*}
\end{align}
\end{Corollary}

This work does not intend to optimize the bound in~\eqref{st1CorollaryMainResult*}, which can be arbitrarily large as~$P$ approaches infinity or $\varepsilon$ approaches~$0$.

\subsection{When $L$ grows linearly in~$n$} \label{subsubsecMainResult2}
Before presenting the second main result, we make some necessary definitions.
Fix an arbitrary $\lambda\in (0, 1]$ and assume that
\begin{equation}
L=\lfloor \lambda n \rfloor. \label{defL}
\end{equation}
Define
\begin{equation}
q\triangleq \left\lfloor\frac{1}{\lambda}\right\rfloor \label{defQ}
\end{equation}
and
\begin{equation}
d\triangleq 1-q\lambda \label{defD}
\end{equation}
to be the quotient and remainder respectively resulting from dividing $1$ by $\lambda$. The following theorem is our second main result, which provides lower and upper bounds on $C_\varepsilon$. The proof of the lower and upper bounds will be given in Sections~\ref{subsecAchProofLinearIn} and~\ref{subsecCvProofLinearIn} respectively.
\begin{Theorem}\label{thmArrivalBlockLinearInN}
Suppose $L$ grows linearly in~$n$ according to~\eqref{defL} for some constant $\lambda\in(0,1)$. Let $q$ and~$d$ be as defined in~\eqref{defQ} and~\eqref{defD} respectively, and recall that $p_{E_1}=p_{E_2}=\ldots = p_{E_n}$. Define
\begin{equation}
\underline{R}_\varepsilon \triangleq \sup_{\delta>0} \sup\left\{r\in \mathbb{R}_+ \left| \, \Pr_{\prod_{\ell=1}^{q+1}p_{E_\ell}}\left\{\sum_{\ell=1}^{q}\lambda\mathrm{C}(E_{\ell}) + d\mathrm{C}(E_{q+1})\ge r\right\} \ge 1-\varepsilon+\delta \right.\right\} \label{st1ThmArrivalLinearInN}
\end{equation}
and
\begin{equation}
\overline{R}_\varepsilon \triangleq \inf_{\delta>0} \inf\left\{r\in \mathbb{R}_+ \left| \, \Pr_{\prod_{\ell=1}^{q+1}p_{E_\ell}}\left\{\sum_{\ell=1}^{q}\lambda\mathrm{C}(E_{\ell}) + d\mathrm{C}(E_{q+1}) \ge r\right\} \le 1-\varepsilon-\delta \right.\right\}.
\label{st2ThmArrivalLinearInN}
\end{equation}
Then, we have
\begin{equation}
\underline{R}_\varepsilon \le C_\varepsilon \le \overline{R}_\varepsilon \label{st3ThmArrivalLinearInN}
\end{equation}
for all $\varepsilon\in(0,1)$.
\end{Theorem}

The following corollary identifies a sufficient condition under which $C_\varepsilon$ can be fully characterized by Theorem~\ref{thmArrivalBlockLinearInN}. The proof of Corollary~\ref{corollaryArrivalBlockLinearInN} is straightforward and hence relegated to Appendix~\ref{appendixA-}.
\begin{Corollary}\label{corollaryArrivalBlockLinearInN}
Under the setting of Theorem~\ref{thmArrivalBlockLinearInN}, if we further assume that $E_1$ has a continuous and strictly increasing cdf, then
 \begin{equation}
 C_\varepsilon = R_\varepsilon^{\text{thr}} \label{stCorollaryBlockLinearInN}
 \end{equation}
 holds for all $\varepsilon\in(0,1)$ where $R_\varepsilon^{\text{thr}}$ is the unique threshold that satisfies
 \begin{align}
  \Pr_{\prod_{\ell=1}^{q+1}p_{E_\ell}}\left\{\sum_{\ell=1}^{q}\lambda\mathrm{C}(E_{\ell}) + d\mathrm{C}(E_{q+1}) < R_\varepsilon^{\text{thr}}\right\} =\varepsilon.
 \end{align}
\end{Corollary}

\subsection{Remarks on Theorem~\ref{thmMainResult} and Corollary~\ref{corollaryMainResult}} \label{subsecRemark}
\begin{Remark}\label{remark0}
It is already known~\cite[Remark~1]{FTY15} that
\begin{equation}
C_\varepsilon = \mathrm{C}(P) \label{scStatement}
\end{equation}
for all $\varepsilon\in(0,1)$ under an i.i.d.\ EH process. In other words, the AWGN EH channel admits the \emph{strong converse property} \cite[Ch.~3]{Han10} for $L=1$, meaning that $C_\varepsilon$ does not depend on~$\varepsilon\in(0,1)$.
It follows from \eqref{st0ThmMainResult} in Theorem~\ref{thmMainResult} that~\eqref{scStatement} remains to hold under a block i.i.d.\ EH process when $L=o(n)$. An intuitive explanation about why the strong converse property holds when $L=o(n)$ is as follows. When $L=o(n)$, since the number of energy-arrival blocks $n/L$ grows to infinity, it follows from the strong law of large numbers that the received power $\frac{1}{n}\sum_{k=1}^n E_k$ converges to $\E[E_1]=P$ with probability~$1$, which leads to a strong converse proof.
\end{Remark}
\begin{Remark}\label{remark1*}
Consider the special case where $L=1$ and fix any $\varepsilon\in(0, 1/2)$. Clearly, both $V_\varepsilon^-$ in~\eqref{defV-} and $V_\varepsilon^+$ in~\eqref{defV+} are negative. Therefore, it follows from~\eqref{st2ThmMainResult} in Theorem~\ref{thmMainResult} that the second-order term in the asymptotic expansion of $\frac{1}{n}\log M_{n, \varepsilon}^*$ scales as $-\Theta\left(\sqrt{\frac{1}{n}}\right)$. In particular, it follows from Corollary~\ref{corollaryMainResult} that $V_\varepsilon^{--}\le V_\varepsilon$, meaning that the second-order term scales as $-O\left(\sqrt{\frac{\log(1/\varepsilon)}{n}}\right)$.
 This improves the previous findings in~\cite[Th.~1]{FTY15} and~\cite{ShenoySharma16} which established that the second-order term scales as $-O\Big(\sqrt{\frac{\log n}{n}}\Big)$ and $-O\Big(\frac{1}{\sqrt{n \varepsilon}}\Big)$ respectively.
\end{Remark}
\begin{Remark}\label{remark1}
Suppose $L=o(n)$ and fix any $\varepsilon\in(0, 1/2)$. Clearly, $V_\varepsilon^-$ in~\eqref{defV-} is negative. In addition, the left hand side (LHS) of~\eqref{st2ThmMainResult} which involves $V_\varepsilon^-$ is the rate achievable by the save-and-transmit strategy (whose details can be found in Section~\ref{sectionProofOfMainResult}).
For a fixed~$P$ and a fixed $\E[E_1^2]$, since $V_\varepsilon^-$ is negative,
it follows from the LHS of~\eqref{st2ThmMainResult} that the rate achievable by the save-and-transmit strategy will increase at a slower rate if~$L$ approaches infinity at a faster rate. This can be explained by the fact that block i.i.d.\ EH processes with longer~$L$ result in higher probabilities of ``energy outage" --- the source cannot output the desired codeword due to energy shortage. Similarly for a fixed~$P$, since $|V_\varepsilon^-|$ increases as the variance $\Var[E_1]=\E[E_1^2]-P^2$ increases, it follows that block i.i.d.\ EH processes with larger variance~$\Var[E_1]$ result in higher probabilities of ``energy outage".
\end{Remark}
\begin{Remark}\label{remark2}
Suppose $L=o(n)$ and fix any $\varepsilon\in(0, 1/2)$. Clearly, $V_\varepsilon^+$ in~\eqref{defV+} is negative. For a fixed~$P$, since $V_\varepsilon^+$ is negative,
it follows that the right hand side (RHS) of~\eqref{st2ThmMainResult} increases at a slower rate if the following holds: \begin{itemize}\item[(*)] $L$ approaches infinity at a faster rate or $\Var[E_1]$ is increased.\end{itemize} In addition, it was shown in the previous remark that the LHS of~\eqref{st2ThmMainResult} increases at a slower rate if (*) holds. Consequently, both the LHS and RHS of~\eqref{st2ThmMainResult} increase at slower rates if (*) holds, which implies that the maximum rate achievable by an $(n, M_{n,\varepsilon}^*, \varepsilon)$-code increases at a slower rate if~(*) holds.
\end{Remark}
\begin{Remark} \label{remark5}
Suppose $L=o(n)$. The achievability proof of Theorem~\ref{thmMainResult} is based on analyzing the \emph{save-and-transmit strategy}, which was illustrated in Figure~\ref{figSaveAndTransmit} and will be formally discussed in Section~\ref{sectionSaveAndTransmit}. Equation~\eqref{defV-} in Theorem~\ref{thmMainResult} is indeed a lower bound on the second-order coding rate achieved by the save-and-transmit strategy. By inspecting~\eqref{defV-} we see that the two components that dominate the lower bound achieved by save-and-transmit are the saving period (contributed by the two terms with $-\mathrm{C}(P)$ in~\eqref{defV-}) and the Gaussian noise (contributed by the term with $\varepsilon_2$ in~\eqref{defV-}). If~$L$ is a constant, both components contribute to the rate loss of the lower bound on the second-order coding rate achieved by save-and-transmit because the length of the saving period is $\Theta(\sqrt{L(n+L)})=\Theta(\sqrt{Ln})$ and the minimum rate backoff needed to overcome the Gaussian noise is $\Theta(\sqrt{n})$, which correspond to the quantities $-\mathrm{C}(P)\sqrt{\varrho\log \frac{1}{\varepsilon_1}}$ and $\sqrt{\frac{P(\log \mathrm{e})^2}{L(1+P)}}\Phi^{-1}(\varepsilon_2)$ in~\eqref{defV-} respectively. If $L=\omega(1)$, the term $\sqrt{\frac{P(\log \mathrm{e})^2}{L(1+P)}}\Phi^{-1}(\varepsilon_2)$ vanishes and the resultant lower bound achieved by save-and-transmit is the quantity $- \mathrm{C}(P)\sqrt{\varrho\log \frac{1}{\varepsilon}}$ in~\eqref{defV-}, meaning that the length of the saving period dominates the lower bound.
\end{Remark}
\subsection{Remarks on Theorem~\ref{thmArrivalBlockLinearInN} and Corollary~\ref{corollaryArrivalBlockLinearInN}}
\begin{Remark} \label{remark3}
Suppose $L$ grows linearly in~$n$ and $E_1$ has a continuous and strictly increasing cdf. Using the formula of the~$\varepsilon$-capacity provided by Corollary~\ref{corollaryArrivalBlockLinearInN}, we conclude that
$
 C_\varepsilon $
is strictly increasing on $(0,1)$, implying that
 the strong converse property ceases to hold. An intuitive explanation about why the strong converse property does not hold is as follows: Since the number of energy-arrival blocks $n/L$ remains constant and the cdf of $E_1$ is continuous and strictly increasing, the received power $\frac{1}{n}\sum_{k=1}^n E_k$ does not converge (with probability~$1$) to a constant, which leads to the impossibility of a strong converse.
\end{Remark}

\begin{Remark} \label{remark7}
Consider the special case where $L=n$ and the cdf of $E_1$ is continuous and strictly increasing. Let $F_{E_1}(e)=\Pr\{E_1\le e\}$ be the cdf of $E_1$. It then follows from Corollary~\ref{corollaryArrivalBlockLinearInN} with the identifications $\lambda=1$, $q=1$ and $d=0$ that
\begin{align}
C_\varepsilon =  R_{\varepsilon}^{\text{thr}}
 = F_{E_1}^{-1}(\varepsilon)
\end{align}
for all $\varepsilon\in(0,1)$, which is analogous to the $\varepsilon$-capacities (outage capacities) of slow fading channels as stated in~\cite[Sec.\ 23.3.1]{elgamal}, the $\varepsilon$-capacities of channels with mixed states as stated in~\cite[Example~1]{TomamichelTan14}, and the $\varepsilon$-capacities of mixed channels as stated in~\cite[Example~3.4.2]{Han10}.
\end{Remark}
\begin{Remark} \label{remark10}
Suppose $L$ grows linearly in~$n$. The achievability proof of Theorem~\ref{thmArrivalBlockLinearInN} is based on designing an \emph{adaptive save-and-transmit code} that enables the source to adjust the transmission rate for each energy-arrival block according to the changes of harvested energy across different energy-arrival blocks. The adaptive save-and-transmit code was illustrated in Figure~\ref{figAdaptiveSaveAndTransmit} and will be formally discussed in Section~\ref{subsecAchProofLinearIn}. Equation~\eqref{st1ThmArrivalLinearInN} in Theorem~\ref{thmArrivalBlockLinearInN} is the coding rate achievable by the adaptive save-and-transmit strategy. By inspecting~\eqref{st1ThmArrivalLinearInN}, we see that the main event that dominates the coding rate achievable by adaptive save-and-transmit is the ``slow fading" behavior of the EH process --- the energy-harvesting rate changes slowly such that the number of energy-arrival blocks stays constant as~$n$ increases.
\end{Remark}
\begin{Remark} \label{remark6}
Suppose $L$ grows linearly in~$n$. The converse proof of Theorem~\ref{thmArrivalBlockLinearInN} is proved by considering a typical set of energy-arrival sequences followed by simplifying the conditional type-II errors of some binary hypothesis tests where the type-II errors are conditioned on the sequences in the typical set. In particular, the typical set is defined through~\eqref{defPsi} in the converse proof in Section~\ref{subsecCvProofLinearIn}, and the energy-arrival sequence falls into the set with high probability by~\eqref{defPsi*}.
\end{Remark}
\section{Proof of Theorem~\ref{thmMainResult}} \label{sectionAchConvThm}
The achievability proof of Theorem~\ref{thmMainResult} relies on the following lemma, whose proof will be presented in Section~\ref{sectionSaveAndTransmit}.
\begin{Lemma}\label{lemmaSaveAndTransmit}
Fix any $\varepsilon\in(0,1)$, $\varepsilon_1>0$ and $\varepsilon_2>0$ such that
\begin{equation}
 \varepsilon_1+\varepsilon_2 = \varepsilon.
\end{equation}
Recall the definition of~$\varrho$ in~\eqref{defVarrho}.
Then for all sufficiently large~$n$, there exist a natural number
\begin{align}
m\le \sqrt{\varrho L n\log \frac{1}{\varepsilon_1}} + O(L)
\end{align}
 and an $(n + m, M, \varepsilon)$-code such that
\begin{align}
\log M \ge n \mathrm{C}(P) +  \sqrt{\frac{nP(\log \mathrm{e})^2}{1+P}}\Phi^{-1}(\varepsilon_2) - \frac{1}{2}\log n-\kappa_1 \label{st4ThmSaveAndTransmit}
\end{align}
for some constant $\kappa_1$. More specifically, $\kappa_1$ is defined as
 \begin{align}
 \kappa_1\triangleq \frac{\sqrt{\frac{P}{1+P}}\left(\tau_1+1\right)\log \mathrm{e}}{\mathcal{N}\left(\Phi^{-1}(\min\{\varepsilon_2^2, 1-\varepsilon_2\});0,1\right)}+1 \label{defKappa1}
 \end{align}
 where
 \begin{equation}
\tau_1\triangleq \frac{\left(15^{1/3}\sqrt{P}+\frac{8}{\sqrt{\pi}}\right)^3}{(1+P)^{3/2}}. \label{defTau}
 \end{equation}
 In addition, equation~\eqref{st4ThmSaveAndTransmit} holds for any sufficiently large $n\in\mathbb{N}$ that satisfies
 \begin{equation}
 n > \frac{2L\log(1/\varepsilon_1)}{\E[E_1^2] +P^2}\cdot \max\left\{4P^2,\frac{(\E[E_1^2])^2}{4P^2}\right\}, \label{st1ThmSaveAndTransmit}
 \end{equation}
  \begin{equation}
 n \ge \frac{L(\E[E_1^2])^2\log(1/\varepsilon_1)}{P^4}, \label{st1*ThmSaveAndTransmit}
 \end{equation}
 and
 \begin{equation}
 \varepsilon_2 - \varepsilon_2^2 - \frac{\tau_1}{\sqrt{n}} - \frac{1}{\sqrt{n}}   \ge 0, \label{st2ThmSaveAndTransmit}
 \end{equation}
 and
 $m$ can be chosen to satisfy
\begin{align}
m\le \frac{\Bigg(\E[E_1^2] + \frac{3P^2}{2\left(1-2\sqrt{\frac{L\log(1/\varepsilon_1)}{n}}\right)^{5/2}} -\frac{P^2}{2} +\frac{\E[E_1^2]}{2} \sqrt{\frac{L\log(1/\varepsilon_1)}{n}}\Bigg) \sqrt{(L n+L^2)\log(1/\varepsilon_1)}}{ \left(P - \frac{\E[E_1^2]}{2P}\sqrt{\frac{L\log(1/\varepsilon_1)}{n}}\right)\sqrt{ \frac{\E[E_1^2] +P^2}{2}}} + 2L +1. \label{st3ThmSaveAndTransmit}
\end{align}
\end{Lemma}
\begin{Remark}\label{remarkLemmaSaveAndTransmit}
Lemma~\ref{lemmaSaveAndTransmit} guarantees the existence of a carefully designed save-and-transmit scheme with the saving phase being no greater than the RHS of~\eqref{st3ThmSaveAndTransmit} and the message size being no less than the RHS of~\eqref{st4ThmSaveAndTransmit}. Here, $\varepsilon_1$ specifies the probability of energy outage induced by energy shortage and $\varepsilon_2$ specifies the probability of decoding error induced by noise for the save-and-transmit scheme. As indicated by~\eqref{st3ThmSaveAndTransmit}, the designed saving phase has to be increased as~$\varepsilon_1$ decreases. In addition, as indicated by~\eqref{st4ThmSaveAndTransmit}, the designed message size has to be decreased as~$\varepsilon_2$ decreases.
\end{Remark}

The following corollary is a direct consequence of Lemma~\ref{lemmaSaveAndTransmit}. The proof of Corollary~\ref{corollarySaveTransmit} is given in Appendix~\ref{appendixC} for completeness.
\begin{Corollary} \label{corollarySaveTransmit}
Fix any $\varepsilon\in(0,1)$. For any $\varepsilon_1>0$
and $\varepsilon_2>0$ that satisfy
$ \varepsilon_1+\varepsilon_2 = \varepsilon$,
there exists a sequence of $(n^*, M, \varepsilon)$-codes such that
\begin{align}
\liminf_{n^*\rightarrow \infty}\frac{\log M - n^* \mathrm{C}(P)}{\sqrt{L n^*}} \ge \begin{cases}
- \mathrm{C}(P)\sqrt{\varrho\log \frac{1}{\varepsilon_1}} & \text{if $L=\omega(1)$,}\\
- \mathrm{C}(P)\sqrt{\varrho\log \frac{1}{\varepsilon_1}} + \sqrt{\frac{P(\log \mathrm{e})^2}{L(1+P)}}\Phi^{-1}(\varepsilon_2) &\text{if $L$ is a constant}.
\end{cases}
\label{stCorollarySaveTransmit}
\end{align}
\end{Corollary}
The converse proof of Theorem~\ref{thmMainResult} relies on the following lemma, whose proof will be presented in Section~\ref{sectionConverse}.
\begin{Lemma}\label{lemmaConverse}
Fix any $\varepsilon\in(0,1)$.
For any sufficiently large~$n$ and any $(n, M, \varepsilon)$-code, we have
\begin{align}
\log M \le (n+L) \mathrm{C}(P) + \frac{\sqrt{n+L} \log \mathrm{e}}{2(1+P)}\sqrt{2P(P+2)+L(\E[E_1^2]-P^2)}\,\Phi^{-1}(\varepsilon) + \frac{1}{2}\log (n+L) + \kappa_2 \label{stThmConverse}
\end{align}
for some $\kappa_2=O(L)$. More specifically, $\kappa_2$ is defined as
\begin{align}
\kappa_2\triangleq \frac{\frac{\tau_2}{1+P}\sqrt{2LP(P+2) +L^2(\E[E_1^2]-P^2)}\log \mathrm{e}}{\mathcal{N}\left(\Phi^{-1}(\min\{\varepsilon, \varepsilon(1-\varepsilon)\});0,1\right)} - \log (\tau_2\sqrt{L}) \label{defKappa2}
\end{align}
where
\begin{align}
\tau_2\triangleq \frac{\left(15^{1/3}P+2(2\sqrt{2/\pi})^{1/3}\cdot \left(\E[E_1^{3/2}]\right)^{1/3}+\left(\E[E_1^3]\right)^{1/3}\right)^3}{\left(\frac{2P(P+2)}{L} +\E[E_1^2]-P^2\right)^{3/2}}. \label{defTau2}
\end{align}
In addition, equation~\eqref{stThmConverse} holds for any sufficiently large~$n$ that satisfies
\begin{align}
n \ge \frac{4L\tau_2^2}{(1-\varepsilon)^4}.
 \label{st1ThmConverse}
\end{align}
\end{Lemma}
\begin{Remark} \label{remarkLemmaConverse}
For any $\varepsilon\in(0,1/2)$, since $\Phi^{-1}(\varepsilon)$ is negative, it follows from Corollary~\ref{corollarySaveTransmit} and Lemma~\ref{lemmaConverse} that the second-order term in the asymptotic expansion of $\log M_{n,\varepsilon}^*$ is $-O\bigl(\sqrt{L(n+L)}\bigr)=-O\bigl(\sqrt{Ln}\bigr)$.
\end{Remark}

We are now ready to prove Theorem~\ref{thmMainResult}.
\begin{IEEEproof}[Proof of Theorem~\ref{thmMainResult}]
For any $\varepsilon\in(0,1)$, the left inequality of~\eqref{st2ThmMainResult} follows directly from Corollary~\ref{corollarySaveTransmit} and the definition of $V_\varepsilon^-$ in~\eqref{defV-}. The right inequality of~\eqref{st2ThmMainResult} follows directly from~\eqref{stThmConverse} in Lemma~\ref{lemmaConverse} and the definition of $V_\varepsilon^+$ in~\eqref{defV+}. Using~\eqref{st2ThmMainResult} and Definition~\ref{defAchDispersion}, we obtain~\eqref{st1ThmMainResult} as well as~\eqref{st0ThmMainResult}.
\end{IEEEproof}
\begin{Remark} \label{remark8}
Theorem~\ref{thmMainResult} no longer holds when $L=\lfloor\lambda n\rfloor$ for some $\lambda\in(0,1]$. As we can see above, the proof of Theorem~\ref{thmMainResult} hinges on the achievability and converse results stated in Lemmas~\ref{lemmaSaveAndTransmit} and~\ref{lemmaConverse} respectively. However, when $L=\lfloor\lambda n\rfloor$, both Lemmas~\ref{lemmaSaveAndTransmit} and~\ref{lemmaConverse} do not yield the desired respective achievability and converse bounds. This is due to the fact that the length of the saving period~$m$ guaranteed by~\eqref{st3ThmSaveAndTransmit} in Lemma~\ref{lemmaSaveAndTransmit} grows linearly with~$n$ when $L=\lfloor\lambda n\rfloor$ and hence the overall rate achievable by save-and-transmit $\frac{n\mathrm{C}(P)}{m+n}$ does not converge to the desired $\mathrm{C}(P)$. In addition, the upper bound~\eqref{stThmConverse} in Lemma~\ref{lemmaSaveAndTransmit} does not converge to the desired $\mathrm{C}(P)$ when $L=\lfloor\lambda n\rfloor$.
\end{Remark}

\section{Proof of Lemma~\ref{lemmaSaveAndTransmit} via the Save-and-Transmit Strategy} \label{sectionSaveAndTransmit}
In this section, we investigate the save-and-transmit scheme proposed in \cite[Sec.~IV]{ozel12} in the finite blocklength regime. We will use this achievability scheme to prove Lemma~\ref{lemmaSaveAndTransmit}.
 \subsection{Prerequisites}
 The following lemma is useful for obtaining a lower bound on the length of the energy-saving phase. The proof is deferred to Appendix~\ref{appendixA}.

\begin{Lemma} \label{lemmaCharacteristicFunction}
Let $m$ and $n$ be two natural numbers. Suppose $\{X_k\}_{k=1}^n$ and $\{E_k\}_{k=1}^{m+n}$ are two sequences of i.i.d.\ random variables such that $X^n$ and $E^{m+n}$ are independent,
\begin{equation}
\Pr\{E_1 < 0\}=0, \label{lemmaCharFuncAssump0}
\end{equation}
\begin{equation}
\E[E_1]=\E[X_1^2]=P \label{lemmaCharFuncAssump1}
\end{equation}
and
\begin{equation}
\E[E_1^2] < \infty. \label{lemmaCharFuncAssump2}
\end{equation}
Suppose there exists a sufficiently small $t\in(0,1)$ such that $ \E[X_1^4 \mathrm{e}^{tX_1^2}]<\infty$ and
\begin{align}
a_t\triangleq P - \frac{t\E[E_1^2]}{2} >0, \label{defAt}
\end{align}
and we define
\begin{align}
b_t\triangleq \max\left\{0, \frac{\E[E_1^2] + \E[X_1^4 \mathrm{e}^{tX_1^2}]}{2} -P^2 + \frac{t P}{2}(\E[E_1^2] - \E[X_1^4 \mathrm{e}^{tX_1^2}]) + \frac{t^2 \E[E_1^2]\E[X_1^4 \mathrm{e}^{tX_1^2}]}{2}\right\}. \label{defBt}
\end{align}
Then,
\begin{align}
\Pr_{p_{X^n}p_{E^{m+n}}}\left\{\bigcup_{k=1}^n \left\{\sum_{i=1}^k X_i^2 \ge \sum_{i=1}^{m+k} E_i\right\}\right\}  \le \mathrm{e}^{-a_ttm+b_tt^2n}.
\end{align}
\end{Lemma}

In order to adapt Lemma~\ref{lemmaCharacteristicFunction} to the block energy arrival setting, we define the following quantities for each $t>0$ and each $L\in\mathbb{N}$ (cf.\ \eqref{defAt} and~\eqref{defBt}):
\begin{equation}
\alpha_t \triangleq L\left(P - \frac{tL\E[E_1^2]}{2}\right), \label{defAlphaT}
\end{equation}
\begin{equation}
\beta_0 \triangleq \frac{L^2(\E[E_1^2] +P^2)}{2}, \label{defBeta0}
\end{equation}
and
\begin{align}
\beta_t &\triangleq L^2\max\Bigg\{0, \frac{\E[E_1^2]}{2} + \frac{3P^2}{2(1-2LPt)^{5/2}} -P^2 +\frac{t LP}{2}\left(\E[E_1^2] - \frac{3P^2}{(1-2LPt)^{5/2}} \right)   + \frac{3t^2L^2P^2  \E[E_1^2] }{2(1-2LPt)^{5/2}} \Bigg\}. \label{defBetaT}
\end{align}
The following corollary adapts Lemma~\ref{lemmaCharacteristicFunction} to the block energy arrival setting. Since the proof of the corollary is tedious, it is deferred to Appendix~\ref{appendixB}.
\begin{Corollary}
\label{corollaryBlkCharFunc}
Fix a natural number~$L$. Suppose $\{X_k\}_{k=1}^n$ is a sequence of i.i.d.\ random variables where $X_1\sim\mathcal{N}(x_1; 0, P)$, and suppose $\{E_k\}_{k=1}^{m+n}$ is a sequence of random variables that are distributed according to~\eqref{defDistEk} (in an i.i.d.-block manner with block size~$L$).
  Fix an $\varepsilon_1>0$ and define
 \begin{equation}
 t_n\triangleq \sqrt{\frac{\log(1/\varepsilon_1)}{\lceil n/L\rceil  \beta_0}}. \label{defTnBlk}
 \end{equation}
 If
 \begin{align}
n >  \frac{L^3\log(1/\varepsilon_1)}{\beta_0}\cdot\max\left\{4P^2, \frac{(\E[E_1^2])^2}{4P^2}\right\} \label{st1CorollaryCharFuncBlk}
 \end{align}
 and
 \begin{equation}
 m \ge \frac{\sqrt{(n/L+1)\log(1/\varepsilon_1)}L(\beta_{t_n}+\beta_0)}{\alpha_{t_n}\sqrt{\beta_0}} + 2L,  \label{st2CorollaryCharFuncBlk}
 \end{equation}
 then
\begin{align}
\Pr_{p_{X^n}p_{E^{m+n}}}\left\{\bigcup_{k=1}^n \left\{\sum_{i=1}^k X_i^2 \ge \sum_{i=1}^{m+k} E_i\right\}\right\} \le  \varepsilon_1 . \label{corollaryEqnBlk}
\end{align}
\end{Corollary}
The following lemma~\cite{sha57} is standard for proving achievability results in the finite blocklength regime and its proof can be found in \cite[Th.~3.8.1]{Han10}.
\begin{Lemma}[Implied by Shannon's bound~\cite{sha57}] \label{lemmaFeinstein}
Let $p_{X^n, Y^n}$ be the probability distribution of a pair of random variables $(X^n,Y^n)$. Let $\{X^n(w), Y^n(w)\}_{w=1}^{\infty}$ be a sequence of i.i.d.\ random variables where $(X^n(1), Y^n(1))\sim p_{X^n, Y^n}$.
 For each $\delta>0$ and each~$M\in \mathbb{N}$, we have
\begin{align}
\Pr\bigg\{ \bigcup_{w=2}^M\left\{\log\left(\frac{p_{Y^n|X^n}(Y^n(1)|X^n(w))}{p_{Y^n}(Y^n(1))} \right)> \log M + n\delta \right\}\bigg\} \le \mathrm{e}^{-n\delta}.
\end{align}
\end{Lemma}
\subsection{Proof of Lemma~\ref{lemmaSaveAndTransmit}} \label{sectionProofOfMainResult}
Fix any $\varepsilon\in(0,1)$, and fix any $\varepsilon_1>0$ and $\varepsilon_2>0$ such that
\begin{equation}
\varepsilon=\varepsilon_1 + \varepsilon_2. \label{defEpsilon*}
  \end{equation}
  Define $\alpha_t$, $\beta_0$, $\beta_t$ and $t_n$ as in~\eqref{defAlphaT}, \eqref{defBeta0}, \eqref{defBetaT} and \eqref{defTnBlk} respectively.
 Fix a sufficiently large~$n$ such that \eqref{st1ThmSaveAndTransmit}, \eqref{st1*ThmSaveAndTransmit} and~\eqref{st2ThmSaveAndTransmit} hold. Since~\eqref{st1ThmSaveAndTransmit} holds, it follows from the definition of $\beta_0$ in~\eqref{defBeta0} that \eqref{st1CorollaryCharFuncBlk} also holds. Define
\begin{equation}
m\triangleq \left\lceil \frac{\sqrt{(n/L+1)\log(1/\varepsilon_1)}L(\beta_{t_n}+\beta_0)}{\alpha_{t_n}\sqrt{\beta_0}} + 2L \right\rceil, \label{defm}
\end{equation}
which satisfies~\eqref{st2CorollaryCharFuncBlk} and specifies the number of time slots which are used for saving energy.
Consider the random code that uses the channel $m+n$ times as follows: \\
\textbf{Save-and-Transmit Random Codebook Construction}\\
Let $0^m$ denote the length-$m$ zero tuple. Define the distribution $p_X$ as
 \begin{equation}
 p_{X}(x) \triangleq \mathcal{N}(x; 0, P) \label{defDistX}.
 \end{equation}
 In addition, define the distribution $p_{X^n}$ as $p_{X^n}(x^n)\triangleq \prod_{k=1}^n p_X(x_k)$. Construct~$M$ i.i.d.\ random tuples denoted by $X^n(1), X^n(2), \ldots, X^n(M)$ such that $X^n(1)$ is distributed according to~$p_{X^n}$, where $M$ will be carefully chosen later when we evaluate the probability of decoding error.
 Define
 \begin{equation}
 \tilde X^{m+n}(w)\triangleq (0^m, X^n(w)) \label{defTildeXmn}
 \end{equation}
  for each $w\in\{1, 2, \ldots, M\}$ and construct the random codebook
\begin{equation}
\big\{\tilde X^{m+n}(w) \,\big|\, w\in\{1, 2, \ldots, M\} \big\}. \label{defCodebook}
\end{equation}
The codebook is revealed to both the encoder and the decoder.
To facilitate discussion, we let $X_k(w)$ and $\tilde X_k(w)$ denote the $k^{\text{th}}$ symbols in $X^n(w)$ and $\tilde X^{m+n}(w)$ respectively for each $i$.
Since the first~$m$ symbols of each random codeword $\tilde X^{m+n}(w)$ are zeros by \eqref{defTildeXmn}, the source will just transmit~$0$ with probability~$1$ until time slot $m+1$ when the amount of energy $\sum_{k=1}^{m+1} E_k$ is available for encoding $\tilde X_{m+1}(W)$.
\\
\textbf{Encoding under the EH Constraints}\\
 For each $w\in\{1, 2, \ldots, M\}$, recalling that $\tilde X_k(w)$ is the $k^{\text{th}}$ element of $\tilde X^{m+n}(w)\stackrel{\eqref{defTildeXmn}}{=} (0^m, X^n(w))$, we construct recursively for $k=1, 2, \ldots, m+n$ the random variable
\begin{align}
\hat X_k(w, E^k)\triangleq
\begin{cases}
\tilde X_k(w) &  \text{if $(\tilde X_k(w))^2 \le \sum\limits_{i=1}^k E_i - \sum\limits_{i=1}^{k-1}(\hat X_i(w, E^{i}))^2$,}\\
 0 & \text{otherwise.}
\end{cases} \label{defHatXk}
\end{align}
To send message~$W$ which is uniformly distributed on $\{1, 2, \ldots, M\}$, the source transmits $\hat X_k(W, E^k)$ in time slot~$k$ for each $k\in\{1, 2, \ldots, m+n\}$. Note that the source transmits~$0$ with probability~$1$ in the first~$m$ times slots by \eqref{defTildeXmn} and \eqref{defHatXk}, and the transmitted codeword $(\hat X_1(W, E^1), \hat X_2(W, E^2), \ldots, \hat X_{m+n}(W, E^{m+n}))$ satisfies the EH constraints \eqref{eqn:eh} by \eqref{defHatXk}.
\\
\textbf{Threshold Decoding}\\
Upon receiving
\begin{equation}
\hat Y^{m+n} =\hat X^{m+n}(W, E^{m+n}) + Z^{m+n} \label{defHatYmn}
\end{equation}
where
\begin{align}
\hat X^{m+n}(W, E^{m+n}) \triangleq(\hat X_1(W, E^1), \hat X_2(W, E^2), \ldots, \hat X_{m+n}(W, E^{m+n})) \label{defHatXmn}
 \end{align}
 denotes the transmitted tuple specified in \eqref{defHatXk} and $Z^{m+n}\sim \prod_{k=1}^{m+n} \mathcal{N}(z_k;0, 1)$ by the channel law (cf.\ \eqref{defChannelInDefinition*}), the destination constructs its subtuple denoted by $\bar Y^n$ by keeping only the last~$n$ symbols of $\hat Y^{m+n}$. Recalling that $q_{Y|X}$ denotes the channel law and $p_X(x)\equiv \mathcal{N}(x;0, P)$, we define the joint distribution
 \begin{equation}
p_{X,Y}\triangleq p_Xq_{Y|X}, \label{defJointDistribution}
 \end{equation}
and define the joint distribution $p_{X^n, Y^n}$ as
 \begin{equation}
 p_{X^n, Y^n}(x^n, y^n) \triangleq \prod_{k=1}^n p_{X,Y}(x_k,y_k). \label{factorizationOfPXY}
 \end{equation}
Then, the decoder declares $\varphi(\bar Y^n) \in\{1, 2, \ldots, M\}$ (with a slight abuse of notation, we write $\varphi(\bar Y^n)$ instead of $\varphi(\hat Y^{m+n})$) to be the transmitted message where $\varphi(\bar Y^{n})$ is the decoding function defined as follows:
 If there exists a unique index~$j$ such that
\begin{equation}
\log\left(\frac{p_{Y^n|X^n}(\bar Y^n| X^n(j))}{p_{Y^n}(\bar Y^n)}\right) > \log M + \frac{1}{2}\log n, \label{decodingRule}
\end{equation}
then $\varphi(\bar Y^{n})$ is assigned the value~$j$.
Otherwise, $\varphi(\bar Y^{n})$ is assigned a random value uniformly distributed on $\{1,2, \ldots,M\}$.
\\
\textbf{Calculating the Probability of Violating the EH Constraints}\\
Defining $\bar X^n(W, E^{m+n})$ to be the tuple containing the last~$n$ symbols of $\hat X^{m+n}(W, E^{m+n})$, we obtain from~\eqref{defTildeXmn}, \eqref{defHatXk} and \eqref{defHatXmn} that
\begin{align}
&\Pr\left\{\bar X^n(W, E^{m+n}) = X^{n}(W)\left| \bigcap_{k=1}^{n}\left\{ \sum_{i=1}^{k}(X_{i}(W))^2 \le \sum_{i=1}^{m+k} E_i \right\}\right.\right\}=1. \label{eqn2InCalculationErrorProb*} 
\end{align}
Using Corollary~\ref{corollaryBlkCharFunc} and noting that $E^{m+n}$ and $(W,X^n(W))$ are independent by construction, we obtain
\begin{align}
\Pr\left\{\bigcup_{k=1}^n \left\{\sum_{i=1}^k (X_i(W))^2 > \sum_{i=1}^{m+k} E_i\right\}\right\} \le  \varepsilon_1.  \label{eqn4InCalculationErrorProb}
\end{align}
Using \eqref{eqn2InCalculationErrorProb*} and \eqref{eqn4InCalculationErrorProb}, we have
\begin{equation}
\Pr\left\{ \bar X^n(W, E^{m+n}) = X^{n}(W)\right\} \ge 1- \varepsilon_1. \label{eqn6InCalculationErrorProb*}
\end{equation}
\textbf{Calculating the Probability of Decoding Error}\\
Defining $\bar Z^n$ to be the tuple containing the last~$n$ symbols of $Z^{m+n}$ and recalling $\bar X^n(W, E^{m+n})$ and $\bar Y^n$ are the tuples containing the last~$n$ symbols of $\hat X^{m+n}(W, E^{m+n})$ and $\hat Y^{m+n}$ respectively, we obtain from \eqref{defHatYmn} and \eqref{eqn6InCalculationErrorProb*} that
\begin{equation}
\Pr\left\{ \bar Y^n = X^{n}(W) + \bar Z^n \right\} \ge 1-  \varepsilon_1, \label{eqn6InCalculationErrorProb}
\end{equation}
where $X^n(W)$ and $\bar Z^n$ are independent and $\bar Z^n\sim \prod_{k=1}^n\mathcal{N}(\bar z_k; 0, 1)$.
Following \eqref{decodingRule} and \eqref{eqn6InCalculationErrorProb}, we define the events
 \begin{equation}
 \mathcal{E}_{j|w} \!\triangleq\! \left\{ \log\!\left(\frac{p_{Y^n|X^n}(X^n(w) \!+\! \bar Z^n| X^n(j))}{p_{Y^n}(X^n(w) + \bar Z^n)}\right) \!\le\! \log M \!+\! \frac{1}{2}\log n \right\} \label{defEiw}
 \end{equation}
and consider the following chain of inequalities for each $w\in \{1, 2, \ldots, M\}$:
\begin{align}
& \Pr_{p_W(\prod_{\bar w=1}^M p_{X^n(\bar w)})p_{\bar Z^n}}\!\!\left\{\!\left.\mathcal{E}_{w|w}  \cup \bigcup_{j\in \{1,2, \ldots, M\}\setminus\{w\}}  \mathcal{E}_{j|w}^c\right|\!W\!=\!w\right\}  \\
&\stackrel{\text{(a)}}{\le} \Pr_{p_W(\prod_{\bar w=1}^M p_{X^n(\bar w)})p_{\bar Z^n}}\left\{\left.\mathcal{E}_{1|1} \right|W=1\right\} + \Pr_{p_W(\prod_{\bar w=1}^M p_{X^n(\bar w)})p_{\bar Z^n}}\left\{\left.\cup_{j=2}^M \, \mathcal{E}_{j|1}^c\,\right|W=1\right\} \\
& \stackrel{\text{(b)}}{\le} \Pr_{p_Wp_{X^n(1)}p_{\bar Z^n}}\left\{\left.  \mathcal{E}_{1|1} \right|W=1\right\} + \frac{1}{\sqrt{n}} \\
& \stackrel{\text{(c)}}{=}\Pr_{\prod_{k=1}^n p_{X_k(1)}p_{\bar Z_k}}\left\{\mathcal{E}_{1|1}\right\} + \frac{1}{\sqrt{n}}, \label{eqn6*InCalculationErrorProb}
\end{align}
where
\begin{enumerate}
\item[(a)] follows from symmetry of the random codebook construction and the union bound.
\item[(b)] follows from Lemma~\ref{lemmaFeinstein} and \eqref{defEiw}.
\item[(c)] follows from the fact that $X^n(1)$ and $\bar Z^n$ are independent copies of $X_1(1)$ and $\bar Z_1$ respectively by construction.
\end{enumerate}
\textbf{Applying the Berry-Ess\'een Theorem}\\
Using~\eqref{defJointDistribution}, \eqref{defDistX} and \eqref{defChannelInDefinition*} we conclude that~$X\sim\mathcal{N}(x; 0, P)$ and
$
Z\triangleq Y-X
$
are independent, $Z\sim \mathcal{N}(z;0, 1)$, and
\begin{align}
\log\left(\frac{p_{Y|X}(Y|X)}{p_{Y}(Y)}\right) = \frac{1}{2}\log(1+P) + \frac{(-PZ^2 + 2XZ + X^2)\log \mathrm{e}}{2(1+P)}. \label{defInfoSpectrum}
\end{align}
In order to ensure the first term in \eqref{eqn6*InCalculationErrorProb} can be bounded above by a simple term, we first define the mean~$\mu$, the variance~$\sigma^2$ and the third absolute moment~$T$ of $\log\left(\frac{p_{Y|X}(Y|X)}{p_{Y}(Y)}\right)$ as follows:
$
\mu \triangleq \frac{1}{2}\log\left(1+P\right)$,
$
\sigma \triangleq \sqrt{\frac{P(\log \mathrm{e})^2}{1+P}} $
and
\begin{align}
T^{1/3}&\triangleq \left(\E_{p_{X,Y}}\left[\left|\log\left(\frac{p_{Y|X}(Y|X)}{p_{Y}(Y)}\right)-\mu\right|^3\right]\right)^{1/3} \label{defT}\\
&\le \frac{\log \mathrm{e}}{1+P}\left(15^{1/3}P+\frac{8\sqrt{P}}{\pi}\right) \label{defT*}
\end{align}
where the derivation of the last inequality is relegated to Appendix~\ref{appendixB+}. Clearly,
\begin{align}
\frac{T}{\sigma^3} \le \frac{\left(15^{1/3}\sqrt{P}+\frac{8}{\sqrt{\pi}}\right)^3}{(1+P)^{3/2}} \stackrel{\eqref{defTau}}{=} \tau_1. \label{eqnTau}
\end{align}
After defining~$\mu$, $\sigma$ and~$T$, we choose~$M$ to be the unique integer that satisfies
\begin{align}
\log (M+1)\ge n \mu + \sqrt{n \sigma^2}\Phi^{-1}\left(\varepsilon_2-\frac{T}{\sigma^3\sqrt{n}}-\frac{1}{\sqrt{n}}\right) - \frac{1}{2}\log n
 > \log M \label{defChoiseOfM}
\end{align}
where
\begin{align}
\varepsilon_2-\frac{T}{\sigma^3\sqrt{n}}-\frac{1}{\sqrt{n}}
\stackrel{\eqref{eqnTau}}{\ge} \varepsilon_2-\frac{\tau_1}{\sqrt{n}}-\frac{1}{\sqrt{n}}
 \stackrel{\eqref{st2ThmSaveAndTransmit}}{>}0.
\end{align}
Following \eqref{eqn6*InCalculationErrorProb}, we obtain the following inequality where the random variables are distributed according to $\prod_{k=1}^n p_{X_k(1)}p_{\bar Z_k}$:
\begin{align}
\Pr\left\{\mathcal{E}_{1|1}\right\}
&\stackrel{\text{(a)}}{\le} \Pr\left\{\sum_{k=1}^n \log\left(\frac{p_{Y|X}(X_k(1) + \bar Z_k| X_k(1))}{p_{Y}(X_k(1) + \bar Z_k)}\right) \le n \mu + \sqrt{n \sigma^2}\Phi^{-1}\left(\varepsilon_2-\frac{T}{\sigma^3\sqrt{n}}-\frac{1}{\sqrt{n}}\right)\right\}\\
&\stackrel{\text{(b)}}{\le}\varepsilon_2 - \frac{1}{\sqrt{n}} \label{eqn8*InCalculationErrorProb}
\end{align}
where
\begin{enumerate}
\item[(a)] follows from \eqref{defEiw} and \eqref{defChoiseOfM}.
\item[(b)] follows from the Berry-Ess\'een theorem for i.i.d\ random variables \cite{KorolevShevtsova10}, i.e.,
$\left|\Pr\left\{\frac{\sum_{k=1}^n V_k - n\mu}{\sqrt{n\sigma^2}}\le a\right\}-\Phi(a)\right|\le \frac{T}{\sigma^3\sqrt{n}}$ for all $a\in\mathbb{R}$ where $\mu$, $\sigma^2$ and $T$ denote the mean, the variance and the third absolute moment of $V_k$ respectively.
\end{enumerate}
We are ready to compute the probability of decoding error as follows, where the random variables are distributed according to~$p_{W, X^n(W)}p_{\bar Z^n}p_{\bar Y^n|W, X^n(W), \bar Z^n}$:
\begin{align}
 \Pr\left\{\varphi(\bar Y^{n})\ne W \right\}
& \stackrel{\eqref{eqn6InCalculationErrorProb}}{\le}  \Pr\left\{\left\{\varphi(\bar Y^{n})\ne W\right\}\cap \{\bar Y^n=X^n(W) + \bar Z^n\} \right\} + \varepsilon_1 \\
& \le \Pr\left\{\varphi(X^n(W) + \bar Z^n)\ne W \right\} + \varepsilon_1\\
& \stackrel{\text{(a)}}{\le}\varepsilon_1+\varepsilon_2 \\
&\stackrel{\eqref{defEpsilon*}}{=} \varepsilon \label{eqn8InCalculationErrorProb}
\end{align}
where (a) follows from the threshold decoding rule (cf.\ \eqref{decodingRule} and \eqref{defEiw}), \eqref{eqn6*InCalculationErrorProb} and \eqref{eqn8*InCalculationErrorProb}.\\
 \textbf{Obtaining a Lower Bound on the Message Size~$\boldsymbol{M}$}\\
 Using \eqref{defChoiseOfM}, \eqref{eqn8InCalculationErrorProb} and the simple fact that $\log (M+1)\le \log M + 1$, we conclude that the constructed code is an $(n+ m, M, \varepsilon)$-code that satisfies
\begin{equation}
\log M\ge \frac{n}{2}\log (1+P)+ \sqrt{ \frac{nP(\log \mathrm{e})^2}{1+P}}\Phi^{-1}\left(\varepsilon_2-\frac{T}{\sigma^3\sqrt{n}}-\frac{1}{\sqrt{n}}\right) - \frac{1}{2}\log n -1. \label{eqn10InCalculationErrorProb}
\end{equation}
Using Taylor's theorem together with the fact by \eqref{st2ThmSaveAndTransmit} that $\left[\varepsilon_2 -\frac{T}{\sigma^3\sqrt{n}}-\frac{1}{\sqrt{n}},\varepsilon_2\right]\subseteq\left[\varepsilon_2^2,\varepsilon_2\right]$,
we obtain
\begin{align}
\Phi^{-1}\left(\varepsilon_2-\frac{T}{\sigma^3\sqrt{n}}-\frac{1}{\sqrt{n}}\right) &\ge \Phi^{-1}(\varepsilon_2) - \frac{\left(\frac{T}{\sigma^3\sqrt{n}}+\frac{1}{\sqrt{n}}\right)}{\mathcal{N}\left(\Phi^{-1}(\min\{\varepsilon_2^2, 1-\varepsilon_2\});0,1\right)} \label{eqn11*InCalculationErrorProb}\\
& \stackrel{\eqref{eqnTau}}{\ge}   \Phi^{-1}(\varepsilon_2) - \frac{\tau_1+1}{\sqrt{n}\mathcal{N}\left(\Phi^{-1}(\min\{\varepsilon_2^2, 1-\varepsilon_2\});0,1\right)} \label{eqn11**InCalculationErrorProb}
\end{align}
where the derivation of~\eqref{eqn11*InCalculationErrorProb} is relegated to Appendix~\ref{appendixB++}. Combining~\eqref{eqn10InCalculationErrorProb} and~\eqref{eqn11**InCalculationErrorProb} and recalling the definition of $\kappa_1$ in~\eqref{defKappa1}, we have
\begin{align}
\log M &\ge \frac{n}{2}\log (1+P) +  \sqrt{\frac{nP(\log \mathrm{e})^2}{1+P}}\Phi^{-1}(\varepsilon_2) - \frac{1}{2}\log n-\kappa_1. \label{eqn13InCalculationErrorProb**}
\end{align}
 \textbf{Obtaining an Upper Bound on the Length of Saving Phase $\boldsymbol{m}$}\\
Since the constructed $(n+ m, M, \varepsilon)$-code satisfies~\eqref{st4ThmSaveAndTransmit} by~\eqref{eqn13InCalculationErrorProb**}, it remains to show that~$m$ satisfies~\eqref{st3ThmSaveAndTransmit}. To this end, recall the definition of~$m$ in~\eqref{defm} and consider the following bounds on $t_n$, $\alpha_{t_n}$ and $\beta_{t_n}$:
\begin{align}
t_n &\stackrel{\eqref{defTnBlk}}{\le}  \sqrt{\frac{\log(1/\varepsilon_1)}{(n/L) \beta_0}}\\
& \stackrel{\eqref{defBeta0}}{=}            \sqrt{\frac{2\log(1/\varepsilon_1)}{n L(\E[E_1^2] +P^2)}} \\
& \stackrel{\eqref{expectationE1}}{\le} \frac{1}{P}\sqrt{\frac{\log(1/\varepsilon_1)}{L n}}\label{defTnBlk*}\\*
&\stackrel{\eqref{st1*ThmSaveAndTransmit}}{\le} \frac{P}{L\E[E_1^2]}, \label{defTnBlk**}
\end{align}
\begin{align}
\alpha_{t_n} &\stackrel{\eqref{defAlphaT}}{=} L\left(P - \frac{t_nL\E[E_1^2]}{2}\right)\\
& \stackrel{\eqref{defTnBlk*}}{\ge}  L\left(P - \frac{\E[E_1^2]}{2P}\sqrt{\frac{L\log(1/\varepsilon_1)}{n}}\right) \label{defAlphaT*}
\end{align}
and
\begin{align}
\beta_{t_n} &\stackrel{\eqref{defBetaT}}{=}L^2\max\Bigg\{0, \frac{\E[E_1^2]}{2} + \frac{3P^2}{2(1-2LPt_n)^{5/2}} -P^2 +\frac{t_n LP}{2}\left(\E[E_1^2] - \frac{3P^2}{(1-2LPt_n)^{5/2}} \right)  \notag\\*
&\qquad \qquad + \frac{3t_n^2 L^2 P^2\E[E_1^2]}{2(1-2LPt_n)^{5/2}}\Bigg\}\\
&\stackrel{\eqref{defTnBlk**}}{\le}  L^2\left(\frac{\E[E_1^2]}{2} + \frac{3P^2}{2(1-2LPt_n)^{5/2}} -P^2 +\frac{t_n LP\E[E_1^2]}{2}\right)\\
&\stackrel{\eqref{defTnBlk*}}{\le}  L^2\Bigg(\frac{\E[E_1^2]}{2} + \frac{3P^2}{2\left(1-2\sqrt{\frac{L\log(1/\varepsilon_1)}{n}}\right)^{5/2}} -P^2 + \frac{\E[E_1^2]}{2} \sqrt{\frac{L\log(1/\varepsilon_1)}{n}}\Bigg). \label{defBetaT*}
\end{align}
In order to obtain an upper bound on~$m$, consider the following chain of inequalities:
\begin{align}
m &\stackrel{\eqref{defm}}{\le} \frac{\sqrt{(n/L +1)\log(1/\varepsilon_1)}L(\beta_{t_n}+\beta_0)}{\alpha_{t_n}\sqrt{\beta_0}} + 2L +1 \\
& \stackrel{\text{(a)}}{\le}  \frac{L^3\Bigg(\E[E_1^2] + \frac{3P^2}{2\left(1-2\sqrt{\frac{L\log(1/\varepsilon_1)}{n}}\right)^{5/2}} -\frac{P^2}{2} +\frac{\E[E_1^2]}{2} \sqrt{\frac{L\log(1/\varepsilon_1)}{n}}\Bigg) \sqrt{(n/L+1)\log(1/\varepsilon_1)}}{\alpha_{t_n}\sqrt{\beta_0}} + 2L +1 \\
& \stackrel{\text{(b)}}{\le}   \frac{\Bigg(\E[E_1^2] + \frac{3P^2}{2\left(1-2\sqrt{\frac{L\log(1/\varepsilon_1)}{n}}\right)^{5/2}} -\frac{P^2}{2} +\frac{\E[E_1^2]}{2} \sqrt{\frac{L\log(1/\varepsilon_1)}{n}}\Bigg) \sqrt{(L n+L^2)\log(1/\varepsilon_1)}}{ \left(P - \frac{\E[E_1^2]}{2P}\sqrt{\frac{L\log(1/\varepsilon_1)}{n}}\right)\sqrt{ \frac{\E[E_1^2] +P^2}{2}}} + 2L +1  \label{eqn14InCalculationErrorProb}
\end{align}
where
\begin{enumerate}
\item[(a)] follows from~\eqref{defBetaT*} and the definition of~$\beta_0$ in~\eqref{defBeta0}.
\item[(b)] follows from~\eqref{defAlphaT*} and the definition of~$\beta_0$ in~\eqref{defBeta0}.
\end{enumerate}
Consequently, the constructed $(n+ m, M, \varepsilon)$-code satisfies~\eqref{st4ThmSaveAndTransmit} and~\eqref{st3ThmSaveAndTransmit}  by~\eqref{eqn13InCalculationErrorProb**} and~\eqref{eqn14InCalculationErrorProb} respectively. This completes the proof.
%

\section{Proof of Lemma~\ref{lemmaConverse} via Binary Hypothesis Testing} \label{sectionConverse}
\subsection{Prerequisites} \label{sectionPrerequisites}
The following definition concerning the non-asymptotic fundamental limits of a simple binary hypothesis test is standard. See for example \cite[Section~III-E]{PPV10}.
\medskip
\begin{Definition}\label{defBHTDivergence}
Let $p_{X}$ and $q_{X}$ be two probability distributions defined on some common alphabet $\mathcal{X}$. Let
\begin{equation}
\mathcal{Q}(\{0,1\}|\mathcal{X})\triangleq \{
r_{Z|X}: \text{$Z$ and $X$ assume values in $\{0,1\}$ and $\mathcal{X}$ respectively}\}
\end{equation}
be the set of randomized binary hypothesis tests between $p_{X}$ and $q_{X}$ where $\{Z=0\}$ indicates the test chooses $q_X$, and let $\delta\in [0,1]$ be a real number. The minimum type-II error in a simple binary hypothesis test between $p_{X}$ and $q_{X}$ with type-I error no larger than $1-\delta$ is defined as
\begin{align}
 \beta_{\delta}(p_X\|q_X) \triangleq
\inf\limits_{\substack{r_{Z|X} \in \mathcal{Q}(\{0,1\}|\mathcal{X}): \\ \int_{\mathcal{X}}r_{Z|X}(1|x)p_X(x)\, \mathrm{d}x\ge \delta}} \int_{\mathcal{X}}r_{Z|X}(1|x)q_X(x)\, \mathrm{d}x.\label{eqDefISDivergence}
\end{align}
\end{Definition}
\medskip
The existence of a minimizing test $r_{Z|X}$ is guaranteed by the Neyman-Pearson lemma.

We state in the following lemma and proposition some important properties of $\beta_{\delta}(p_X\|q_X)$, which are crucial for the proof of Theorem~\ref{thmMainResult}. The proof of the two statements in the following lemma can be found in~\cite[Lemma~1]{Wang2009} and~\cite[Sec.~2.3]{Pol10} respectively.
\medskip
\begin{Lemma}\label{lemmaDPI} Let $p_{X}$ and $q_{X}$ be two probability distributions defined on some $\mathcal{X}$, and let $g$ be a function whose domain contains $\mathcal{X}$. Then, the following two statements hold:
\begin{enumerate}
\item[1.] (Data processing inequality (DPI)) $\beta_{\delta}(p_X\|q_X) \le \beta_{\delta}(p_{g(X)}\|q_{g(X)})$.
\item[2.] For all $\xi>0$, $\beta_{\delta}(p_X\|q_X)\ge \frac{1}{\xi}\left(\delta - \int_{\mathcal{X}}p_X(x) \boldsymbol{1}\left\{ \frac{p_X(x)}{q_X(x)} \ge \xi \right\} \mathrm{d}x\right) $.
\end{enumerate}
\end{Lemma}
\medskip
 The proof of the following proposition is similar to Lemma~3 in \cite{Wang2009} and therefore omitted.
\medskip
\begin{Proposition} \label{propositionBHTLowerBound}
Let $p_{U,V}$ and $s_V$ be two probability distributions defined on $\mathcal{W}\times \mathcal{W}$ and $\mathcal{W}$ respectively for some $\mathcal{W}$, and let $p_U$ be the marginal distributions of $p_{U,V}$. Suppose $p_{U}$ is the uniform distribution, and let
\begin{equation}
\alpha = \Pr\{U\ne V\} \label{defAlpha}
\end{equation}
be a real number in $[0, 1)$. Then,
\begin{equation}
\beta_{1-\alpha}(p_{U,V}\|p_{U} s_V) \le \frac{1}{|\mathcal{W}|}. \label{propositionBHTLowerBoundEq1}
\end{equation}
\end{Proposition}
\subsection{Proof of Lemma~\ref{lemmaConverse}} \label{subsecLemmaConverse}
Fix an $\varepsilon\in (0,1)$, an $\bar n\in\mathbb{N}$ which is larger than the RHS of~\eqref{st1ThmConverse} and an $(\bar n, M, \varepsilon)$-code for the AWGN EH channel. Using Definition~\ref{defCode}, we have
\begin{equation}
\Pr\left\{\sum_{k=1}^{\bar n}X_{k}^2 \le \sum_{k=1}^{\bar n}E_k\right\} = 1 \label{powerConstraintInProof}
\end{equation}
for the $(\bar n, M, \varepsilon)$-code. Define
 \begin{align}
 \Delta\triangleq
 \begin{cases}
 L & \text{if $\bar n$ is a multiple of~$L$,}\\
 L\left\lceil\frac{\bar n}{L}\right\rceil - \bar n & \text{otherwise}
 \end{cases} \label{defDelta}
 \end{align}
 to be the smallest positive integer such that $\bar n +  \Delta$ is a multiple of~$L$. Then, we can always construct an $(\bar n +  \Delta, M, \varepsilon)$-code by appending carefully chosen $X_{\bar n+1}, X_{\bar n+2}, \ldots, X_{\bar n+ \Delta}$ to each transmitted sequence $X^{\bar n}$ generated by the $(\bar n, M, \varepsilon)$-code such that
\begin{equation}
\Pr\left\{\sum_{k=1}^{\bar n+ \Delta}X_{k}^2 =  \sum_{k=1}^{\bar n+ \Delta}E_k\right\} = 1. \label{powerConstraintInProofnBar}
\end{equation}
The technique of transforming the peak power inequality constraint \eqref{powerConstraintInProof} to a power equality constraint \eqref{powerConstraintInProofnBar} by appending an extra symbol has been employed in \cite[Lemma~39]{PPV10} and \cite[Theorem~4.4]{Tan_FnT} (and is called the Yaglom-map trick).
To simplify notation, we let
\begin{equation}
n\triangleq\bar n+ \Delta  \label{n=barN+Delta}
\end{equation}
where~$n$ is a multiple of~$L$ and satisfies~\eqref{st1ThmConverse}.
\\
\textbf{Obtaining a Lower Bound on the Error Probability in Terms of the Type-II Error of a Hypothesis Test}\\
 Let $p_{W, E^n, X^n, Y^n, \hat W}$ be the probability distribution induced by the $(n, M, \varepsilon)$-code constructed above, where $p_{W, E^n, X^n, Y^n, \hat W}$ can be expressed according to \eqref{memorylessStatement}. In view of \eqref{powerConstraintInProofnBar}, we assume without loss of generality that
 \begin{equation}
\int_{\mathcal{A}}p_{W,E^n, X^n, Y^n}(w,e^n, x^n, y^n) =\int_{\mathcal{A}} p_{W,E^n, X^n, Y^n}(w,e^n, x^n, y^n)\mathbf{1}\left\{\sum_{k=1}^{n}x_k^2 = \sum_{k=1}^{n}e_k\right\} \label{powerConstraintCharacteristicFunction}
 \end{equation}
 for all Borel measurable~$\mathcal{A}\subseteq \mathcal{W}\times \mathbb{R}_+^n\times \mathbb{R}^n\times\mathbb{R}^n$. All the probability and expectation terms in the rest of this proof are evaluated according to $p_{W,E^n, X^n, Y^n, \hat W}$ unless specified otherwise. Define
\begin{align}
s_{Y^{n}, \hat W}\triangleq \left(\prod_{k=1}^{n} s_{Y_{k}} \right)p_{\hat W|Y^{n}} \label{defDistTildeS}
\end{align}
where
\begin{align}
s_{Y_{k}}(y_{k})\triangleq \mathcal{N}\left(y_{k}; 0, P+1\right). \label{defTildeSy3}
\end{align}
It follows from Proposition~\ref{propositionBHTLowerBound} and Definition~\ref{defCode}
with the identifications $U\equiv W$, $V\equiv \hat W$, $p_{U,V}\equiv  p_{W,  \hat W}$, $|\mathcal{W}|\equiv  M$ and
$
\alpha\equiv \Pr\{W\ne  \hat W\} \le \varepsilon 
$
 that
 \begin{align}
\beta_{1-\varepsilon}(p_{W,\hat W}\|p_W s_{\hat W}) \le \beta_{1-\alpha}(p_{W,\hat W}\|p_W s_{\hat W}) \le \frac{1}{M}. \label{eqnBHTReverseChain}
 \end{align}
\textbf{Using the DPI to Introduce the Channel Inputs and Outputs}\\
Using the DPI of $\beta_{1- \varepsilon}$ in Lemma~\ref{lemmaDPI}, we have
 \begin{align}
& \beta_{1- \varepsilon}(p_{W,\hat W}\|p_{W} s_{\hat W})\notag\\*
& \ge \beta_{1- \varepsilon}\left(p_{W, E^n, X^n, Y^n, \hat W}\left\|p_{W}p_{E^n} \left(\prod_{k=1}^n p_{X_k|W, E^k}\right) s_{Y^n,\hat W}\right.\right). \label{converseProofEq1}
\end{align}
Fix a $\xi_n>0$ to be specified later. Since
\begin{align}
p_{W,E^n, X^n, Y^n, \hat W} &\stackrel{\eqref{memorylessStatement}}{=} p_W p_{E^n} \left(\prod_{k=1}^n p_{X_k|W, E^k} p_{Y_k|X_k}\right)p_{\hat W |Y^n},
 \end{align}
it follows from \eqref{eqnBHTReverseChain}, the definition of $s_{Y^n,\hat W}$ in~\eqref{defDistTildeS}, \eqref{converseProofEq1} and Lemma~\ref{lemmaDPI} that
\begin{align}
\log M \le \log \xi_n - \log\left(1-\varepsilon - \Pr\left\{\sum_{k=1}^n \log\left( \frac{p_{Y_{k}|X_{k}}(Y_{k}|X_{k})}{s_{Y_{k}}(Y_{k})}\right)\ge \log\xi_n\right\}\right). \label{converseProofEq2}
\end{align}
\\
\textbf{Simplifying the Non-Asymptotic Bound}\\
Combining~\eqref{defChannelInDefinition*} and~\eqref{defTildeSy3}, we have
\begin{equation}
\log\frac{p_{Y_k|X_k}(Y_{k}|X_{k})}{s_{Y_{k}}(Y_{k})} = \frac{1}{2}\log(1+P)+ \frac{\log \mathrm{e}}{2(1+P)}\left(-P(Y_k-X_k)^2 + X_k^2 + 2X_k(Y_k-X_k)\right) \label{defLogLikelihood}
\end{equation}
for each $k\in\{1, 2, \ldots, n\}$. Due to the power equality constraint imposed on the codewords, we have
\begin{equation}
\Pr\left\{ \sum_{k=1}^n  X_k^2 = \sum_{k=1}^n  E_k \right\} \stackrel{\eqref{powerConstraintInProofnBar}}{=}1. \label{powerConstraintInProofn}
\end{equation}
 Letting
 \begin{equation}
U_k \triangleq \frac{\log \mathrm{e}}{2(1+P)}(-P(Y_k-X_k)^2 + 2X_k(Y_k-X_k)+E_k) \label{defUkConv}
 \end{equation}
 for each $k\in\{1, 2, \ldots, n\}$, we obtain from \eqref{defLogLikelihood} and \eqref{powerConstraintInProofn} that
\begin{equation}
\Pr\left\{\sum_{k=1}^n \log\frac{p_{Y_k|X_k}(Y_{k}|X_{k})}{s_{Y_{k}}(Y_{k})} =\frac{n}{2}\log(1+P) + \sum_{k=1}^n U_k\right\}=1.\label{expectationSumUk}
\end{equation}
Combining \eqref{converseProofEq2} and \eqref{expectationSumUk}, we have
\begin{align}
\log M \le \log\xi_n - \log\left(\Pr\left\{\sum_{k=1}^n U_k < \log\xi_n - \frac{n}{2}\log(1+P) \right\}-\varepsilon\right). \label{eqnBHT3rdChain}
\end{align}
\textbf{Evaluating the Distribution of the Sum of Random Variables $\sum_{k=1}^n U_k$}\\
In order to simplify \eqref{eqnBHT3rdChain}, we now investigate the distribution of the sum of random variables $\sum_{k=1}^n U_k$. We will show in the following that the distribution of $\sum_{k=1}^n U_k$ can be evaluated in closed form.
Define the function
$
\lambda: \mathbb{R}_+\times \mathbb{R}\times \mathbb{R}\rightarrow \mathbb{R}
$
\begin{equation}
\lambda(e, x, y)= -P(y-x)^2 + 2 x(y-x) + e. \label{defLambda}
\end{equation}
We begin evaluating the distribution of $\sum_{k=1}^n U_k$ by examining the distribution of
$\sum_{k=1}^n \lambda(E_k, X_k, Y_k)$ (cf.\ \eqref{defUkConv}) as follows.
Let
\begin{equation}
\E\left[\mathrm{e}^{it\sum_{k=1}^n \lambda(E_k, X_k, Y_k)}\right] \label{momentGenPartialUk}
\end{equation}
 be the characteristic function of $\sum_{k=1}^n \lambda(E_k, X_k, Y_k)$ where $i$ denotes the imaginary unit. In order to evaluate a closed-form expression for \eqref{momentGenPartialUk}, we write
 \begin{align}
 &\E\left[\mathrm{e}^{it\sum_{k=1}^n \lambda(E_k, X_k, Y_k)}\right]  \notag\\
   & \stackrel{\eqref{powerConstraintCharacteristicFunction}}{=}\E\left[\mathrm{e}^{it\sum\limits_{k=1}^n \lambda(E_k, X_k, Y_k)-\frac{2t^2}{1+2itP}\sum\limits_{k=1}^{n}(E_k - X_k^2)}\right]\\
   & = \E\left[\left.\E\left[\mathrm{e}^{it\sum\limits_{k=1}^n \lambda(E_k, X_k, Y_k)-\frac{2t^2}{1+2itP}\sum\limits_{k=1}^{n}(E_k - X_k^2)}\right|E^n\right]\right].
   \label{eqnBHT4thChain}
\end{align}
In order to simplify the RHS of~\eqref{eqnBHT4thChain}, we consider the following chain of equalities for each $r\in\{2, 3, \ldots, n\}$:
\begin{align}
&\E\Bigg[\mathrm{e}^{it\sum\limits_{k=1}^r \lambda(E_k, X_k, Y_k)-\frac{2t^2}{1+2itP}\sum\limits_{k=1}^{r}(E_k - X_k^2)}\Bigg|E^n\Bigg]\notag\\
& = \E\Bigg[\E\Bigg[ \mathrm{e}^{it\sum\limits_{k=1}^r \lambda(E_k, X_k, Y_k)-\frac{2t^2}{1+2itP}\sum\limits_{k=1}^{r}(E_k - X_k^2)}   \Bigg| E^n,X^{r-1}, Y^{r-1}\Bigg]\Bigg|E^n\Bigg]\\
& = \E\Bigg[\E\Bigg[ \mathrm{e}^{it\sum\limits_{k=1}^{r-1} \lambda(E_k, X_k, Y_k)-\frac{2t^2}{1+2itP}\sum\limits_{k=1}^{r-1}(E_k - X_k^2)} \notag\\*
&\quad \times\E\left[\left. \mathrm{e}^{it \lambda(E_r, X_r, Y_r)-\frac{2t^2}{1+2itP}(E_r - X_r^2)}   \right| E^n,X^{r-1}, Y^{r-1}\right]\Bigg| E^n,X^{r-1}, Y^{r-1}\Bigg]\Bigg|E^n\Bigg].  \label{eqnBHT5thChain}
\end{align}
Since $Y_r-X_r$ is a standard normal random variable which is independent of $(E^n,X^{r-1}, Y^{r-1})$ by the channel law in~\eqref{defChannelInDefinition*}, it follows from straightforward calculations based on~\eqref{defLambda} that the following equality holds with probability~$1$ for each $i\in\{2, 3, \ldots, n\}$:
\begin{align}
&\E\left[\left. \mathrm{e}^{it \lambda(E_r, X_r, Y_r)-\frac{2t^2}{1+2itP}(E_r - X_r^2)}   \right| E^n,X^{r-1}, Y^{r-1}\right]\notag\\
& = \frac{1}{\sqrt{1+2itP}}\,\E\left[\left. \mathrm{e}^{itE_r-\frac{2t^2 E_r}{1+2itP}}   \right| E^n,X^{r-1}, Y^{r-1}\right] \\
& = \frac{1}{\sqrt{1+2itP}}\, \mathrm{e}^{itE_r-\frac{2t^2 E_r}{1+2itP}}. \label{eqnBHT6thChain}
\end{align}
Using~\eqref{eqnBHT5thChain} and~\eqref{eqnBHT6thChain}, we have the following equality that holds with probability~$1$ for each $r\in\{2, 3, \ldots, n\}$:
\begin{align}
&\E\Bigg[\mathrm{e}^{it\sum\limits_{k=1}^r \lambda(E_k, X_k, Y_k)-\frac{2t^2}{1+2itP}\sum\limits_{k=1}^{r}(E_k - X_k^2)}\Bigg|E^n\Bigg]\notag\\*
& =  \frac{1}{\sqrt{1+2itP}}\, \mathrm{e}^{itE_r-\frac{2t^2 E_r}{1+2itP}} \cdot\E\Bigg[\mathrm{e}^{it\sum\limits_{k=1}^{r-1} \lambda(E_k, X_k, Y_k)-\frac{2t^2}{1+2itP}\sum\limits_{k=1}^{r-1}(E_k - X_k^2)}\Bigg|E^n\Bigg]. \label{eqnBHT7thChain*}
\end{align}
Combining~\eqref{eqnBHT4thChain} and~\eqref{eqnBHT7thChain*}, we obtain
\begin{align}
\E\left[\mathrm{e}^{it\sum_{k=1}^n \lambda(E_k, X_k, Y_k)}\right]  ={(1+2itP)}^{-\frac{n}{2}} \E\left[\mathrm{e}^{\sum_{k=1}^n itE_k-\frac{2t^2 E_k}{1+2itP}}\right]. \label{momentGenSumXkZk}
\end{align}
Let $\{Z_k\}_{k=1}^n$ be $n$ independent copies of the standard normal random variable. Straightforward calculations reveal that
\begin{equation}
\E_{p_{E^n}\prod_{k=1}^n p_{Z_k}}\left[\mathrm{e}^{it\sum_{k=1}^n (-PZ_k^2 + 2\sqrt{E_k}Z_k+E_k)}\right] = {(1+2itP)}^{-\frac{n}{2}} \E\left[\mathrm{e}^{\sum_{k=1}^n itE_k-\frac{2t^2 E_k}{1+2itP}}\right]. \label{momentGenSumXkZk*}
\end{equation}
Therefore,
\begin{equation}
\E_{p_{W,E^n, X^n, Y^n}}\left[\mathrm{e}^{it\sum_{k=1}^n \lambda(E_k, X_k,Y_k)}\right] = \E_{p_{E^n}\prod_{k=1}^n p_{Z_k}}\left[\mathrm{e}^{it\sum_{k=1}^n (-PZ_k^2 + 2\sqrt{E_k}Z_k+E_k)}\right] \label{momentGenSumXkZk**}
\end{equation}
 by \eqref{momentGenSumXkZk} and \eqref{momentGenSumXkZk*}, i.e., the characteristic functions of $\sum_{k=1}^n \lambda(E_k, X_k,Y_k)$ and $\sum_{k=1}^n(-PZ_k^2 + 2\sqrt{E_k}Z_k+E_k)$ are equal. Consequently, the probability distributions of
$\sum_{k=1}^n \lambda(E_k, X_k,Y_k)$ and $\sum_{k=1}^n(-PZ_k^2 + 2\sqrt{E_k}Z_k+E_k)$ are equal almost everywhere, which implies from \eqref{defUkConv} and \eqref{defLambda} that the probability distributions of $\sum_{k=1}^n\frac{\log \mathrm{e}}{2(1+P)}(-PZ_k^2 + 2\sqrt{E_k}Z_k+E_k)$ and $\sum_{k=1}^n U_k$ are equal almost everywhere, which then implies from \eqref{eqnBHT3rdChain} that
\begin{align}
&\log M \notag\\
&\le \log\xi_n - \log\Bigg(\Pr_{p_{E^n}\prod_{k=1}^n p_{Z_k}} \left\{\sum_{k=1}^n\frac{\log \mathrm{e}}{2(1+P)}(-PZ_k^2 + 2\sqrt{E_k}Z_k+E_k) < \log\xi_n - \frac{n}{2}\log(1+P) \right\}-\varepsilon \Bigg). \label{eqnBHT7thChain**}
\end{align}
We recall from~\eqref{defDelta} that $n=\bar n+\Delta$ is a multiple of~$L$ and define
\begin{align}
\tilde V_\ell = \frac{\log \mathrm{e}}{2(1+P)}\sum_{i=1}^{L} (-PZ_{b_\ell+i}^2 + 2\sqrt{E_{b_\ell+i}}Z_{b_\ell+i}+E_{b_\ell+i}) \label{defVell}
\end{align}
for each $\ell\in\{1, 2, \ldots, n/L\}$
(cf.\ the definition of $b_\ell$ in~\eqref{blockIndex}). Then, equation~\eqref{eqnBHT7thChain**} can be rewritten as
\begin{align}
\log M  \le  \log\xi_n - \log\Bigg(\Pr_{p_{E^n}\prod_{k=1}^n p_{Z_k}} \left\{\sum_{\ell=1}^{n/L}\tilde V_\ell < \log\xi_n - \frac{n}{2}\log(1+P) \right\}-\varepsilon \Bigg). \label{eqnBHT8thChain}
\end{align}
\textbf{Applying the Berry-Ess\'een Theorem}\\
Using the facts that $\{Z_k\}_{k=1}^n$ are i.i.d., $\{E_{b_\ell+1}\}_{\ell=1}^{n/L}$ are i.i.d.\ and
 \begin{equation}
E_{b_\ell+1}=E_{b_\ell+2}=\ldots = E_{b_{\ell}+L} \label{assumptionEHConv}
 \end{equation}
 for each $\ell\in\{1, 2, \ldots, n/L\}$, we conclude in view of~\eqref{defVell} that $\{\tilde V_\ell\}_{\ell=1}^{n/L}$ are i.i.d.\ where
 \begin{align}
\tilde V_1 = \frac{\log \mathrm{e}}{2(1+P)} \left(-P\sum_{i=1}^L Z_i^2 + 2\sqrt{E_1}\sum_{i=1}^L Z_i + L E_1\right). \label{defV1*}
 \end{align}
In order to invoke the Berry-Ess\'een Theorem to bound the probability term in~\eqref{eqnBHT8thChain}, we define the following quantities related to~$\tilde V_1$:
\begin{align}
\mu &\triangleq \E_{p_{E_1}\prod_{i=1}^Lp_{Z_i}}[\tilde V_1] \\*
& \stackrel{\eqref{defV1*}}{=}0,
\end{align}
\begin{align}
\sigma & \triangleq \sqrt{\Var_{p_{E_1}\prod_{i=1}^Lp_{Z_i}}[\tilde V_1]} \notag\\*
& \stackrel{\eqref{defV1*}}{=} \frac{L\log \mathrm{e}}{2(1+P)}\sqrt{\frac{2P(P+2)}{L}+\E[E_1^2]-P^2}\label{defSigmaConv}
\end{align}
and
\begin{align}
T^{1/3}&\triangleq \left(\E_{p_{E_1}\prod_{i=1}^Lp_{Z_i}}\big[\big|\tilde V_1\big|^3\big]\right)^{1/3} \label{defTConv}\\
&\le \frac{L\log \mathrm{e}}{2(1+P)}\left(15^{1/3}P+2(2\sqrt{2/\pi})^{1/3}\cdot \left(\E[E_1^{3/2}]\right)^{1/3}+\left(\E[E_1^3]\right)^{1/3}\right)\label{defT*Conv}
\end{align}
where the derivation of the last inequality is relegated to Appendix~\ref{appendixC+}.
 Recalling the definition of~$\tau_2$ in~\eqref{defTau2}, we use the Berry-Ess\'een theorem for i.i.d.\ random variables \cite{KorolevShevtsova10} to obtain
  \begin{align}
&\Pr_{p_{E^n}\prod_{k=1}^n p_{Z_k}}\left\{\frac{1}{\sigma\sqrt{\frac{n}{L}}}\sum_{\ell=1}^{\frac{n}{L}} \frac{\log \mathrm{e}}{2(1+P)}\sum_{i=1}^{L} (-PZ_{b_\ell+i}^2 + 2\sqrt{E_{b_\ell+i}}Z_{b_\ell+i}+E_{b_\ell+i}) \le \Phi^{-1}\left(\varepsilon + 2\tau_2\sqrt{\frac{L}{n}}\right)\right\} \notag\\*
&\ge \varepsilon + 2\tau_2\sqrt{\frac{L}{n}}- \frac{T}{\sigma^3 \sqrt{\frac{n}{L}}}\\
& \stackrel{\eqref{defTau2}}{\ge} \varepsilon +\tau_2\sqrt{\frac{L}{n}}\,, \label{BerryEsseenSt1}
\end{align}
where the argument of $\Phi^{-1}$ satisfies
\begin{equation}
\varepsilon +2\tau_2\sqrt{\frac{L}{n}} \stackrel{\eqref{st1ThmConverse}}{\le} \varepsilon + (1-\varepsilon)^2 < 1. \label{defTau2*}
\end{equation}
Following \eqref{eqnBHT8thChain} and letting
\begin{equation}
\xi_n \triangleq \frac{n}{2}\log(1+P) + \sigma\sqrt{\frac{n}{L}}\,\Phi^{-1}\left(\varepsilon + 2\tau_2\sqrt{\frac{L}{n}}\right),
\end{equation}
we can express~\eqref{eqnBHT8thChain} as
\begin{align}
\log M  &\le  \frac{n}{2}\log(1+P) + \sigma\sqrt{\frac{n}{L}}\,\Phi^{-1}\left(\varepsilon + 2\tau_2\sqrt{\frac{L}{n}}\right) \notag\\*
 &\qquad- \log\Bigg(\Pr_{p_{E^n}\prod_{k=1}^n p_{Z_k}} \Bigg\{\sum_{\ell=1}^{n/L}\tilde V_\ell < \sigma\sqrt{\frac{n}{L}}\,\Phi^{-1}\Bigg(\varepsilon + 2\tau_2\sqrt{\frac{L}{n}}\Bigg) \Bigg\}-\varepsilon \Bigg),
\end{align}
which implies from the definition of~$\sigma$ in~\eqref{defSigmaConv}, the inequality in~\eqref{BerryEsseenSt1} and the definition of~$\tilde V_\ell$ in~\eqref{defVell} that
\begin{align}
\log M \le \frac{n}{2}\log(1+P) + \frac{\sqrt{n}\log \mathrm{e}}{2(1+P)}\sqrt{2P(P+2)+L(\E[E_1^2]-P^2)}\,\Phi^{-1}\left(\varepsilon + 2\tau_2\sqrt{\frac{L}{n}}\right) - \log \left(\tau_2\sqrt{\frac{L}{n}}\right)\,. \label{eqnBHT8thChain*}
\end{align}
Using Taylor's theorem together with the fact by \eqref{st2ThmSaveAndTransmit} that $\left[\varepsilon, \varepsilon + 2\tau_2\sqrt{\frac{L}{n}} \right]\stackrel{\eqref{defTau2*}}{\subseteq}\left[\varepsilon,\varepsilon+(1-\varepsilon)^2\right]$,
we obtain
\begin{align}
\Phi^{-1}\left(\varepsilon +2\tau_2\sqrt{\frac{L}{n}}\right) \le \Phi^{-1}(\varepsilon) + \frac{2\tau_2\sqrt{\frac{L}{n}}}{ \mathcal{N}\left(\Phi^{-1}(\min\{\varepsilon, \varepsilon(1-\varepsilon)\});0,1\right)}, \label{eqnBHT9thChain*}
\end{align}
whose derivation is relegated to Appendix~\ref{appendixC++}. Combining \eqref{eqnBHT8thChain*} and \eqref{eqnBHT9thChain*} and recalling the definition of $\kappa_2$ in~\eqref{defKappa2}, we have
\begin{equation}
\log M \le \frac{n}{2}\log(1+P) +\frac{\sqrt{n}\log \mathrm{e}}{2(1+P)}\sqrt{2P(P+2)+L(\E[E_1^2]-P^2)}\,\Phi^{-1}(\varepsilon)+ \frac{1}{2}\log n + \kappa_2. \label{eqnBHT10thChain*}
\end{equation}
Using~\eqref{eqnBHT10thChain*} and the fact by~\eqref{defDelta} and~\eqref{n=barN+Delta} that $n\le \bar n + L$, we have
\begin{equation}
\log M \le \frac{\bar n+L}{2}\log(1+P) +\frac{\sqrt{\bar n+L}\log \mathrm{e}}{2(1+P)}\sqrt{2P(P+2)+L(\E[E_1^2]-P^2)}\,\Phi^{-1}(\varepsilon)+ \frac{1}{2}\log (\bar n+L) + \kappa_2.
\end{equation}
This completes the proof.


\section{When the Length of Each Energy Arrival Block Grows Linearly in Blocklength} \label{sectionArrivalBlockLinearInN}
This section focuses on the scenario $L=\lfloor\lambda n\rfloor$ for some real constant $\lambda\in (0, 1]$. Define
\begin{equation}
\rho\triangleq \left\lfloor\frac{n}{L}\right\rfloor \label{defRho}
\end{equation}
to be the number of length-$L$ energy-arrival blocks. The total number of energy-arrival blocks is $\rho +1$ where
the length of each of the first $\rho$ energy-arrival blocks equals~$L$ and the length of the $(\rho+1)^{\text{th}}$ energy-arrival block equals
\begin{align}
n-\rho L.
\end{align}
 The following proposition gives us a lower bound and an upper bound on the length of the $(\rho+1)^{\text{th}}$ energy-arrival block, which are useful for the achievability and the converse proofs of Theorem~\ref{thmArrivalBlockLinearInN} respectively. The proof of Proposition~\ref{propositionBoundOnLastBlock} is straightforward and is deferred to Appendix~\ref{appendixLastBlock}.
\begin{Proposition} \label{propositionBoundOnLastBlock}
For all sufficiently large $n\in\mathbb{N}$,
\begin{equation}
\rho = q =\left\lfloor\frac{1}{\lambda}\right\rfloor\label{st*PropBoundOnLastBlock}
\end{equation}
and
\begin{equation}
 nd \le n-\rho L \le nd + \left\lfloor\frac{1}{\lambda}\right\rfloor, \label{stPropBoundOnLastBlock}
\end{equation}
where $n-\rho L$ is the length of the $(\rho+1)^{\text{th}}$ energy-arrival block and $q$ and $d$ were defined in~\eqref{defQ} and~\eqref{defD} respectively.
\end{Proposition}
\subsection{Achievability proof of Theorem~\ref{thmArrivalBlockLinearInN}} \label{subsecAchProofLinearIn}
In this section, we propose an \emph{adaptive save-and-transmit code} which will be used to prove the achievability part of Theorem~\ref{thmArrivalBlockLinearInN}. The adaptive save-and-transmit code enables the source to transmit information at a rate close to $\mathrm{C}(E_{b_{\ell+1}})$ for each block~$\ell\in\{1, 2, \ldots, \rho+1\}$. For each block~$\ell$, since the destination does not know the EH random variables $E_{b_{\ell+1}}$, the source first needs to quantize $E_{b_{\ell+1}}$ and convey the quantized version to the destination before adjusting the transmission rate.
 To facilitate discussion, we let \begin{equation}
\Delta\triangleq \frac{\sqrt{4P+2}}{L^{1/6}} \label{defDeltaLinearInN}
 \end{equation}
 and define the set of quantization points
 \begin{equation}
 \Gamma \triangleq \{2 \upsilon\Delta\left|\, \text{$\upsilon$ is a non-negative integer}\right.\}.\label{defGamma}
  \end{equation}
  In addition, define the quantization mapping $g^\Delta: \mathbb{R}_+\rightarrow \Gamma$ such that $g^\Delta(a)$ is the unique quantization point that satisfies
\begin{equation}
g^\Delta(a)\le a <g^\Delta(a) + 2\Delta. \label{defFunctionG}
\end{equation}
In order to enable communication at a rate close to~$\mathrm{C}(g^\Delta(E_{b_\ell+1}))$ and with error probability $O(\frac{1}{L^{1/6}})$ in block~$\ell$ for each $\ell\in\{1, 2, \ldots, \rho+1\}$, we propose to use an \emph{adaptive} save-and-transmit code in each block so that node~$\mathrm{s}$ can adapt the coding rate to the EH process.
\medskip
\begin{Definition}\label{definitionAdaptiveCode}
An $(L, \Delta, \varepsilon)$-adaptive code consists of the following:
\begin{enumerate}
\item A message alphabet $\mathcal{U}^\infty\triangleq\{0,1\}^\infty$. The message $U^\infty$ is a sequence of i.i.d.\ uniform bits.
\item An adaptive encoding function $f: \Gamma\times \mathcal{U}^\infty\rightarrow \mathbb{R}^L$ which depends on~$g^\Delta(E_1)$ such that
\begin{equation}
X^L = f(g^\Delta(E_1),U^\infty)
\end{equation}
and
\begin{equation}
\Pr\left\{\sum_{i=1}^k X_i^2 \le k E_1\right\}=1
\end{equation}
for each $k\in\{1, 2, \ldots, L\}$.
\item A decoding function $\varphi: \mathbb{R}^{L} \rightarrow \mathcal{U}^\infty
$
for decoding $W$ at node~$\mathrm{d}$ such that the message estimate $\hat U^\infty$ is produced by setting
$
 \hat U^\infty \triangleq \varphi(Y^{L})$. Define the mapping $\gamma:\mathbb{N}\times \mathbb{R}_+$ as
  \begin{equation}
  \gamma(\ell, e)\triangleq \left\lfloor \left(\ell-\left\lceil \ell^{2/3}\right\rceil\right)\mathrm{C}(g^\Delta(e)) - 2\ell^{3/4} \right\rfloor. \label{defFunctionGamma}
  \end{equation}
  Then, the probability of decoding error adapted to~$g^\Delta(E_1)$, which is defined as
 \begin{align}
\Pr\{\hat U^{\gamma(L,E_1)}\ne U^{\gamma(L,E_1)}\},
 \end{align}
 is no larger than~$\varepsilon$.
 \end{enumerate}
\end{Definition}

By Definition~\ref{definitionAdaptiveCode}, node~$\mathrm{s}$ can use an $(L, \Delta, \varepsilon)$-adaptive code to transmit $2^{\gamma(L,E_1)}$ bits to node~$\mathrm{d}$ with small error probability in each length-$L$ energy-arrival block. We use ``adaptive" to describe the code because the number of bits that can be conveyed by the code changes with $E_1$. We will prove the achievability part of Theorem~\ref{thmArrivalBlockLinearInN} by using an adaptive code that has the following two features for every $\ell\in\{1, 2, \ldots, \rho\}$:
 \begin{enumerate}
 \item[(i)]
     Each of the first $\left\lceil\sqrt{L}\right\rceil$ symbols in the $\ell^{\text{th}}$ block sent by node~$\mathrm{s}$ is the constant symbol $\sqrt{g^\Delta(E_{b_\ell+1})}$ so that with probability larger than $1-\frac{1}{L^{1/6}}$, the destination can estimate $g^\Delta(E_{b_\ell+1})$ correctly.
      \item[(ii)] In the remaining $L-\left\lceil\sqrt{L}\right\rceil$ symbols in the $\ell^{\text{th}}$ block, node~$\mathrm{s}$ intends to use a Gaussian codebook with average power $g^\Delta(E_{b_\ell+1})$ to transmit i.i.d.\ uniform bits at a rate close to~$\mathrm{C}(g^\Delta(E_{b_\ell+1}))$ and with error probability $\le \frac{1}{L^{1/6}}$.
\end{enumerate}
Feature~(i) is based on Proposition~\ref{propositionKeyStep1} to be presented later. Feature~(ii) will be established through proving the existence of an adaptive code with the desired properties in Lemma~\ref{lemmaAdaptiveCode}. The proof of the following proposition is simple and thus deferred to Appendix~\ref{appendixD}.
\medskip
\begin{Proposition} \label{propositionKeyStep1}
\begin{equation}
\Pr_{p_{E_1}p_{Z^L}}\left\{\left|\frac{1}{\Big\lceil\sqrt{L}\Big\rceil}\sum_{i=1}^{\left\lceil\sqrt{L}\right\rceil} \left(\sqrt{g^\Delta(E_{1})}+Z_i\right)^2 -  g^\Delta(E_{1})-1\right|<\Delta\right\} > 1-\frac{1}{L^{1/6}}\, .
\end{equation}
\end{Proposition}
\medskip
The following lemma is useful for proving the achievability part of Theorem~\ref{thmArrivalBlockLinearInN}. Since Lemma~\ref{lemmaConstantEnergyArrival} is a direct consequence of~\cite[Th.~1]{FTY15}, its proof is relegated to Appendix~\ref{appendixCorollaryAdaptiveCode}.
\begin{Lemma} \label{lemmaConstantEnergyArrival}
The following statement holds for any sufficiently large $L\in \mathbb{N}$.
Fix an arbitrary $\tilde P>0$ and suppose $E_1=E_2=\ldots =E_L = \tilde P$ holds with probability~$1$.
Then, there exists an $\left(L-\left\lceil \sqrt{L}\right\rceil, M, \frac{1}{\sqrt{L}}\right)$-code such that
\begin{equation}
\log M \ge \gamma(L, \tilde P).
\end{equation}
To facilitate discussion, we call the $\left(L-\left\lceil \sqrt{L}\right\rceil, M, \frac{1}{\sqrt{L}}\right)$-code an $\left(L-\left\lceil \sqrt{L}\right\rceil, M, \tilde P, \frac{1}{\sqrt{L}}\right)$-code.
\end{Lemma}
\medskip

The following lemma is based on Proposition~\ref{propositionKeyStep1} and Lemma~\ref{lemmaConstantEnergyArrival}.
\begin{Lemma}\label{lemmaAdaptiveCode}
For any sufficiently large~$L\in \mathbb{N}$, there exists an $\left(L, \Delta, \frac{1}{L^{1/6}}+\frac{1}{\sqrt{L}}\right)$-adaptive code.
\end{Lemma}
\begin{IEEEproof}
We construct an $\left(L, \Delta, \frac{1}{L^{1/6}}+\frac{1}{\sqrt{L}}\right)$-adaptive code in two steps as follows.
\begin{enumerate}
\item
 In each of the first $\left\lceil\sqrt{L}\right\rceil$ time slots in the length-$L$ block, node~$\mathrm{s}$ sends the constant symbol $\sqrt{g^\Delta(E_{1})}$, which is always possible because $g^\Delta(E_{1}) \le E_1$ by~\eqref{defFunctionG}. Upon receiving $Y^{\left\lceil\sqrt{L}\right\rceil}$, node~$\mathrm{d}$ produces an estimate of $g^\Delta(E_{1})$, denoted by $\hat g^\Delta(E_{1})$, by setting
 \begin{equation}
\hat g^\Delta(E_{1}) \triangleq  \min \argmin_{v\in \Gamma}\left\{\left|\frac{1}{\Big\lceil\sqrt{L}\Big\rceil}\sum_{i=1}^{\big\lceil\sqrt{L}\big\rceil}Y_i^2 - v - 1\right|\right\}. \label{eq1LemmaAdaptiveCodeProof}
 \end{equation}
 It follows from the definition of~$\Gamma$ in~\eqref{defGamma}, Proposition~\ref{propositionKeyStep1} and~\eqref{eq1LemmaAdaptiveCodeProof} that
 \begin{align}
\Pr_{p_{E_1}}\left\{ \hat g^\Delta(E_1) = g^\Delta(E_1)\right\} > 1-\frac{1}{L^{1/6}}.
 \end{align}
 \item In the remaining $L-\left\lceil\sqrt{L}\right\rceil$ time slots, node~$\mathrm{s}$ will choose an $\left(L-\left\lceil \sqrt{L}\right\rceil, M, \frac{1}{\sqrt{L}}\right)$-code based on the knowledge of $E_1$ as follows: Node~$\mathrm{s}$ calculates $g^\Delta(E_1)$ and transmits $\gamma(L, E_1)$ i.i.d.\ uniform bits $U^{\gamma(L,E_1)}$ using a predetermined
     $\left(L-\left\lceil \sqrt{L}\right\rceil, M, g^\Delta(E_1), \frac{1}{\sqrt{L}}\right)$-code whose existence is guaranteed by Lemma~\ref{lemmaConstantEnergyArrival}. The encoding strategy of $\mathrm{s}$ is known to node~$\mathrm{d}$, which will decode the bits using the decoder of the $\left(L-\left\lceil \sqrt{L}\right\rceil, M, \hat g^\Delta(E_1),\frac{1}{\sqrt{L}}\right)$-code predetermined by node~$\mathrm{s}$ and output the bits estimate $\hat U^{\gamma(L,E_1)}$. By the definition of the codes,
     \begin{align}
     \Pr\left\{\left. \hat U^{\gamma(L,E_1)}\ne U^{\gamma(L,E_1)} \right|\hat g^\Delta(E_1) = g^\Delta(E_1)\right\} \le \frac{1}{\sqrt{L}}. \label{eq2LemmaAdaptiveCodeProof}
     \end{align}
 \end{enumerate}
 For the adaptive code described above, it follows from~\eqref{eq1LemmaAdaptiveCodeProof} and~\eqref{eq2LemmaAdaptiveCodeProof} together with the union bound that
 \begin{equation}
  \Pr\left\{\hat U^{\gamma(L,E_1)}\ne U^{\gamma(L,E_1)} \right\} \le \frac{1}{\sqrt{L}}+ \frac{1}{L^{1/6}},
 \end{equation}
 which implies that the adaptive code is an $\left(L, \Delta, \frac{1}{L^{1/6}}+\frac{1}{\sqrt{L}}\right)$-adaptive code.
\end{IEEEproof}

We are ready to prove the achievability part of Theorem~\ref{thmArrivalBlockLinearInN}.
\begin{IEEEproof}[Achievability proof of Theorem~\ref{thmArrivalBlockLinearInN}] Fix an $\varepsilon\in (0,1)$. Our goal is to prove
\begin{equation}
\underline{R}_\varepsilon \le C_\varepsilon. \label{finalGoalInAchProofLinearInN}
\end{equation}
It suffices to show that
\begin{equation}
\underline{R}_\varepsilon -2\eta \le C_\varepsilon \label{goalInAchProofLinearInN}
\end{equation}
for all $\eta>0$. Fix an arbitrary $\eta>0$. By the definition of $\underline{R}_\varepsilon$ in~\eqref{st1ThmArrivalLinearInN}, we have
\begin{align}
 \Pr_{\prod_{\ell=1}^{q+1}p_{E_\ell}}\left\{\sum_{\ell=1}^{q}\lambda\mathrm{C}(E_{\ell}) + d\mathrm{C}(E_{q+1})\ge \underline{R}_\varepsilon-\eta\right\} \ge 1-\varepsilon+\delta \label{st1thmArrivalBlockLinearInN}
\end{align}
for some $\delta>0$. Let $\chi_\delta>0$ be a sufficiently large number such that
\begin{equation}
\Pr\left\{\max_{\ell\in\{1, 2, \ldots, q+1\}}\mathrm{C}(E_{\ell})\ge \chi_\delta\right\}\le \frac{\delta}{3}. \label{defChi}
\end{equation}
We want to show that there exists a sequence of $(n, M_n, \varepsilon)$-codes such that
\begin{align}
\liminf_{n\rightarrow \infty}\frac{1}{n}\log M_n \ge \underline{R}_\varepsilon -2\eta, \label{intermediateGoalAchProof}
\end{align}
which will then imply~\eqref{goalInAchProofLinearInN}.
To this end, fix a sufficiently large~$n\in\mathbb{N}$ such that~\eqref{st*PropBoundOnLastBlock} holds, \eqref{stPropBoundOnLastBlock} holds,
\begin{align}
\sum_{\ell=1}^{q+\mathbf{1}\{d\ne 0\}}\left(\frac{1}{L_\ell^{1/6}}+\frac{1}{\sqrt{L_\ell}}\right) \le \frac{\delta}{3} \label{st0*thmArrivalBlockLinearInN}
\end{align}
and
\begin{align}
n\eta \ge (q+1)\left(L \Delta + \left\lceil L^{2/3}\right\rceil (\chi_\delta+\Delta)+2L^{3/4}+\chi_\delta +1\right) \label{nXiLowerBound}
\end{align}
where $\Delta$ is as defined in~\eqref{defDeltaLinearInN}.
The number of i.i.d.\ uniform bits that can be transmitted by the code is chosen to be
\begin{align}
\log M_n\triangleq \lfloor n\underline{R}_\varepsilon -2n\eta \rfloor. \label{defLogMn}
\end{align}
In the rest of the proof, we are devoted to constructing an $(n, M_n)$-code followed by showing that the error probability is bounded above by~$\varepsilon$.
\\
\textbf{Construction of an $(n, M_n)$-code:}\\
Recall that the length of each of the first~$\rho$ blocks is $L=\lfloor\lambda n\rfloor$. To facilitate discussion, define
\begin{equation}
L_\ell\triangleq L = \lfloor\lambda n\rfloor \label{defLl}
 \end{equation}
 to be the length of the $\ell^{\text{th}}$ block for each $\ell\in\{1, 2, \ldots, \rho\}$, and define
 \begin{equation}
 L_{\rho+1} \triangleq \lfloor d n\rfloor \label{defLrho+1}
  \end{equation}
  to be a lower bound on the length of the $(\rho+1)^{\text{th}}$ block (due to~\eqref{stPropBoundOnLastBlock}). Since $\rho L + L_{\rho+1}\le n$ by construction, we will construct an $(n, M_n)$-code by concatenating $\rho$ blocks of length-$L$ adaptive codes and one block of length-$L_{\rho+1}$ adaptive code as described below. The message of the $(n, M_n)$-code is a sequence of $\log M_n$ i.i.d.\ uniform bits denoted by $U^{\log M_n}$. Then for each block $\ell\in\{1, 2, \ldots, \rho+1\}$, node~$\mathrm{s}$ uses an $\left(L_\ell, \Delta, \frac{1}{L_\ell^{1/6}}+\frac{1}{\sqrt{L_\ell}}\right)$-adaptive code to transmit $\gamma(L, E_{b_\ell+1})$ i.i.d.\ uniform bits. A decoding error is declared if one of the following cases occurs:
\begin{itemize}
\item[(i)] The total number of transmitted bits is less than $\log M_n$, i.e., the following event occurs:
\begin{align}
\mathcal{F}\triangleq\left\{\sum_{\ell=1}^{\rho+1} \gamma(L_\ell, E_{b_\ell+1}) < \log M_n\right\}. \label{errorEvent1}
\end{align}

\item[(ii)] Provided that $\mathcal{F}^c$ occurs, the bits estimates output by~$\mathrm{d}$ denoted by $\hat U^{\log M_n}$ are not equal to the transmitted bits, i.e., the following event occurs:
\begin{align}
\left\{\hat U^{\log M_n} \ne U^{\log M_n} \right\}. \label{errorEvent2}
\end{align}
\end{itemize}
\textbf{Analysis of the Error Probability:}\\
In the rest of the proof,  all the probability terms are evaluated according to the distribution induced by the $(n, M_n)$-code constructed above. Since the $(n, M_n)$-code is a concatenation of $\rho$ blocks of $\left(L, \Delta, \frac{1}{L^{1/6}}+\frac{1}{\sqrt{L}}\right)$ adaptive codes and one block of $\left(L_{\rho+1}, \Delta, \frac{1}{L_{\rho+1}^{1/6}}+\frac{1}{\sqrt{L_{\rho+1}}}\right)$ adaptive code, it follows from Definition~\ref{definitionAdaptiveCode} and the union bound that
\begin{align}
\Pr\left\{\left. \hat U^{\log M_n} \ne U^{\log M_n} \right|\mathcal{F}^c\right\}\le
\begin{cases}
\sum_{\ell=1}^{\rho}\left(\frac{1}{L_\ell^{1/6}}+\frac{1}{\sqrt{L_\ell}}\right) & \text{if $d=0$,} \\
\sum_{\ell=1}^{\rho+1}\left(\frac{1}{L_\ell^{1/6}}+\frac{1}{\sqrt{L_\ell}}\right) & \text{otherwise,}
\end{cases}
\end{align}
which together with~\eqref{st*PropBoundOnLastBlock} and~\eqref{errorEvent1} implies that the error probability of the $(n, M_n)$-code is bounded above as
\begin{align}
\Pr\left\{\hat U^{\log M_n} \ne U^{\log M_n}\right\} \le \sum_{\ell=1}^{q+\mathbf{1}\{d\ne 0\}}\left(\frac{1}{L_\ell^{1/6}}+\frac{1}{\sqrt{L_\ell}}\right)  + \Pr\left\{\sum_{\ell=1}^{q+1} \gamma(L_\ell, E_{b_\ell+1}) < \log M_n\right\}. \label{st0thmArrivalBlockLinearInN}
\end{align}
In order to obtain an upper bound on the last term in~\eqref{st0thmArrivalBlockLinearInN} in terms of $\mathrm{C}(E_{b_\ell+1})$, we consider
\begin{align}
&\Pr\left\{\sum_{\ell=1}^{q+1} \gamma(L_\ell, E_{b_\ell+1}) < \log M_n\right\}\notag\\
& \stackrel{\eqref{defFunctionGamma}}{\le} \Pr\left\{\sum_{\ell=1}^{q+1}  L_\ell\mathrm{C}(g^\Delta(E_{b_\ell+1}))   < \log M_n +\sum_{\ell=1}^{q+1}\left(\left\lceil L_\ell^{2/3}\right\rceil \mathrm{C}(g^\Delta(E_{b_\ell+1}))+2L_\ell^{3/4}+1\right)\right\}\\
&\stackrel{\text{(a)}}{\le} \Pr\left\{\sum_{\ell=1}^{q+1}  L_\ell(\mathrm{C}(E_{b_\ell+1})-\Delta)   < \log M_n +\sum_{\ell=1}^{q+1}\left(\left\lceil L_\ell^{2/3}\right\rceil (\mathrm{C}(E_{b_\ell+1})+\Delta)+2L_\ell^{3/4}+1\right)\right\}\\
&=\Pr\left\{\sum_{\ell=1}^{q+1}  L_\ell \mathrm{C}(E_{b_\ell+1})  < \log M_n +\sum_{\ell=1}^{q+1}\left(L_\ell \Delta + \left\lceil L_\ell^{2/3}\right\rceil (\mathrm{C}(E_{b_\ell+1})+\Delta)+2L_\ell^{3/4}+1\right)\right\}\\
&\stackrel{\text{(b)}}{\le}\Pr\left\{\sum_{\ell=1}^{q+1}  L_\ell \mathrm{C}(E_{b_\ell+1})  < \log M_n +\sum_{\ell=1}^{q+1}\left(L_\ell \Delta + \left\lceil L_\ell^{2/3}\right\rceil (\chi_\delta+\Delta)+2L_\ell^{3/4}+1\right)\right\}+\frac{\delta}{3}\\
&\le\Pr\left\{\sum_{\ell=1}^{q+1}  L_\ell \mathrm{C}(E_{b_\ell+1})  < \log M_n +(q+1)\left(L \Delta + \left\lceil L^{2/3}\right\rceil (\chi_\delta+\Delta)+2L^{3/4}+1\right)\right\}+\frac{\delta}{3}, \label{eq5MainThmLinearInN}
\end{align}
where
\begin{enumerate}
\item[(a)] follows from the definition of $g^{\Delta}$ in~\eqref{defFunctionG} and the fact that
\begin{align}
\log(1+a+2b)- 2b\le \log(1+a)\le \log(1+a+2b).
\end{align}
\item[(b)] follows from~\eqref{defChi} and the union bound.
\end{enumerate}
On the other hand, 
combining~\eqref{st1thmArrivalBlockLinearInN} with the fact that $\{E_{b_\ell+1}\}_{\ell=1}^{n/L}$ are i.i.d., we have
\begin{align}
 \Pr\left\{\sum_{\ell=1}^{q}\lambda n\mathrm{C}(E_{b_\ell+1}) + dn\mathrm{C}(E_{b_{q+1}+1})< n\underline{R}_\varepsilon-n\eta\right\} \le \varepsilon-\delta,
\end{align}
which implies that
\begin{align}
 \Pr\left\{\sum_{\ell=1}^{q}\lfloor\lambda n \rfloor\mathrm{C}(E_{b_\ell+1}) + \lfloor dn\rfloor\mathrm{C}(E_{b_{q+1}+1})< n\underline{R}_\varepsilon-n\eta-\sum_{\ell=1}^{q+1}\mathrm{C}(E_{b_\ell+1})\right\} \le \varepsilon-\delta,
\end{align}
which then together with~\eqref{defLl}, \eqref{defLrho+1} and~\eqref{defLogMn} implies  that
\begin{align}
 \Pr\left\{\sum_{\ell=1}^{q+1} L_\ell\mathrm{C}(E_{b_\ell+1}) < \log M_n +n\eta-\sum_{\ell=1}^{q+1}\mathrm{C}(E_{b_\ell+1})\right\} \le \varepsilon-\delta. \label{eq6MainThmLinearInN}
\end{align}
Using~\eqref{eq6MainThmLinearInN}, \eqref{defChi} and the union bound, we obtain
\begin{align}
 \Pr\left\{\sum_{\ell=1}^{q+1} L_\ell\mathrm{C}(E_{b_\ell+1}) < \log M_n +n\eta-(q+1)\chi_\delta\right\} \le \varepsilon-\frac{2\delta}{3}. \label{eq7MainThmLinearInN}
\end{align}
Combining~\eqref{eq5MainThmLinearInN}, \eqref{eq7MainThmLinearInN} and~\eqref{nXiLowerBound}, we have
\begin{equation}
\Pr\left\{\sum_{\ell=1}^{q+1} \gamma(L_\ell, E_{b_\ell+1}) < \log M_n\right\} \le \varepsilon - \frac{\delta}{3} \label{eq8MainThmLinearInN}
\end{equation}
Using~\eqref{st0thmArrivalBlockLinearInN}, \eqref{st0*thmArrivalBlockLinearInN} and~\eqref{eq8MainThmLinearInN}, we have
\begin{equation}
\Pr\left\{\sum_{\ell=1}^{q+1} \gamma(L_\ell, E_{b_\ell+1}) < \log M_n\right\} \le \varepsilon.
\end{equation}
Therefore, the constructed $(n, M_n)$-code is an $(n, M_n, \varepsilon)$-code where $M_n$ satisfies~\eqref{defLogMn}. Consequently, for any $\eta>0$, there exists a sequence of $(n, M_n, \varepsilon)$-codes where $M_n$ satisfies~\eqref{defLogMn} such that~\eqref{intermediateGoalAchProof} holds, which then implies \eqref{goalInAchProofLinearInN}. Since $\eta>0$ is arbitrary, we have~\eqref{finalGoalInAchProofLinearInN}.
\end{IEEEproof}

\subsection{Converse Proof of Theorem~\ref{thmArrivalBlockLinearInN}} \label{subsecCvProofLinearIn}
Fix an $\varepsilon\in (0,1)$. Our goal is to prove
\begin{equation}
 C_\varepsilon \le \overline{R}_\varepsilon. \label{goalInConvProofLinear}
\end{equation}
It suffices to show that
\begin{equation}
 C_\varepsilon\le  \overline{R}_\varepsilon +\eta
\end{equation}
for all $\eta>0$. Fix an arbitrary $\eta>0$ and an $\varepsilon$-achievable rate~$R$. By the definition of $\overline{R}_\varepsilon$ in~\eqref{st2ThmArrivalLinearInN},
\begin{align}
\Pr_{\prod_{\ell=1}^{q+1}p_{E_\ell}}\left\{\sum_{\ell=1}^{q}\lambda\mathrm{C}(E_{\ell}) + d\mathrm{C}(E_{q+1})\ge \overline{R}_\varepsilon+\eta\right\} \le 1-\varepsilon-2\delta \label{st1thmArrivalBlockLinearInNConv}
\end{align}
for some $\delta>0$. Let $\chi_\delta$ be a sufficiently large number such that
\begin{equation}
\Pr_{\prod_{\ell=1}^{q+1}p_{E_\ell}}\left\{\max_{\ell\in\{1, 2, \ldots, q+1\}}(E_{\ell}^2+2E_\ell)\ge \chi_\delta\right\}\le \delta. \label{defChiDeltaLinear}
\end{equation}

In addition, since~$R$ is $\varepsilon$-achievable, it follows from Definition~\ref{defAchievableRate} that there exists a sequence of $(n, M, \varepsilon)$-codes such that
\begin{equation}
\liminf_{n\rightarrow \infty}\frac{1}{n}\log M \ge R. \label{converseProofEq0Linear}
\end{equation}
Fix a sufficiently large~$n\in \mathbb{N}$ such that \eqref{st*PropBoundOnLastBlock} holds, \eqref{stPropBoundOnLastBlock} holds and
 \begin{align}
\frac{(\log \mathrm{e})^2 \chi_\delta}{\sqrt{n}}\le \frac{\delta}{2}, \label{sufficientlyLargeNConvLinear}
 \end{align}
and fix the corresponding $(n, M, \varepsilon)$-code.
Let $p_{W, E^n, X^n, Y^n, \hat W}$ be the probability distribution induced by the $(n, M, \varepsilon)$-code. Unless specified otherwise, all the probability, expectation and variance terms are evaluated according to~$p_{W, E^n, X^n, Y^n, \hat W}$.
Since $\{E_{b_\ell+1}\}_{\ell=1}^{\infty}$ are i.i.d.\ by assumption, it follows from~\eqref{defChiDeltaLinear} that
\begin{equation}
\Pr\left\{\max_{\ell\in\{1, 2, \ldots, q+1\}}(E_{b_\ell+1}^2+2E_{b_\ell+1})\ge \chi_\delta\right\}\le \delta. \label{defChiConv}
\end{equation}
Define
\begin{align}
\varepsilon(e^n)\triangleq \Pr\left\{\left. \hat W \ne W \right|E^n=e^n\right\}
\end{align}
and
\begin{align}
\Psi_\delta\triangleq \left\{e^n\in\mathbb{R}^n \left|\, \varepsilon(e^n) < 1-\delta , \max_{\ell\in\{1, 2, \ldots, q\}}(e_{b_\ell+1}^2+2e_{b_\ell+1})< \chi_\delta\right.\right\}. \label{defPsi}
\end{align}
Since the average error probability of the code is no larger than $\varepsilon$, we have
\begin{align}
\int_{\mathbb{R}_+^n} p_{E^n}(e^n)\varepsilon(e^n) \mathrm{d} e^n \le \varepsilon. \label{averageEpsilon}
\end{align}
Consider
\begin{align}
\Pr\{E^n \notin \Psi_\delta\}&\stackrel{\eqref{defPsi}}{\le} \Pr\{\varepsilon(E^n) \ge 1-\delta\} + \Pr\left\{ \max_{\ell\in\{1, 2, \ldots, q\}}(E_{b_\ell+1}^2+2E_{b_\ell+1})\ge \chi_\delta\right\}\\
& \stackrel{\eqref{defChiConv}}{\le} \Pr\{\varepsilon(E^n) \ge 1-\delta\} + \delta \\
&\stackrel{\text{(a)}}{\le} \frac{\varepsilon}{1-\delta} + \delta
\end{align}
where (a) follows from~\eqref{averageEpsilon} and Markov's inequality, which implies that
\begin{align}
\Pr\{E^n \in \Psi_\delta\}\ge 1-\varepsilon-2\delta. \label{defPsi*}
\end{align}
\textbf{Obtaining a Lower Bound on the Error Probability in Terms of the Type-II Error of a Hypothesis Test}\\
 Define
\begin{align}
s_{Y^{n}, \hat W|E^n=e^n}\triangleq \left(\prod_{k=1}^{n} s_{Y_k|E_k=e_k} \right)p_{\hat W|Y^{n}} \label{defDistTildeSLinear}
\end{align}
for all $e^n\in \Psi_\delta$ where
\begin{align}
s_{Y_{k}|E_k=e_k}(y_{k})\triangleq \mathcal{N}\left(y_{k}; 0, e_k+1\right). \label{defTildeSy3Linear}
\end{align}
It follows from Proposition~\ref{propositionBHTLowerBound} and Definition~\ref{defCode}
with the identifications $U\equiv W$, $V\equiv \hat W$, $p_{U,V}\equiv  p_{W,  \hat W|E^n=e^n}$, $|\mathcal{W}|\equiv  M$ and
$
\alpha\equiv \Pr\left\{\left.W\ne  \hat W\right.|E^n=e^n\right\} \le \varepsilon(e^n) 
$
 that
 \begin{align}
\beta_{1-\varepsilon(e^n)}(p_{W,\hat W|E^n=e^n}\|p_{W|E^n=e^n} s_{\hat W|E^n=e^n}) \le \beta_{1-\alpha}(p_{W,\hat W|E^n=e^n}\|p_{W|E^n=e^n} s_{\hat W|E^n=e^n}) \le \frac{1}{M} \label{eqnBHTReverseChainLinear}
 \end{align}
 for all $e^n\in\Psi_\delta$.
\\
\textbf{Using the DPI to Introduce the Channel Inputs and Outputs}\\
Using the DPI of $\beta_{1-\varepsilon(e^n)}$ in Lemma~\ref{lemmaDPI}, we have
 \begin{align}
& \beta_{1- \varepsilon(e^n)}(p_{W,\hat W|E^n=e^n}\|p_{W|E^n=e^n} s_{\hat W|E^n=e^n})\notag\\*
& \ge \beta_{1- \varepsilon(e^n)}\left(p_{W, E^n, X^n, Y^n, \hat W|E^n=e^n}\left\|p_{W|E^n=e^n}\left(\prod_{k=1}^n p_{X_k|W, E^k=e^k} \right) s_{Y^n,\hat W|E^n=e^n}\right.\right) \label{converseProofEq1Linear}
\end{align}
 for all $e^n\in\Psi_\delta$.
For each $e^n\in\Psi_\delta$, fix a $\xi(e^n)>0$ to be specified later. Since
\begin{align}
p_{W,E^n, X^n, Y^n, \hat W|E^n=e^n} &\stackrel{\eqref{memorylessStatement}}{=} p_{W|E^n=e^n} \left(\prod_{k=1}^n p_{X_k|W, E^k=e^k} p_{Y_k|X_k}\right)p_{\hat W |Y^n},
 \end{align}
it follows from \eqref{eqnBHTReverseChainLinear}, the definition of $s_{Y^n,\hat W|E^n=e^n}$ in~\eqref{defDistTildeSLinear}, \eqref{converseProofEq1Linear} and Lemma~\ref{lemmaDPI} that
\begin{align}
\log M &\le \log \xi(e^n) - \log\left(1-\varepsilon(e^n) - \Pr\left\{\left.\sum_{k=1}^n \log\left( \frac{p_{Y_{k}|X_{k}}(Y_{k}|X_{k})}{s_{Y_{k}|E_k}(Y_{k}|E_k)}\right)\ge \log\xi(e^n) \right|E^n=e^n\right\}\right)\\
& = \log \xi(e^n) - \log\left(1-\varepsilon(e^n) - \Pr\left\{\left.\sum_{k=1}^n \log\left( \frac{p_{Y_{k}|X_{k}}(Y_{k}|X_{k})}{s_{Y_{k}|E_k}(Y_{k}|e_k)}\right)\ge \log\xi(e^n) \right|E^n=e^n\right\}\right) \label{converseProofEq2Linear}
\end{align}
for all $e^n\in\Psi_\delta$.
\\
\textbf{Simplifying the Non-Asymptotic Bound}\\
Combining~\eqref{defChannelInDefinition*} and~\eqref{defTildeSy3Linear}, we have
\begin{equation}
\log\frac{p_{Y_k|X_k}(Y_{k}|X_{k})}{s_{Y_{k}|E_k}(Y_{k}|e_k)} = \frac{1}{2}\log(1+e_k)+ \frac{\log \mathrm{e}}{2(1+e_k)}\left(-e_k(Y_k-X_k)^2 + X_k^2 + 2X_k(Y_k-X_k)\right) \label{defLogLikelihoodLinear}
\end{equation}
for each $e^n\in\Psi_\delta$ and each $k\in\{1, 2, \ldots, n\}$.
Let
 \begin{equation}
U_k(e_k) \triangleq \frac{\log \mathrm{e}}{2(1+e_k)}(-e_k(Y_k-X_k)^2 + 2X_k(Y_k-X_k)+e_k) \label{defUkConvLinear}
 \end{equation}
 for each $k\in\{1, 2, \ldots, n\}$. Then, it follows from \eqref{defLogLikelihoodLinear} and the energy-harvesting constraints~\eqref{eqn:eh} that
\begin{equation}
\Pr\left\{\left.\sum_{k=1}^n \log\frac{p_{Y_k|X_k}(Y_{k}|X_{k})}{s_{Y_{k}|E_k}(Y_{k}|e_k)} \le \sum_{k=1}^n\frac{1}{2}\log(1+e_k) + \sum_{k=1}^n U_k(e_k)\right|E^n=e^n\right\}=1.\label{expectationSumUkLinear}
\end{equation}
Combining \eqref{converseProofEq2Linear} and \eqref{expectationSumUkLinear}, we have for each $e^n\in\Psi_\delta$
\begin{align}
\log M \le \log\xi(e^n) - \log\left(1-\varepsilon(e^n)-\Pr\left\{\left.\sum_{k=1}^n U_k(e_k) \ge \log\xi(e^n) - \sum_{k=1}^n\frac{1}{2}\log(1+e_k)\right|E^n=e^n \right\}\right). \label{eqnBHT3rdChainLinear}
\end{align}
In order to simplify the RHS of~\eqref{eqnBHT3rdChainLinear}, we choose
\begin{equation}
\log\xi(e^n)\triangleq  \sum_{k=1}^n\frac{1}{2}\log(1+e_k) + n^{3/4},
\end{equation}
recall the definition of~$\Psi_\delta$ in~\eqref{defPsi} and rewrite \eqref{eqnBHT3rdChainLinear} as
\begin{align}
\log M \le \sum_{k=1}^n\frac{1}{2}\log(1+e_k) + n^{3/4}  - \log\left(\delta-\Pr\left\{\left.\sum_{k=1}^n U_k(e_k) \ge n^{3/4}\right|E^n=e^n \right\}\right). \label{eqnBHT4thChainLinear}
\end{align}
\\
\textbf{Applying Chebyshev's inequality}\\
Following~\eqref{eqnBHT4thChainLinear}, we evaluate for each $e^n\in \Psi_\delta$
\begin{align}
\E\left[\left.\sum_{k=1}^nU_k(e_k)\right|E^n=e^n \right] \stackrel{\text{(a)}}{=} 0 \label{eqnBHT5thChainLinear}
\end{align}
and
\begin{align}
\Var\left[\left.\sum_{k=1}^nU_k(e_k)\right|E^n=e^n \right] &\stackrel{\text{(b)}}{=} (\log \mathrm{e})^2\sum_{k=1}^n\frac{e_k^2 + 2\E[X_k^2|E^n=e^n]}{2(1+e_k)^2}\\
& \le (\log \mathrm{e})^2\sum_{k=1}^n (e_k^2 + 2\E[X_k^2|E^n=e^n])\\
& \stackrel{\eqref{eqn:eh}}{\le} (\log \mathrm{e})^2\sum_{k=1}^n (e_k^2 + 2e_k)\\
&\stackrel{\eqref{defPsi}}{\le} (\log \mathrm{e})^2 n \chi_\delta \label{eqnBHT6thChainLinear}
\end{align}
where (a) and (b) are due to the definition of~$U_k(e_k)$ in~\eqref{defUkConvLinear} and the following fact: $\{Y_k-X_k\}_{k=1}^n$ are i.i.d.\ standard normal random variables that are independent of $(E^n, X^n)$. Using~\eqref{eqnBHT5thChainLinear}, \eqref{eqnBHT6thChainLinear} and Chebyshev's inequality, we have for each $e^n\in \Psi_\delta$
\begin{align}
\Pr\left\{\left.\sum_{k=1}^n U_k(e_k) \ge n^{3/4}\right|E^n=e^n \right\} &\le \frac{(\log \mathrm{e})^2 \chi_\delta}{\sqrt{n}}\\
&\stackrel{\eqref{sufficientlyLargeNConvLinear}}{\le} \delta/2. \label{eqnBHT7thChainLinear}
\end{align}
Combining~\eqref{eqnBHT4thChainLinear} and~\eqref{eqnBHT7thChainLinear}, we have
\begin{align}
\log M \le \inf_{e^n\in \Psi_\delta}\sum_{k=1}^n\frac{1}{2}\log(1+e_k) + n^{3/4}  - \log \delta + 1.  \label{eqnBHT8thChainLinear}
\end{align}
Since the number of energy-arrival blocks of length~$L$ equals $q$ by~\eqref{st*PropBoundOnLastBlock} and the length of the last block is no larger than $ nd + \left\lfloor \frac{1}{\lambda}\right\rfloor$ by~\eqref{stPropBoundOnLastBlock}, it follows from~\eqref{eqnBHT8thChainLinear} that
\begin{align}
\log M &\le \inf_{e^n\in \Psi_\delta}\left\{\sum_{\ell=1}^q\frac{n\lambda}{2}\mathrm{C}(e_{b_\ell+1}) + \left(\frac{nd}{2}+\left\lfloor \frac{1}{\lambda}\right\rfloor\right)\mathrm{C}(e_{b_{q+1}+1})\right\} + n^{3/4}  - \log \delta + 1\\
& \stackrel{\eqref{defPsi}}{\le} n\inf_{e^n\in \Psi_\delta}\left\{\sum_{\ell=1}^q \lambda\mathrm{C}(e_{b_\ell+1}) + d\mathrm{C}(e_{b_{q+1}+1})\right\} + n^{3/4}  - \log \delta + 1 +\left\lfloor \frac{1}{\lambda}\right\rfloor\mathrm{C}(\chi_\delta). \label{eqnBHT9thChainLinear}
\end{align}
In order to simplify the first term in~\eqref{eqnBHT9thChainLinear}, we define $\phi: \mathbb{R}_+^n \rightarrow \mathbb{R}$ as
 \begin{align}
 \phi(e^n)\triangleq \sum_{\ell=1}^q \lambda\mathrm{C}(e_{b_\ell+1}) + d\mathrm{C}(e_{b_{q+1}+1}) \label{defPhi}
 \end{align}
 and consider the following chain of inequalities where the sets $\Psi$ and $\Omega$ are assumed to be Borel measurable:
\begin{align}
\inf_{e^n\in \Psi_\delta}\phi(e^n)
&\stackrel{\eqref{defPsi*}}{\le} \sup\limits_{\substack{\Psi\subseteq\mathbb{R}_+^n:\\ \Pr\{E^n\in \Psi\}\ge 1-\varepsilon-2\delta}} \inf_{e^n\in \Psi}\phi(e^n)\\
& = \sup\limits_{\substack{\Psi\subseteq\mathbb{R}_+^n:\\ \Pr\{E^n\in \Psi\}\ge 1-\varepsilon-2\delta,\\ \Pr\{\phi(E^n)\in \phi(\Psi)\}\ge 1-\varepsilon-2\delta }} \inf\phi(\Psi)\\
& \le \sup\limits_{\substack{\Omega\subseteq\mathbb{R}_+}}\left\{\inf\Omega\left|\, \Pr\left\{\phi(E^n)\in \Omega\right\}\ge 1-\varepsilon-2\delta\right.\right\}\\
&\stackrel{\eqref{defPhi}}{=} \sup\left\{r\in \mathbb{R}_+\left|\, \Pr\left\{\sum_{\ell=1}^q \lambda\mathrm{C}(E_{b_\ell+1}) + d\mathrm{C}(E_{b_{q+1}+1})\ge r\right\}\ge 1-\varepsilon-2\delta\right.\right\}\\
&\stackrel{\eqref{st1thmArrivalBlockLinearInNConv}}{\le}  \overline{R}_\varepsilon+\eta. \label{eqnBHT10thChainLinear}
\end{align}
Combining~\eqref{eqnBHT8thChainLinear} and~\eqref{eqnBHT10thChainLinear}, we obtain
\begin{align}
\log M  \le  n( \overline{R}_\varepsilon+\eta) + n^{3/4}  - \log \delta + 1 +\left\lfloor \frac{1}{\lambda}\right\rfloor\mathrm{C}(\chi_\delta), \label{eqnBHT11thChainLinear}
\end{align}
which together with~\eqref{converseProofEq0Linear} implies that
\begin{align}
R \le  \overline{R}_\varepsilon+\eta. \label{eqnBHT12thChainLinear} 
\end{align}
Since $\eta>0$ is arbitrary and $R$ is an arbitrary $\varepsilon$-achievable rate, the inequality~\eqref{goalInConvProofLinear} follows from~\eqref{eqnBHT12thChainLinear}.
 \section{Concluding Remarks and Future Work} \label{sec:conclusion}
This paper studies the $\varepsilon$-capacity and the second-order coding rate for the AWGN EH channel with an infinite battery under the assumption that the error probabilities do not vanish as the blocklength increases. The EH process is assumed to be block i.i.d.\ where the blocks have length~$L$.

For the case where $L$ is a constant or grows sublinearly in the blocklength~$n$, we have the following two findings stated in Theorem~\ref{thmMainResult}: (i) The $\varepsilon$-capacity is the same for all $\varepsilon\in(0,1)$, i.e., the strong converse holds; (ii) A lower bound and an upper bound on the second-order coding rate have been obtained. where the lower bound is obtained by analyzing the conventional save-and-transmit strategy~\cite{ozel12}.

For the case where $L$ grows linearly in~$n$, we prove in Theorem~\ref{thmArrivalBlockLinearInN} a lower bound and an upper bound on the $\varepsilon$-capacity.

Two interesting directions for future research are obtaining the full characterization of the $\varepsilon$-capacity and good approximations on the second-order coding rate for $L=\lambda n$, i.e., a strengthening of Theorem~\ref{thmArrivalBlockLinearInN}. In addition, while this work investigates only optimal codes which have high decoding complexity, future research may compare the performances between an industrial low-complexity yet (first-order) optimal code for the AWGN channel and its (adaptive) save-and-transmit counterpart for the AWGN EH channel (as performed for the binary-input EH channel in~\cite{FongTanJSAC16}). Finally, one may explore analogies among AWGN EH channels, slow fading channels and channels with mixed states for the case in which $L$ grows linearly in $n$.
\appendices
\section{Proof of Corollary~\ref{corollaryMainResult}}\label{appendixA---}
In view of Theorem~\ref{thmMainResult}, it suffices to prove that $V_\varepsilon^{--}\le V_\varepsilon^{-}$ for the case where~$L$ is a constant. To this end, we let $L$ be a fixed constant and consider the following chain of inequalities:
\begin{align}
V_\varepsilon^-&\stackrel{\eqref{defV-}}{=}\sup\limits_{\substack{(\varepsilon_1, \varepsilon_2)\in (0,1)^2: \\\varepsilon_1+ \varepsilon_2=\varepsilon }}\left\{- \mathrm{C}(P)\sqrt{\varrho\log \frac{1}{\varepsilon_1}} + \sqrt{\frac{P(\log \mathrm{e})^2}{1+P}}\Phi^{-1}(\varepsilon_2)\right\} \\
&  \ge - \mathrm{C}(P)\sqrt{\varrho \log \frac{1}{\big(1-\frac{1}{\sqrt{2\pi \mathrm{e}}}\big)\varepsilon}} + \sqrt{\frac{P(\log \mathrm{e})^2}{1+P}}\Phi^{-1}\bigg(\frac{\varepsilon}{\sqrt{2\pi \mathrm{e}}}\bigg)\\
&\stackrel{\text{(a)}}{\ge} - \mathrm{C}(P)\sqrt{\varrho \log \frac{1}{\big(1-\frac{1}{\sqrt{2\pi \mathrm{e}}}\big)\varepsilon}} - \sqrt{\frac{2P(\log \mathrm{e})^2}{1+P}\ln\frac{\sqrt{\mathrm{e}}}{\varepsilon}}\\
& \stackrel{\text{(b)}}{\ge} - \mathrm{C}(P)\sqrt{\varrho \log \frac{1}{\varepsilon^2}} - \sqrt{\frac{2P(\log \mathrm{e})^2}{1+P}\ln\frac{1}{\varepsilon^2}}\\
& = -\left( \mathrm{C}(P)\sqrt{2\varrho} + \sqrt{\frac{4P\log \mathrm{e}}{1+P}}\right) \sqrt{\log\frac{1}{\varepsilon}}\\
& = V_\varepsilon^{--},
 \end{align}
where
\begin{enumerate}
\item[(a)] is due to the easily verified fact that
 \begin{equation}
 \Phi\left(-\sqrt{2\ln\frac{\sqrt{\mathrm{e}}}{\varepsilon}}\right) \le \frac{1}{\sqrt{2\pi}}\,\mathrm{e}^{-\ln\frac{\sqrt{\mathrm{e}}}{\varepsilon}}= \frac{\varepsilon}{\sqrt{2\pi\mathrm{e}}}.
 \end{equation}
 \item[(b)] is due to the fact that $\varepsilon < 1/2 < \min\big\{1-\frac{1}{\sqrt{2\pi\mathrm{e}}}, \frac{1}{\sqrt{e}} \big\}$.
 \end{enumerate}
\section{Proof of Corollary~\ref{corollaryMainResult*}}\label{appendixA--}
Fix any $\varepsilon\in (0, \Phi(-1))$. Using the easily verified fact that $\Phi^{-1}(\varepsilon) \le -\sqrt{2\ln\frac{1}{\varepsilon}}$, we obtain that
\begin{equation}
V_\varepsilon^+ \le -\sqrt{\frac{(2P^2+\E[E_1^2])\log \mathrm{e}}{2(1+P)^2}}\times \sqrt{\log\frac{1}{\varepsilon}},
\end{equation}
which together with~\eqref{st1CorollaryMainResult} and~\eqref{defV--} implies that~\eqref{st1CorollaryMainResult*} holds. The rest of the proof is dedicated to showing $\Phi^{-1}(\varepsilon) \le -\sqrt{2\ln\frac{1}{\varepsilon}}$. Let $a=\Phi^{-1}(\varepsilon)$. Since $a\le -1$ due to the assumption that $\varepsilon \le \Phi(-1)$, we have $\varepsilon=\Phi(a) \le \mathrm{e}^{-a^2/2}$, which then implies that $a\le -\sqrt{2\ln\frac{1}{\varepsilon}}$.
\section{Proof of Corollary~\ref{corollaryArrivalBlockLinearInN}} \label{appendixA-}
%
Suppose $E_1$ has a continuous and strictly increasing cdf (i.e., the mapping $a\mapsto\Pr\{E_1\le a\}$ is continuous and strictly increasing on $[0, \infty)$). It follows that \begin{equation}
\Pr_{\prod_{\ell=1}^{q+1}p_{E_\ell}}\left\{\sum_{\ell=1}^{q}\lambda\mathrm{C}(E_{\ell}) + d\mathrm{C}(E_{q+1}) \ge r\right\} \label{continuousFunctionInRemark}
 \end{equation}
  is continuous in~$r$ and strictly increasing, which then implies that
 \begin{align}
 \underline{R}_\varepsilon  
 &=\sup\left\{r\in \mathbb{R}_+ \left| \, \Pr_{\prod_{\ell=1}^{q+1}p_{E_\ell}}\left\{\sum_{\ell=1}^{q}\lambda\mathrm{C}(E_{\ell}) + d\mathrm{C}(E_{q+1})< r\right\} = \varepsilon \right.\right\}\\*
 & = R_\varepsilon^{\text{thr}} \\
 & = \inf\left\{r\in \mathbb{R}_+ \left| \, \Pr_{\prod_{\ell=1}^{q+1}p_{E_\ell}}\left\{\sum_{\ell=1}^{q}\lambda\mathrm{C}(E_{\ell}) + d\mathrm{C}(E_{q+1})< r\right\} = \varepsilon \right.\right\}\\*
 & =  \overline{R}_\varepsilon,\label{stCorollaryBlockLinearInN*}
 \end{align}
 which together with Theorem~\ref{thmArrivalBlockLinearInN} implies that \eqref{stCorollaryBlockLinearInN}
 holds for all $\varepsilon\in(0,1)$.
\section{Proof of Corollary~\ref{corollarySaveTransmit}} \label{appendixC}
To facilitate discussion, let $m_n$ denote the RHS of~\eqref{st3ThmSaveAndTransmit}, and simple calculations reveal that
 \begin{align}
 \lim_{n\rightarrow \infty}\frac{m_n}{\sqrt{L n}}& = \sqrt{2\left(\frac{\E[E_1^2]}{P^2} +1\right)\log \frac{1}{\varepsilon_1}}\\
 & =\sqrt{\varrho\log \frac{1}{\varepsilon_1}} \,. \label{eq0CorollarySaveTransmit}
 \end{align}
 For each $n^*\in\mathbb{N}$, let $\tilde n$ be the unique natural number that satisfies
\begin{align}
\tilde n -1 + \lfloor m_{\tilde n -1}\rfloor \le n^* < \tilde n + \lfloor m_{\tilde n}\rfloor. \label{st1CorollarySaveTransmit}
\end{align}
It is clear from~\eqref{eq0CorollarySaveTransmit} and~\eqref{st1CorollarySaveTransmit} that
\begin{align}
\lim_{n^*\rightarrow \infty} \frac{n^*}{\tilde n} = 1.\label{st1*CorollarySaveTransmit}
\end{align}
 By Lemma~\ref{lemmaSaveAndTransmit}, there exists for each sufficiently large~$n\in\mathbb{N}$ an~$(n + \lfloor m_n\rfloor, M, \varepsilon)$-code such that~\eqref{st4ThmSaveAndTransmit} holds, which implies from the left inequality in~\eqref{st1CorollarySaveTransmit} that for each sufficiently large~$n^*\in\mathbb{N}$, there exists  an~$(n^*, M, \varepsilon)$-code such that
 \begin{align}
 \log M \ge (\tilde n -1 )\mathrm{C}(P) + \sqrt{\frac{(\tilde n -1 )P(\log \mathrm{e})^2}{1+P}}\Phi^{-1}(\varepsilon_2) - \frac{1}{2}\log (\tilde n -1 )-\kappa_1,
 \end{align}
 which then implies from the right inequality in~\eqref{st1CorollarySaveTransmit} that
 \begin{align}
  \log M \ge (n^* -1 - \lfloor m_{\tilde n}\rfloor)\mathrm{C}(P) + \sqrt{\frac{(\tilde n -1 )P(\log \mathrm{e})^2}{1+P}}\Phi^{-1}(\varepsilon_2) - \frac{1}{2}\log (\tilde n -1 )-\kappa_1. \label{eq1CorollarySaveTransmit}
 \end{align}
Combining the facts that
 \begin{align}
 \frac{ \log M - n^*\mathrm{C}(P)}{\sqrt{L n^*}} \stackrel{\eqref{eq1CorollarySaveTransmit}}{\ge} \frac{( -1 - \lfloor m_{\tilde n}\rfloor)\mathrm{C}(P)}{\sqrt{L n^*}} + \sqrt{\frac{ (\tilde n -1 )P(\log \mathrm{e})^2}{L n^*(1+P)}}\Phi^{-1}(\varepsilon_2) - \frac{1}{2\sqrt{L n^*}}\log (\tilde n -1 )-\frac{\kappa_1}{\sqrt{L n^*}},
 \end{align}
 \begin{align}
 \lim_{n^*\rightarrow \infty} \frac{\tilde n}{n^*} \stackrel{\eqref{st1*CorollarySaveTransmit}}{=} 1 \label{eq1*CorollarySaveTransmit}
 \end{align}
 and
 \begin{align}
\lim_{n^*\rightarrow\infty}\frac{\lfloor m_{\tilde n}\rfloor}{\sqrt{L n^*}} \stackrel{\eqref{eq1*CorollarySaveTransmit}}{=} \lim_{\tilde n\rightarrow\infty}\frac{\lfloor m_{\tilde n}\rfloor}{\sqrt{L\tilde n}} \stackrel{\eqref{eq0CorollarySaveTransmit}}{=} \sqrt{\varrho\log \frac{1}{\varepsilon_1}},
 \end{align}
we conclude that
\begin{align}
\liminf_{n^*\rightarrow\infty}\frac{ \log M - n^*\mathrm{C}(P)}{\sqrt{L n^*}} &\ge -\mathrm{C}(P)\sqrt{\varrho\log \frac{1}{\varepsilon_1}} + \liminf_{n^*\rightarrow\infty}\sqrt{\frac{P(\log \mathrm{e})^2}{L(1+P)}}\Phi^{-1}(\varepsilon_2)\\
& = \begin{cases}
-\mathrm{C}(P)\sqrt{\varrho\log \frac{1}{\varepsilon_1}}& \text{if $L=\omega(1)$,}\\
-\mathrm{C}(P)\sqrt{\varrho\log \frac{1}{\varepsilon_1}} + \sqrt{\frac{P(\log \mathrm{e})^2}{L(1+P)}}\Phi^{-1}(\varepsilon_2) &\text{if $L$ is a constant.}
\end{cases}
\end{align}
This completes the proof.
\section{Proof of Lemma~\ref{lemmaCharacteristicFunction}}\label{appendixA}
In this proof,  all the probability, expectation and variance terms are evaluated according to $p_{X^n}p_{E^{m+n}}$. In order to obtain an upper bound on
$\Pr\left\{\bigcup_{k=1}^n \left\{\sum_{i=1}^k X_i^2 \ge \sum_{i=1}^{m+k} E_i\right\}\right\}$, we construct the following sequence denoted by $\{B_k\}_{k=1}^{m+n}$. For each $k\in\{1, 2, \ldots, m+n\}$, define $B_k$ recursively\footnote{The construction of $\{B_k\}_{k=1}^{m+n}$ is inspired by a standard proof of Kolmogorov's inequality.} as
\begin{equation}
B_k\triangleq
\begin{cases}
E_1 & \text{if $k=1$,}\\
B_{k-1} +  E_k & \text{if $k\in\{2, 3, \ldots, m\}$,}\\
B_{k-1} + E_k - X_{k-m}^2 & \parbox[t]{3 in}{if $k\in\{m+1, m+2, \ldots, m+n\}$ and $B_{k-1} > 0$,}\\
B_{k-1} & \parbox[t]{3 in}{if $k\in\{m+1, m+2, \ldots, m+n\}$ and $B_{k-1} \le 0$.}
\end{cases} \label{defBk}
\end{equation}
By inspecting~\eqref{defBk}, we have
\begin{equation}
\{B_{m+n}\le 0\}=\bigcup_{k=1}^n\left\{\sum_{i=1}^{m+k} E_i-\sum_{i=1}^k X_i^2\le 0\right\},
\end{equation}
which implies that
\begin{align}
\Pr\left\{\bigcup_{k=1}^n \left\{\sum_{i=1}^k X_i^2 \ge \sum_{i=1}^{m+k} E_i\right\}\right\} = \Pr\left\{B_{m+n}\le 0\right\}. \label{appendixLemma1Eq1}
\end{align}
Define for each $k\in\{1, 2, \ldots, m+n\}$
 \begin{align}
 U_k &\triangleq
 \begin{cases}
 B_1 & \text{if $k=1$,}\\
 B_k-B_{k-1} & \text{otherwise}
 \end{cases}  \label{defUk}\\
 & \stackrel{\eqref{defBk}}{=}
 \begin{cases}
E_k & \text{if $k\in\{1, 2, \ldots, m\}$,}\\
E_k - X_{k-m}^2 & \parbox[t]{3 in}{if $k\in\{m+1, m+2, \ldots, m+n\}$ and $B_{k-1} > 0$,}\\
0 & \parbox[t]{3 in}{if $k\in\{m+1, m+2, \ldots, m+n\}$ and $B_{k-1} \le 0$.}
\end{cases}  \label{defUk*}
 \end{align}
Following~\eqref{appendixLemma1Eq1}, we consider the following chain of inequalities for any $t>0$:
 \begin{align}
 \Pr\left\{B_{m+n}\le 0\right\} &\stackrel{\eqref{defUk}}{=}  \Pr\left\{\sum_{k=1}^{m+n}U_k\le 0\right\} \\
 &= \Pr\left\{\mathrm{e}^{-t\sum_{k=1}^{m+n}U_k} \ge 1\right\}\\
 &\stackrel{\text{(a)}}{\le} \E\left[\mathrm{e}^{-t\sum_{k=1}^{m+n}U_k}\right] \label{appendixLemma1Eq2}
 \end{align}
 where (a) follows from Markov's inequality. In order to simplify the RHS of \eqref{appendixLemma1Eq2}, we consider the following chain of inequalities for each $i\in\{1, \ldots, n\}$:
 \begin{align}
 \E\left[\mathrm{e}^{-t\sum_{k=1}^{m+i}U_k}\right] & =  \E\left[ \E\left[\left.\mathrm{e}^{-t\sum_{k=1}^{m+i}U_k}\right| U^{m+i-1} \right] \right] \\
 & =   \E\left[ \E\left[ \E\left[\left.\mathrm{e}^{-tU_{m+i}}\right| U^{m+i-1}\right]\left.\mathrm{e}^{-t\sum_{k=1}^{m+i-1}U_k}\right| U^{m+i-1} \right] \right]  \\
 &\stackrel{\eqref{defUk*}}{\le} \E\left[ \E\left[\left.\max\left\{\E\left[\left. \mathrm{e}^{-t(E_{m+i} - X_i^2)}\right|U^{m+i-1}\right] , 1\right\} \mathrm{e}^{-t\sum_{k=1}^{m+i-1}U_k} \right| U^{m+i-1}\right]                 \right]\\
 &\stackrel{\text{(a)}}{=}  \max\left\{\E\left[ \mathrm{e}^{-t(E_{m+i} - X_i^2)}\right] , 1\right\}  \E\left[\mathrm{e}^{-t\sum_{k=1}^{m+i-1}U_k}\right] \\
 & =  \max\left\{\E\left[ \mathrm{e}^{-t(E_1 - X_1^2)}\right] , 1\right\}  \E\left[\mathrm{e}^{-t\sum_{k=1}^{m+i-1}U_k}\right]  \label{appendixLemma1Eq3}
 \end{align}
 where (a) follows from the independence between $(E_{m+i}, X_{i})$ and $U^{m+i-1}$ due to the independence between $(E_{m+i}, X_{i})$ and $(E^{m+i-1}, X^{i-1})$. Combining~\eqref{appendixLemma1Eq2} and \eqref{appendixLemma1Eq3}, we have
 \begin{align}
  \Pr\left\{B_{m+n}\le 0\right\} \le \max\left\{ \left(\E\left[ \mathrm{e}^{-t(E_1 - X_1^2)}\right]\right)^{n},1\right\} \E\left[\mathrm{e}^{-t\sum_{k=1}^{m}U_k}\right].  \label{appendixLemma1Eq4}
 \end{align}
 Since
 \begin{equation}
 \E\left[\mathrm{e}^{-t\sum_{k=1}^{m}U_k}\right] = \E\left[\mathrm{e}^{-t\sum_{k=1}^{m}E_k}\right]
 \end{equation}
by~\eqref{defUk*}, it follows from~\eqref{appendixLemma1Eq4} that
 \begin{align}
   \Pr\left\{B_{m+n}\le 0\right\} \le \max\left\{ \left(\E\left[ \mathrm{e}^{-t(E_1 - X_1^2)}\right]\right)^{n},1\right\}\left(\E\left[\mathrm{e}^{-tE_1}\right]\right)^m.  \label{appendixLemma1Eq5}
 \end{align}
 Since $X_1$ and $E_1$ are independent, we can rewrite~\eqref{appendixLemma1Eq5} as
 \begin{align}
   \Pr\left\{B_{m+n}\le 0\right\} \le \max\left\{\left(\E\left[\mathrm{e}^{-tE_1}\right]\right)^{m+n}\left(\E\left[\mathrm{e}^{tX_1^2}\right]\right)^{n},\left(\E\left[\mathrm{e}^{-tE_1}\right]\right)^m\right\}. \label{appendixLemma1Eq6}
 \end{align}
 In order to simplify the RHS of~\eqref{appendixLemma1Eq6}, we use the following two facts, whose proofs can be found in~\cite[Appendix]{FTY15}:
 For any $y\ge 0$,
\begin{equation}
1+y\le \mathrm{e}^y \le 1+ y+ \frac{y^2 \mathrm{e}^y}{2} \label{propositionStatement1}
\end{equation}
and
\begin{equation}
1-y \le \mathrm{e}^{-y} \le 1 - y+ \frac{y^2}{2}\,. \label{propositionStatement2}
\end{equation}
Fix a sufficiently small $t>0$ such that $ \E[X_1^4 \mathrm{e}^{tX_1^2}]<\infty$ and $P - \frac{t\E[E_1^2]}{2} >0$.
Following~\eqref{appendixLemma1Eq6}, we use~\eqref{propositionStatement1}, \eqref{propositionStatement2} and~\eqref{lemmaCharFuncAssump1} to obtain
\begin{align}
\E\left[\mathrm{e}^{-tE_1}\right] \le 1-tP + \frac{t^2\E[E_1^2]}{2}   \label{appendixLemma1Eq7}
\end{align}
and
\begin{align}
\E\left[\mathrm{e}^{tX_1^2}\right] \le 1+tP + \frac{t^2\E[X_1^4 \mathrm{e}^{tX_1^2}]}{2},
\end{align}
which implies that
\begin{align}
&\E\left[\mathrm{e}^{-tE_1}\right]\E\left[\mathrm{e}^{tX_1^2}\right] \notag\\
&\le 1+t^2\left(\frac{\E[E_1^2] + \E[X_1^4 \mathrm{e}^{tX_1^2}]}{2} -P^2\right) + \frac{t^3 P}{2}(\E[E_1^2] - \E[X_1^4 \mathrm{e}^{tX_1^2}]) + \frac{t^4\E[E_1^2]\E[X_1^4 \mathrm{e}^{tX_1^2}]}{4}.   \label{appendixLemma1Eq8}
\end{align}
Define $a_t$ and $b_t$ as in~\eqref{defAt} and~\eqref{defBt} respectively. It then follows from~\eqref{appendixLemma1Eq7} and~\eqref{appendixLemma1Eq8} that
\begin{align}
\E\left[\mathrm{e}^{-tE_1}\right] \le 1-a_tt   \label{appendixLemma1Eq9}
\end{align}
and
\begin{align}
\E\left[\mathrm{e}^{-tE_1}\right]\E\left[\mathrm{e}^{tX_1^2}\right] \le 1+b_tt^2.   \label{appendixLemma1Eq10}
\end{align}
Combining~\eqref{appendixLemma1Eq1}, \eqref{appendixLemma1Eq6}, \eqref{appendixLemma1Eq9}, \eqref{appendixLemma1Eq10}, \eqref{propositionStatement1} and~\eqref{propositionStatement2}, we obtain
\begin{align}
 \Pr\left\{\bigcup_{k=1}^n \left\{\sum_{i=1}^k X_i^2 \ge \sum_{i=1}^{m+k} E_i\right\}\right\} &\le  \mathrm{e}^{-a_ttm}\max\left\{   \mathrm{e}^{b_t t^2n} ,1\right\} \\
 & = \mathrm{e}^{-a_ttm+b_tt^2n}.
\end{align}
\section{Proof of Corollary~\ref{corollaryBlkCharFunc}}\label{appendixB}
Since $\{E_k\}_{k=1}^{m+n}$ is distributed according to an i.i.d.-block manner with block size~$L$, we cannot apply Lemma~\ref{lemmaCharacteristicFunction} directly for $L>1$ to bound $\Pr_{p_{X^n}p_{E^{m+n}}}\left\{\bigcup_{k=1}^n \left\{\sum_{i=1}^k X_i^2 \ge \sum_{i=1}^{m+k} E_i\right\}\right\}$. In the following, we will construct two sequences based on $\{E_k\}_{k=1}^{m+n}$ and $\{X_k\}_{k=1}^{n}$ so that Lemma~\ref{lemmaCharacteristicFunction} can be applied to the resultant sequences.
Define
\begin{equation}
\bar n\triangleq \left\lceil \frac{n}{L} \right\rceil L \label{defBarN}
 \end{equation}
 and
 \begin{equation}
 \bar m\triangleq \left\lfloor \frac{m}{L}\right\rfloor L. \label{defBarM}
  \end{equation}
  Let $\{X_k\}_{k=1}^{\bar n}$ be a sequence of i.i.d.\ random variables where $X_1\sim\mathcal{N}(x_1; 0, P)$, and let $\{E_k\}_{k=1}^{\bar m+\bar n}$ be a sequence of random variables that are distributed according to~\eqref{defDistEk}. Since $\bar n \ge n$ and $\bar m \le m$, we have
\begin{align}
&\Pr_{p_{X^n}p_{E^{m+n}}}\left\{\bigcup_{k=1}^n \left\{\sum_{i=1}^k X_i^2 \ge \sum_{i=1}^{m+k} E_i\right\}\right\} \notag\\*
& \le \Pr_{p_{X^{\bar n}}p_{E^{{\bar m}+{\bar n}}}}\left\{\bigcup_{k=1}^{\bar n} \left\{\sum_{i=1}^k X_i^2 \ge \sum_{i=1}^{{\bar m}+k} E_i\right\}\right\}. \label{corollaryEqn1}
\end{align}
To simplify notation, define
\begin{equation}
\tilde{E}_{\ell}\triangleq \sum_{i=1}^{L}E_{b_{\ell}+i} \label{defTildeE}
\end{equation}
for each $\ell\in\{1, 2, \ldots, (\bar m+\bar n)/L\}$, and define
\begin{equation}
\tilde{X}_{\ell}\triangleq \sum_{i=1}^{L}X_{b_{\ell}+i} \label{defTildeX}
\end{equation}
for each $\ell\in\{1, 2, \ldots, \bar n/L\}$. For each $k\in\{1, 2, \ldots, \bar n\}$, define $\nu(k)$ to be the unique integer in $\{1, 2, \ldots, \bar n/L\}$ that satisfies
$k\in\{b_{\nu(k)}+1, b_{\nu(k)}+2, \ldots, b_{\nu(k)} + L\}$, i.e., $\nu(k)$ is the index of the information block that contains the $k^{\text{th}}$ symbol of $X^{\bar n}$. Then for each $k\in\{1, 2, \ldots, \bar n\}$,
\begin{equation}
\sum_{i=1}^k X_i^2 \le \sum_{i=1}^{\nu(k)}\tilde X_i^2
\end{equation}
and
\begin{equation}
\sum_{i=1}^{\bar m + k} E_i \ge \sum_{i=1}^{\frac{\bar m}{L}+\nu(k)-1}\tilde E_i
\end{equation}
with probability~$1$. Therefore,
 \begin{equation}
 \left\{\sum_{i=1}^k X_i^2 \ge \sum_{i=1}^{{\bar m}+k} E_i\right\}\subseteq \left\{\sum_{i=1}^{\nu(k)} \tilde X_i^2 \ge \sum_{i=1}^{\frac{\bar m}{L}+\nu(k)-1} \tilde E_i\right\}
 \end{equation}
 for each $k\in\{1, 2, \ldots, \bar n\}$,
which implies that
 \begin{align}
&\Pr_{p_{X^{\bar n}}p_{E^{{\bar m}+{\bar n}}}}\left\{\bigcup_{k=1}^{\bar n} \left\{\sum_{i=1}^k X_i^2 \ge \sum_{i=1}^{{\bar m}+k} E_i\right\}\right\}\notag\\
& \le  \Pr_{p_{X^{\bar n}}p_{E^{{\bar m}+{\bar n}}}}\left\{\bigcup_{\ell=1}^{\frac{\bar n}{L}} \left\{\sum_{i=1}^\ell \tilde X_i^2 \ge \sum_{i=1}^{\frac{\bar m}{L}+\ell-1} \tilde E_i\right\}\right\},
\end{align}
which then implies from~\eqref{corollaryEqn1} that
\begin{align}
\Pr_{p_{X^n}p_{E^{m+n}}}\left\{\bigcup_{k=1}^n \left\{\sum_{i=1}^k X_i^2 \ge \sum_{i=1}^{m+k} E_i\right\}\right\}\le  \Pr_{p_{X^{\bar n}}p_{E^{{\bar m}-L+{\bar n}}}}\left\{\bigcup_{\ell=1}^{\frac{\bar n}{L}} \left\{\sum_{i=1}^\ell \tilde X_i^2 \ge \sum_{i=1}^{\frac{\bar m}{L}-1+\ell} \tilde E_i\right\}\right\}.\label{corollaryEqn1*}
\end{align}
%
By construction, $\left\{\tilde{E}_{\ell}\right\}_{\ell=1}^{(\bar m -L+\bar n)/L}$ is a sequence of i.i.d.\ random variables with
\begin{equation}
\E\left[\tilde E_1\right] =  \E\left[L E_{b_{1}+1}\right]=  LP \label{corollaryEqn2}
 \end{equation}
 and
 \begin{equation}
 \E\left[\tilde E_1^2\right]= \E\left[\left(LE_{b_{1}+1}\right)^2\right]=L^2\E[E_1^2], \label{corollaryEqn2*}
  \end{equation}
  and $\left\{\tilde{X}_{\ell}\right\}_{\ell=1}^{\bar n/L}$ is a sequence of i.i.d.\ Gaussian random variables with
  \begin{equation}
  \E\left[\tilde X_1\right]=0,
  \end{equation}
    \begin{equation}
  \E\left[\tilde X_1^2\right]=LP, \label{corollaryEqn3}
  \end{equation}
  and
    \begin{equation}
  \E\left[\tilde X_1^4 \mathrm{e}^{t\tilde X_1^2}\right]\stackrel{\eqref{eqnNormalDist}}{=}\frac{3L^2P^2}{(1-2LPt)^{5/2}} \qquad \forall t\in\left(0, \frac{1}{2LP}\right). \label{corollaryEqn4}
  \end{equation}
 Combining~\eqref{corollaryEqn2} and \eqref{corollaryEqn3} and~\eqref{corollaryEqn4}, we have
  \begin{equation}
   \E\left[\tilde E_1\right]=\E\left[\tilde X_1^2\right]=LP \label{corollaryEqn5}
  \end{equation}
  and
      \begin{equation}
   \E\left[\tilde X_1^4 \mathrm{e}^{t\tilde X_1^2}\right]<\infty \qquad \forall t\in\left(0, \frac{1}{2LP}\right).\label{corollaryEqn6}
  \end{equation}
 Since~\eqref{corollaryEqn5} holds, we can apply Lemma~\ref{lemmaCharacteristicFunction} to $\left\{\tilde{X}_{\ell}\right\}_{\ell=1}^{\bar n/L}$ and $\left\{\tilde{E}_{\ell}\right\}_{\ell=1}^{(\bar m -L+\bar n)/L}$ if the following two statements hold for $t_n$ (which was defined in~\eqref{defTnBlk}):
  \begin{equation}
   \E\left[\tilde X_1^4 \mathrm{e}^{t_n\tilde X_1^2}\right]<\infty \label{assumption1InProof}
\end{equation}
and
  \begin{align}
  LP - \frac{t_n\E[\tilde X_1^2]}{2}>0. \label{assumption2InProof}
  \end{align}
  To this end, we suppose $n$ and $m$ satisfy~\eqref{st1CorollaryCharFuncBlk} and~\eqref{st2CorollaryCharFuncBlk} respectively. Recalling the definition of $\beta_0$ in~\eqref{defBeta0}, we obtain from the definition of~$t_n$ in~\eqref{defTnBlk} and~\eqref{st1CorollaryCharFuncBlk} that
   \begin{align}
   t_n\in\left(0, \min\left\{\frac{1}{2LP}, \frac{2P}{L\E[E_1^2]}\right\}\right), \label{defTnInterval}
   \end{align}
   which implies from~\eqref{corollaryEqn6} that
\begin{equation}
   \E\left[\tilde X_1^4 \mathrm{e}^{t_n\tilde X_1^2}\right]<\infty. \label{corollaryEqn7}
\end{equation}
In addition,
\begin{align}
LP - \frac{t_n\E[\tilde E_1^2]}{2}&\stackrel{\eqref{corollaryEqn5}}{=}LP - \frac{t_nL^2\E[E_1^2]}{2} \\*
&\stackrel{\eqref{defTnInterval}}{>}0.   \label{corollaryEqn8}
\end{align}
Consequently, $\left\{\tilde{E}_{\ell}\right\}_{\ell=1}^{(\bar m -L+\bar n)/L}$ and $\left\{\tilde{X}_{\ell}\right\}_{\ell=1}^{\bar n/L}$ are two sequences of i.i.d.\ random variables that satisfy~\eqref{corollaryEqn5}, \eqref{corollaryEqn7} and~\eqref{corollaryEqn8}. Therefore, we have the following inequality due to Lemma~\ref{lemmaCharacteristicFunction} together with the definitions of $\alpha_t$ and $\beta_t$ in~\eqref{defAlphaT} and~\eqref{defBetaT} respectively and the equalities~\eqref{corollaryEqn2}, \eqref{corollaryEqn2*} and \eqref{corollaryEqn4}:
\begin{align}
\Pr_{p_{X^{\bar n}}p_{E^{{\bar m}-L+{\bar n}}}}\left\{\bigcup_{\ell=1}^{\frac{\bar n}{L}} \left\{\sum_{i=1}^\ell \tilde X_i^2 \ge \sum_{i=1}^{\frac{\bar m -L}{L}+\ell} \tilde E_i\right\}\right\} \le \mathrm{e}^{-\alpha_{t_n}t_n (\bar m/L - 1)+\beta_{t_n}t_n^2\bar n/L}. \label{corollaryEqn9}
\end{align}
In order to simplify the RHS of~\eqref{corollaryEqn9}, we consider
 \begin{align}
 \bar m/L -1 &\stackrel{\eqref{defBarM}}{>} m/L-2 \\
  & \stackrel{\eqref{st2CorollaryCharFuncBlk}}{>}\frac{\sqrt{ (n/L+1) \log(1/\varepsilon_1)}(\beta_{t_n}+\beta_0)}{\alpha_{t_n}\sqrt{\beta_0}}\\
  &\stackrel{\eqref{defBarN}}{>}\frac{\sqrt{ (\bar n/L) \log(1/\varepsilon_1)}(\beta_{t_n}+\beta_0)}{\alpha_{t_n}\sqrt{\beta_0}}. \label{defBarM*}
 \end{align}
Following~\eqref{corollaryEqn9}, we consider
\begin{align}
-\alpha_{t_n}t_n (\bar m/L - 1)+\beta_{t_n}t_n^2\bar n/L &\stackrel{\eqref{defBarM*}}{<} t_n\left(\frac{-\sqrt{ (\bar n /L) \log(1/\varepsilon_1)}(\beta_{t_n}+\beta_0)}{\sqrt{\beta_0}}+\frac{\beta_{t_n}t_n \bar n}{L}\right) \\
& \stackrel{\text{(a)}}{=} -t_n\sqrt{ \beta_0(\bar n /L) \log(1/\varepsilon_1)}\\
& \stackrel{\text{(b)}}{=} \log(\varepsilon_1) \label{corollaryEqn10}
\end{align}
where (a) and (b) follow from the fact due to~\eqref{defTnBlk} and~\eqref{defBarN} that $t_n =  \sqrt{\frac{\log(1/\varepsilon_1)}{(\bar n/L)  \beta_0}}$.
Combining~\eqref{corollaryEqn9} and~\eqref{corollaryEqn10}, we have
\begin{align}
\Pr_{p_{X^{\bar n}}p_{E^{{\bar m}-L+{\bar n}}}}\left\{\bigcup_{\ell=1}^{\frac{\bar n}{L}} \left\{\sum_{i=1}^\ell \tilde X_i^2 \ge \sum_{i=1}^{\frac{\bar m -L}{L}+\ell} \tilde E_i\right\}\right\} \le \varepsilon_1,
\end{align}
which implies from~\eqref{corollaryEqn1*} that~\eqref{corollaryEqnBlk} holds.

\section{Simple derivations in the proof of Lemma~\ref{lemmaSaveAndTransmit}}
\subsection{Derivation of~\eqref{defT*}}\label{appendixB+}
Consider the following chain of inequalities:
\begin{align}
& \left(\E_{p_{X,Y}}\left[\left|\log\left(\frac{p_{Y|X}(Y|X)}{p_{Y}(Y)}\right)-\mu\right|^3\right]\right)^{1/3} \notag\\
& \stackrel{\text{(a)}}{\le} \frac{\log \mathrm{e}}{2(1+P)}\left(P\left(\E_{p_{Z}}\left[Z^6\right]\right)^{1/3}+2\left(\E_{p_{X}}\left[|X|^3\right]\E_{p_{Z}}\left[|Z|^3\right]\right)^{1/3}+ \left(\E_{p_{X}}\left[X^6\right]\right)^{1/3}\right)\\
& = \frac{\log \mathrm{e}}{2(1+P)}\left(15^{1/3}P+\frac{16\sqrt{P}}{\pi} +15^{1/3}P\right)\\
&= \frac{\log \mathrm{e}}{1+P}\left(15^{1/3}P+\frac{8\sqrt{P}}{\pi}\right).
\end{align}
where (a) follows from \eqref{defInfoSpectrum} and the triangle inequality for the $3$-norm.
\subsection{Derivation of \eqref{eqn11*InCalculationErrorProb}}\label{appendixB++}
By Taylor's theorem, we have
\begin{equation}
\Phi^{-1}\left(\varepsilon_2-\frac{T}{\sigma^3\sqrt{n}}-\frac{1}{\sqrt{n}}\right) = \Phi^{-1}(\varepsilon_2) - \left(\frac{T}{\sigma^3\sqrt{n}}+\frac{1}{\sqrt{n}}\right) \left(\Phi^{-1}\right)^\prime(c) \label{eqn11InCalculationErrorProb}
\end{equation}
for some real number
\begin{equation}
c\in \left[\varepsilon_2 -\frac{T}{\sigma^3\sqrt{n}}-\frac{1}{\sqrt{n}},\varepsilon_2\right]\stackrel{\eqref{st2ThmSaveAndTransmit}}{\subseteq}\left[\varepsilon_2^2,\varepsilon_2\right]. \label{eqn12InCalculationErrorProb}
 \end{equation}
Since
\begin{equation}
\left(\Phi^{-1}\right)^\prime(c) = \frac{1}{\Phi^\prime(\Phi^{-1}(c))}= \frac{1}{\mathcal{N}\left(\Phi^{-1}(c);0,1\right)}
\end{equation}
and
\begin{equation}
\mathcal{N}\left(\Phi^{-1}(c);0,1\right) \ge \mathcal{N}\left(\Phi^{-1}(\min\{\varepsilon_2^2, 1-\varepsilon_2\});0,1\right)
\end{equation}
by~\eqref{eqnNormalDist} and~\eqref{eqn12InCalculationErrorProb} respectively, it follows that
 \begin{align}
 \left(\Phi^{-1}\right)^\prime(c) \le \frac{1}{\mathcal{N}\left(\Phi^{-1}(\min\{\varepsilon_2^2, 1-\varepsilon_2\});0,1\right)}. \label{eqn12**InCalculationErrorProb}
 \end{align}
 Consequently, \eqref{eqn11*InCalculationErrorProb} follows from~\eqref{eqn11InCalculationErrorProb} and~\eqref{eqn12**InCalculationErrorProb}.
\section{Simple derivations in the proof of Lemma~\ref{lemmaConverse}}
\subsection{Derivation of~\eqref{defT*Conv}}\label{appendixC+}
Consider the following chain of inequalities:
\begin{align}
T^{1/3}& \stackrel{\eqref{defV1*}}{=} \left(\E_{p_{E_1}\prod_{i=1}^Lp_{Z_i}}\left[\left|-P\sum_{i=1}^L Z_i^2 + 2\sqrt{E_1}\sum_{i=1}^L Z_i + L E_1\right|^3\right]\right)^{1/3}\notag\\
& \stackrel{\text{(a)}}{\le}  \frac{\log \mathrm{e}}{2(1+P)}\left(LP\left(\E_{p_{Z_1}}\left[Z_1^6\right]\right)^{1/3}+2L \left(\E[E_1^{3/2}]\right)^{1/3}\cdot\left(\E_{p_{Z_1}}[\left|Z_1\right|^3]\right)^{1/3}+L\left(\E[E_1^3]\right)^{1/3}\right)\\
& = \frac{L\log \mathrm{e}}{2(1+P)}\left(15^{1/3}P+2(2\sqrt{2/\pi})^{1/3}\cdot \left(\E[E_1^{3/2}]\right)^{1/3}+\left(\E[E_1^3]\right)^{1/3}\right)
\end{align}
where (a) follows from the triangle inequality for the $3$-norm.
\subsection{Derivation of~\eqref{eqnBHT9thChain*}} \label{appendixC++}

By Taylor's theorem, we have
\begin{equation}
\Phi^{-1}\left(\varepsilon +2\tau_2\sqrt{\frac{L}{n}}\right) = \Phi^{-1}(\varepsilon) + 2\tau_2\sqrt{\frac{L}{n}} \left(\Phi^{-1}\right)^\prime(c) \label{eqnBHT9thChain}
\end{equation}
for some real number
\begin{equation}
c\in \left[\varepsilon, \varepsilon + 2\tau_2\sqrt{\frac{L}{n}} \right]\stackrel{\eqref{defTau2*}}{\subseteq}\left[\varepsilon,\varepsilon+(1-\varepsilon)^2\right]. \label{eqnBHT10thChain}
 \end{equation}
Since
\begin{equation}
\left(\Phi^{-1}\right)^\prime(c) = \frac{1}{\Phi^\prime(\Phi^{-1}(c))}= \frac{1}{\mathcal{N}\left(\Phi^{-1}(c);0,1\right)}
\end{equation}
by~\eqref{eqnNormalDist}
and
\begin{align}
\mathcal{N}\left(\Phi^{-1}(c);0,1\right) &\ge \mathcal{N}\left(\Phi^{-1}(\min\{\varepsilon, 1-\varepsilon - (1-\varepsilon)^2\});0,1\right)\\
& = \mathcal{N}\left(\Phi^{-1}(\min\{\varepsilon, \varepsilon(1-\varepsilon)\});0,1\right)
\end{align}
by~\eqref{eqnBHT10thChain}, it follows that
 \begin{align}
 \left(\Phi^{-1}\right)^\prime(c) \le \frac{1}{\mathcal{N}\left(\Phi^{-1}(\min\{\varepsilon, \varepsilon(1-\varepsilon)\});0,1\right)}. \label{eqnBHT11thChain}
 \end{align}
\section{Proof of Proposition~\ref{propositionBoundOnLastBlock}} \label{appendixLastBlock}
Fix any sufficiently large~$n$ such that
\begin{equation}
\left\lfloor\frac{1}{\lambda}\right\rfloor = \left\lfloor\frac{1}{\lambda-\frac{1}{n}}\right\rfloor. \label{eqn1PropositionBoundLastBlockProof}
\end{equation}
Then,
\begin{align}
\rho & \stackrel{\eqref{defRho}}{=}  \left\lfloor\frac{n}{\lfloor \lambda n \rfloor}\right\rfloor\\
&\ge \left\lfloor\frac{n}{ \lambda n }\right\rfloor\\
& \stackrel{\eqref{defQ}}{=}  q. \label{eqn1*PropositionBoundLastBlockProof}
\end{align}
In addition,
\begin{align}
\rho & \stackrel{\eqref{defRho}}{=}  \left\lfloor\frac{n}{\lfloor \lambda n \rfloor}\right\rfloor\\
&\le \left\lfloor\frac{n}{ \lambda n -1}\right\rfloor\\
&\stackrel{\eqref{eqn1PropositionBoundLastBlockProof}}{=} \left\lfloor\frac{1}{\lambda}\right\rfloor \\
& \stackrel{\eqref{defQ}}{=}  q. \label{eqn1**PropositionBoundLastBlockProof}
\end{align}
Combining~\eqref{eqn1*PropositionBoundLastBlockProof} and~\eqref{eqn1**PropositionBoundLastBlockProof}, we have
\begin{equation}
\rho=q. \label{eqn1***PropositionBoundLastBlockProof}
\end{equation}

It remains to prove~\eqref{stPropBoundOnLastBlock}. Using~\eqref{eqn1***PropositionBoundLastBlockProof} and the definition of $L$ in~\eqref{defL}, we have
\begin{align}
n-\rho L &= n-q  \lfloor \lambda n \rfloor,
\end{align}
which together with the definition of $d$ in~\eqref{defD} implies~\eqref{stPropBoundOnLastBlock}.

\section{Proof of Proposition~\ref{propositionKeyStep1}} \label{appendixD}
Consider the following chain of inequalities where all the probability and expectation terms are evaluated with respect to~$p_{E_1}p_{Z^L}$:
\begin{align}
&\Pr_{p_{E_1}p_{Z^L}}\left\{\left|\frac{1}{\Big\lceil\sqrt{L}\Big\rceil}\sum_{i=1}^{\left\lceil\sqrt{L}\right\rceil} \left(\sqrt{g^\Delta(E_1)}+Z_i\right)^2 -  g^\Delta(E_1)-1\right|\ge\Delta\right\}\notag\\
&\stackrel{\text{(a)}}{\le} \frac{4\E\left[g^\Delta(E_1)\right]+2}{\Delta^2 \Big\lceil\sqrt{L}\Big\rceil}\\
&\stackrel{\eqref{defFunctionG}}{\le}\frac{4\E\left[E_1\right]+2}{\Delta^2 \Big\lceil\sqrt{L}\Big\rceil}\\
&\stackrel{\eqref{defDeltaLinearInN}}{\le} \frac{1}{L^{1/6}}
\end{align}
where (a) follows from Chebyshev's inequality.

\section{Proof of Lemma~\ref{lemmaConstantEnergyArrival}} \label{appendixCorollaryAdaptiveCode}
Since $\Var_{p_{E_1}}[E_1]=0$ by assumption, it follows that $\E_{p_{E_1}}[E_1^2]=\tilde P^2$. Therefore, for any $\varepsilon>0$ and any sufficiently large~$L$ that satisfies
 \begin{equation}
L\ge \left(\log\left(\frac{2+\varepsilon}{\varepsilon^2}\right)\right)^4\label{assum1ProofOfLemmaConstantEH}
\end{equation}
and
\begin{equation}
 \frac{L}{\log L} \ge \max\left\{12\sqrt{2}, \frac{\mathrm{e}^{0.4}(2+\varepsilon)}{\varepsilon}\right\}, \label{assum2ProofOfLemmaConstantEH}
\end{equation}
we can use \cite[Th.~1]{FTY15} to conclude that there exists an~$\left(L- \left\lceil L^{2/3}\right\rceil + m, M, \varepsilon\right)$-code such that
\begin{align}
\log M &\ge  \left(L-\left\lceil L^{2/3}\right\rceil\right)\mathrm{C}(\tilde P) - \sqrt{\frac{(2+\varepsilon)\left(L-\left\lceil L^{2/3}\right\rceil\right) \tilde P}{\varepsilon(\tilde P+1)}} - \left(L-\left\lceil L^{2/3}\right\rceil\right)^{\frac{1}{4}}-1 \\
& \ge \left(L-\left\lceil L^{2/3}\right\rceil\right)\mathrm{C}(\tilde P) - \sqrt{\frac{3L}{\varepsilon}} - \left(L-\left\lceil L^{2/3}\right\rceil\right)^{\frac{1}{4}}-1\label{thmMainResultStatement**proof1}
\end{align}
where
\begin{equation}
m\triangleq \left\lceil 12\sqrt{3\sqrt{2}  \left(L-\left\lceil L^{2/3}\right\rceil\right)\log \left(L-\left\lceil L^{2/3}\right\rceil\right)} \right\rceil \label{thmMainResultStatement**proof2}
\end{equation}
 denotes the length of the initial saving period before any transmission occurs and~$L- \left\lceil L^{2/3}\right\rceil$ denotes the length of the actual transmission period. Let $\varepsilon\triangleq \frac{1}{\sqrt{L}}$ and fix a sufficiently large~$L$ that satisfies
\begin{equation}
L\ge \left(\log\left(2L+\sqrt{L}\right)\right)^4, \label{assum1LemmaConstantEnergyArrival}
\end{equation}
\begin{equation}
 \frac{L}{\log L} \ge \max\left\{12\sqrt{2}, \mathrm{e}^{0.4}\left(2\sqrt{L}+1\right)\right\} \label{assum2LemmaConstantEnergyArrival}
\end{equation}
and
\begin{equation}
(2- \sqrt{3})L^{3/4} \ge  \left(L-\left\lceil L^{2/3}\right\rceil\right)^{\frac{1}{4}}+1.\label{assum3LemmaConstantEnergyArrival}
\end{equation}
Since~\eqref{assum1ProofOfLemmaConstantEH} and~\eqref{assum2ProofOfLemmaConstantEH} hold by~\eqref{assum1LemmaConstantEnergyArrival} and~\eqref{assum2LemmaConstantEnergyArrival}, it follows from~\eqref{thmMainResultStatement**proof1} and~\eqref{thmMainResultStatement**proof2} that there exists an~$\left(L-\left\lceil \sqrt{L}\right\rceil, M, \frac{1}{\sqrt{L}}\right)$-code such that
\begin{align}
\log M &\ge \left(L-\left\lceil L^{2/3}\right\rceil\right)\mathrm{C}(\tilde P) - \sqrt{3}L^{3/4} - \left(L-\left\lceil L^{2/3}\right\rceil\right)^{\frac{1}{4}}-1\\
&\stackrel{\eqref{assum3LemmaConstantEnergyArrival}}{\ge} \left(L-\left\lceil L^{2/3}\right\rceil\right)\mathrm{C}(\tilde P) - 2L^{3/4}\\
&\stackrel{\eqref{defFunctionGamma}}{\ge}\gamma(L, \tilde P).
\end{align}


\section*{Acknowledgements}
The authors would like to thank Associate Editor Michele Wigger and the three anonymous reviewers for the useful comments that greatly improve the presentation of this work.

\begin{thebibliography}{10}
\providecommand{\url}[1]{#1}
\csname url@samestyle\endcsname
\providecommand{\newblock}{\relax}
\providecommand{\bibinfo}[2]{#2}
\providecommand{\BIBentrySTDinterwordspacing}{\spaceskip=0pt\relax}
\providecommand{\BIBentryALTinterwordstretchfactor}{4}
\providecommand{\BIBentryALTinterwordspacing}{\spaceskip=\fontdimen2\font plus
\BIBentryALTinterwordstretchfactor\fontdimen3\font minus
  \fontdimen4\font\relax}
\providecommand{\BIBforeignlanguage}[2]{{%
\expandafter\ifx\csname l@#1\endcsname\relax
\typeout{** WARNING: IEEEtran.bst: No hyphenation pattern has been}%
\typeout{** loaded for the language `#1'. Using the pattern for}%
\typeout{** the default language instead.}%
\else
\language=\csname l@#1\endcsname
\fi
#2}}
\providecommand{\BIBdecl}{\relax}
\BIBdecl

\bibitem{ZhangLau14}
F.~Zhang and V.~K.~N. Lau, ``Closed-form delay-optimal power control for energy
  harvesting wireless system with finite energy storage,'' \emph{{IEEE} Trans.
  Signal Process.}, vol.~62, no.~21, pp. 5706--5715, 2014.

\bibitem{davidTseBook}
D.~Tse and P.~Viswanath, \emph{Fundamentals of Wireless Communication}.\hskip
  1em plus 0.5em minus 0.4em\relax Cambridge, U.K.: Cambridge University Press,
  2005.

\bibitem{ozel12}
O.~Ozel and S.~Ulukus, ``Achieving {AWGN} capacity under stochastic energy
  harvesting,'' \emph{{IEEE} Trans. Inf. Theory}, vol.~58, no.~10, pp.
  6471--6483, 2012.

\bibitem{Han10}
T.~S. Han, \emph{Information-Spectrum Methods in Information Theory}.\hskip 1em
  plus 0.5em minus 0.4em\relax Springer Berlin Heidelberg, 2003.

\bibitem{Hayashi09}
M.~Hayashi, ``Information spectrum approach to second-order coding rate in
  channel coding,'' \emph{{IEEE} Trans. Inf. Theory}, vol.~55, no.~11, pp.
  4947--4966, 2009.

\bibitem{DavidIC}
R.~H. Etkin, D.~N.~C. Tse, and H.~Wang, ``Gaussian interference channel
  capacity to within one bit,'' \emph{{IEEE} Trans. Inf. Theory}, vol.~54,
  no.~12, pp. 5534--5562, Dec. 2008.

\bibitem{Ayfermono}
\BIBentryALTinterwordspacing
A.~\"{O}zg\"{u}r, O.~L\'{e}v\^{e}que, and D.~Tse, ``Operating regimes of large
  wireless networks,'' \emph{Foundations and Trends® in Networking}, vol.~5,
  no.~1, pp. 1--107, 2011. [Online]. Available:
  \url{http://dx.doi.org/10.1561/1300000016}
\BIBentrySTDinterwordspacing

\bibitem{RSV14}
R.~Rajesh, V.~Sharma, and P.~Viswanath, ``Capacity of {Gaussian} channels with
  energy harvesting and processing cost,'' \emph{{IEEE} Trans. Inf. Theory},
  vol.~60, no.~5, pp. 2563--2575, 2014.

\bibitem{shavivOzgur16-1}
D.~Shaviv and A.~\"{O}zg\"{u}r, ``Online power control for block i.i.d.
  {Bernoulli} energy harvesting channels,'' in \emph{Proc.\ IEEE Wireless
  Commun.\ and Networking Conference}, San Francisco, CA, USA, Mar. 2017.

\bibitem{ShavivOzgur16-2}
------, ``Online power control for block i.i.d. energy harvesting channels,''
  in \emph{to be presented in IEEE GLOBECOM}, Singapore, Dec. 2017.

\bibitem{MaoHassibi2013}
W.~Mao and B.~Hassibi, ``On the capacity of a communication system with energy
  harvesting and a limited battery,'' in \emph{Proc. IEEE Intl. Symp.
  Inf.~Theory}, Istanbul, Turkey, Jul. 2013, pp. 1789--1793.

\bibitem{Jog:ISIT:2014}
V.~Jog and V.~Anantharam, ``An energy harvesting {AWGN} channel with a finite
  battery,'' in \emph{Proc. IEEE Intl. Symp. Inf.~Theory}, Honolulu, HI, USA,
  Jun. 2014, pp. 806--810.

\bibitem{SNOzgur2016}
D.~Shaviv, P.-M. Nguyen, and A.~\"{O}zg\"{u}r, ``Capacity of the energy
  harvesting channel with a finite battery,'' \emph{{IEEE} Trans. Inf. Theory},
  vol.~62, no.~11, pp. 6436 -- 6458, 2016.

\bibitem{SOzgur2017}
D.~Shaviv and A.~\"{O}zg\"{u}r, ``A communication channel with random battery
  recharges,'' \emph{{IEEE} Trans. Inf. Theory}, vol.~63, no.~5, 2017.

\bibitem{FTY15}
S.~L. Fong, V.~Y.~F. Tan, and J.~Yang, ``Non-asymptotic achievable rates for
  energy-harvesting channels using save-and-transmit,'' \emph{{IEEE} J. Sel.
  Areas Commun.}, vol.~34, no.~12, pp. 3499 -- 3511, 2016.

\bibitem{ShenoySharma16}
K.~G. Shenoy and V.~Sharma, ``Finite blocklength achievable rates for energy
  harvesting {AWGN} channels with infinite buffer,'' in \emph{Proc. IEEE Intl.
  Symp. Inf.~Theory}, Barcelon, Spain, Jul. 2016, pp. 465 -- 469.

\bibitem{sha57}
C.~E. Shannon, ``Certain results in coding theory for noisy channels,''
  \emph{Information and Control}, vol.~1, pp. 6--25, 1957.

\bibitem{TomamichelTan14}
M.~Tomamichel and V.~Y.~F. Tan, ``Second-order coding rates for channels with
  state,'' \emph{{IEEE} Trans. Inf. Theory}, vol.~60, no.~8, pp. 4427--4448,
  2014.

\bibitem{elgamal}
A.~{El~Gamal} and Y.-H. Kim, \emph{Network Information Theory}.\hskip 1em plus
  0.5em minus 0.4em\relax Cambridge, U.K.: Cambridge University Press, 2012.

\bibitem{KorolevShevtsova10}
V.~Y. Korolev and I.~G. Shevtsova, ``On the upper bound for the absolute
  constant in the {Berry-Ess\'een} inequality,'' \emph{Theory of Probability
  and Its Applications}, vol.~54, no.~4, pp. 638–--658, 2010.

\bibitem{PPV10}
Y.~Polyanskiy, H.~V. Poor, and S.~Verd\'{u}, ``Channel coding rate in the
  finite blocklength regime,'' \emph{{IEEE} Trans. Inf. Theory}, vol.~56,
  no.~5, pp. 2307--2359, 2010.

\bibitem{Wang2009}
L.~Wang, R.~Colbeck, and R.~Renner, ``Simple channel coding bounds,'' in
  \emph{Proc. IEEE Intl. Symp. Inf.~Theory}, Seoul, Korea, Jul. 2009, pp. 1804
  -- 1808.

\bibitem{Pol10}
Y.~Polyanskiy, ``Channel coding: {Non}-asymptotic fundamental limits,'' Ph.D.
  dissertation, Princeton University, 2010.

\bibitem{Tan_FnT}
V.~Y.~F. Tan, ``Asymptotic estimates in information theory with non-vanishing
  error probabilities,'' \emph{Foundations and Trends in Communications and
  Information Theory}, vol.~11, no. 1-2, pp. 1--183, 2014.

\bibitem{FongTanJSAC16}
S.~L. Fong and V.~Y.~F. Tan, ``On the scaling exponent of polar codes for
  binary-input energy-harvesting channels,'' \emph{{IEEE} J. Sel. Areas
  Commun.}, vol.~34, no.~12, pp. 3540 -- 3551, 2016.

\end{thebibliography}


%
%
%
%
%




\end{document}